\documentclass[twocolumn]{aastex701}

\usepackage{amsmath}
\usepackage{amssymb}
\usepackage{graphicx}
\usepackage{natbib}
\usepackage{xcolor}
    \usepackage{siunitx}
\usepackage{enumitem}
\newcommand{\apogee}{\textsc{APOGEE}}
\newcommand{\gaia}{\textit{Gaia}}

\newcommand{\zsun}{\ensuremath{z_\odot}}

\newcommand{\abund}[2]{\ensuremath{[\mathrm{#1/#2}]}}
\newcommand{\mgfe}{\abund{Mg}{Fe}}
\newcommand{\feh}{\abund{Fe}{H}}
\newcommand{\mgmn}{\abund{Mg}{Mn}}
\newcommand{\alfe}{\abund{Al}{Fe}}
\newcommand{\mgh}{\abund{Mg}{H}}

\newcommand{\dd}{\ensuremath{\mathrm{d}}}
\newcommand{\teff}{\ensuremath{T_\mathrm{eff}}}
\newcommand{\logg}{\ensuremath{\log g}}
\newcommand{\snr}{S/N}

\definecolor{blue}{RGB}{0,0,150}

\begin{document}
\title{Using Element Abundances to Map the Midplane and Warp of the Milky Way Disk}

\author[0009-0003-4196-8659]{Emma Tasso}
\affiliation{The Graduate Center, City University of New York,
365 Fifth Avenue, New York, NY 10016, USA}
\email{emma.tasso99@gc.cuny.edu}

\author[0000-0003-0872-7098]{Adrian M. Price-Whelan}
\affiliation{Center for Computational Astrophysics, Flatiron Institute,
162 Fifth Avenue, New York, NY 10010, USA}
\email{aprice-whelan@flatironinstitute.org}

\author[0000-0003-3793-8505]{Eloisa Poggio}
\affiliation{INAF - Osservatorio Astrofisico di Torino, via Osservatorio 20, 10025 Pino Torinese (TO), Italy}
\email{eloisa.poggio@inaf.it}

\author[0000-0001-8917-1532]{Jason A. S. Hunt}
\affiliation{School of Mathematics \& Physics, University of Surrey, Stag Hill, Guildford GU2 7XH, UK}
\email{j.a.hunt@surrey.ac.uk}

\author[0009-0004-5306-4734]{Mauro Cabrera-Gadea}
\affiliation{Departamento de Astronomía, Instituto de Física, Universidad de la República, Iguá 4225, CP 11400 Montevideo, Uruguay}
\affiliation{Programa de Desarrollo de las Ciencias Físicas (PEDECIBA), Isidoro de María 1614, Montevideo, Uruguay}
\email{mauro.cabrera@pedeciba.edu.uy}

\author[0000-0003-1856-2151]{Danny~Horta}
\affiliation{Institute for Astronomy, University of Edinburgh, Royal Observatory, Blackford Hill, Edinburgh, EH9 3HJ, UK}
\email{dhorta@roe.ac.uk}

\author[0000-0002-4818-7885]{Jamie Tayar}
\affiliation{University of Florida, Department of Astronomy, Gainesville, FL, 32611 USA}
\email{jtayar@ufl.edu}



\author[0000-0003-3526-5052]{Jose G. Fernandez Trincado}
\affiliation{Instituto de Astronomía, Universidad Católica del Norte, Avenida Angamos 0610, Antofagasta, Chile}
\email{jose.fernandez@ubo.cl}

\author[0000-0001-5522-5029]{Carrie Filion}
\affiliation{Center for Computational Astrophysics, Flatiron Institute,
162 Fifth Avenue, New York, NY 10010, USA}
\email{cfilion@flatironinstitute.org}

\author[0000-0002-1423-2174]{Keith Hawkins}
\affiliation{Department of Astronomy, The University of Texas at Austin, 2515 Speedway Boulevard, Austin, TX 78712, USA}
\email{keithhawkins@utexas.edu}

\author[0000-0001-9852-1610]{Juna A. Kollmeier}
\affiliation{The Observatories of the Carnegie Institution for Science, 813 Santa Barbara Street, Pasadena, CA 91101, USA}
\email{jak@carnegiescience.edu}

\author[0000-0002-6164-5051]{Alexandre Roman Lopes}
\affiliation{Departamento de Física, Facultad de Ciencias, Universidad de La Serena, Cisternas 1200, La Serena, Chile}
\email{aroman@userena.cl}





\author[0000-0002-6972-6411]{José Eduardo Méndez Delgado}
\affiliation{ Instituto de Astronom\'ia, Universidad Nacional Aut\'onoma de M\'exico (UNAM), Apdo Postal 70-264, Ciudad de M\'exico, M\'exico}
\email{jmendez@astro.unam.mx}

\author[0000-0003-2178-8792]{Allyson Sheffield}
\affiliation{The Graduate Center, City University of New York,
365 Fifth Avenue, New York, NY 10016, USA}
\affiliation{City University of New York (CUNY)—LaGuardia Community College, Department of Natural Sciences, Long Island City, NY 11101, USA}
\email{asheffield@lagcc.cuny.edu}



\author[0000-0003-3769-8812]{Eleonora Zari}
\affiliation{INAF--Osservatorio Astrofisico di Arcetri, Largo Enrico Fermi 5, I-50125 Firenze, Italy}
\email{eleonora.zari@inaf.it}


\correspondingauthor{Emma Tasso}

\begin{abstract}
The Milky Way disk is warped and out of equilibrium.
The structure of the disk is often mapped using the density distribution of young stars or gas, which trace recent star formation and are sensitive to survey selection effects.
We instead define a ``chemical'' midplane from the dependence of mean stellar abundance with Galactic vertical position, which is much less sensitive to selection effects.
Because abundances label stellar orbits, the mean \mgfe\ at a given vertical position depends only on the vertical potential, so the optimum (here, minimum) corresponds to the potential midplane.
We apply this method to low-$\alpha$ red giant stars from SDSS-V/\apogee\ and map the chemical midplane out to Galactocentric radius $R = 18.5$~kpc, within $60^\circ$ of the Sun in azimuth.
Beyond $R \approx 10$~kpc, the midplane reaches $\sim\pm$0.5~kpc above and below $z = 0$.
We fit the midplane with a single $m=1$ warp model (twisted and constant phase) and find that our inferred line of nodes is offset by $\sim$16$^\circ$ from that of young tracers.
We also apply the method to two simulated Milky Way-like galaxies, one evolved in isolation and one with mergers that mimic the LMC and Sagittarius.
In the merger simulation, the chemical midplane tracks the density midplane, and young and old stars develop offset lines of nodes.
Because it only needs stellar positions and abundances, and no model of the survey selection function, this method can be applied to future spectroscopic surveys to map the warp across the full Milky Way disk.

\end{abstract}
\keywords{}


\section{Introduction}
\label{sec:intro}

The Milky Way disk is a dynamically complex structure, exhibiting large-scale departures from a simple, flat equilibrium configuration. Observations have revealed coherent vertical and radial distortions across the disk, including warps, flares, and corrugations \citep[e.g.,][]{Widrow:2012, Price-Whelan:2015, Xu:2015, Morganson:2016, Li:2017, Sheffield:2018, Romero-Gomez_2019, Laporte_2022, Drimmel:2023}.

The traditional view of the Galactic disk as a stable, time-independent system has been fundamentally challenged by data from the \gaia\ mission \citep{Gaia:2016, DR2-DPACP-36, GaiaDR3:2023}, which have revealed a broad range of substructures in the phase-space distribution of stars. Snail features \citep{Antoja:2018, Hunt:2022, Antoja:2023}, ridges \citep{Antoja:2018, Kawata:2018, Laporte:2020, Khanna:2019} and streaming motions \citep{Drimmel:2023, Vislosky:2024} have been mapped not only in the Solar Neighborhood, but across a significant portion of the Galactic disk. The lopsided and azimuthally complex nature of the large-scale warp was demonstrated by \citet{Romero-Gomez_2019} using kinematics, revealing that a simple symmetric warp model is insufficient to describe the observed vertical structure. Stellar vertical motions have been found consistent with a prograde precession of the Galactic warp, suggesting that it is a dynamically evolving feature \citep{Poggio:2020,Cheng_2020,Dehnen_2023,JonssonMcMillan2024}. \citet{Laporte_2022} used \gaia\ proper motions to uncover a hierarchy of previously unknown filamentary overdensities embedded in the Galactic anticentre at heliocentric distances $d \geq 10$ kpc, interpreted as excited disk material kicked up by satellite impacts.

Observations have shown that the Galactic midplane exhibits not only a large-scale warp, but also vertical corrugations. At low Galactic latitudes, a ring of stars has been mapped towards the Galactic anticentre \citep{Newberg:2002}; this feature is alternatively referred to as the Monoceros ring, Monoceros stream, or Galactic anti-centre stellar stream \citep{Rocha-Pinto:2003}. Using data from SDSS, \citet{Xu:2015} found an oscillating asymmetry in the main sequence star counts for patches of sky above and below the Galactic plane in the direction of the Galactic anticentre. \citet{Morganson:2016} mapped the three-dimensional structure of the so-called Monoceros ring using the Pan-STARRS1 survey, finding that it is significantly closer to the Sun in the South (6 kpc) than in the North (9 kpc). At even larger distances, the Triangulum Andromeda stream has been detected \citep{Majewski:2004, Rocha-Pinto:2003, Price-Whelan:2015}. Using young giant stars and classical Cepheids, \citet{Poggio_2025} detected a large-scale vertical corrugation on top of the Galactic warp, extending over at least 10 kpc at Galactocentric radii of $R \sim 10$--$14$ kpc, with vertical heights of $\sim$150--200 pc. The kinematics of this feature are consistent with a wave propagating outward through the disk. Vertical corrugations on top of the Galactic warp have also been detected in the outer CO disk, using molecular clouds from the Milky Way Imaging Scroll Painting survey \citep{Sun:2026}. Together, these results describe a disk that is constantly reshaped by internal physics and external interactions, presumably in disequilibrium on large scales. Characterizing these features requires a well-defined reference frame, such as disk midplane, yet the location and structure of the midplane itself remain poorly constrained across the full extent of the Galaxy.


Studies of the Galactic midplane and warp have used both young and old tracers, but the warp is most clearly seen in young populations. For older stellar populations, the shape of the midplane is much less well constrained, because of extinction, distance uncertainties and selection effects. Classical Cepheids have proved especially powerful, offering precise distances ($\sim$3\%) and cold kinematics that closely reflect the gas from which they formed \citep{Cabrera-Gadea_2024, Skowron_2019a, Skowron_2019b, Chen_2019, Dehnen_2023, Poggio_2025}. Using Fourier decomposition of both the vertical height and vertical velocity of Cepheids, \citet{Cabrera-Gadea_2024} characterize the warp's structure, asymmetry, and prograde pattern speed, revealing a lopsided, twisted line of nodes that a simple tilted rings model cannot reproduce. \citet{Cabrera-Gadea_2025} extended this to RR Lyrae stars with thin-disk kinematics, showing that this population also traces a warped disk, with a warp signal systematically offset from that of Cepheids, opening a new window on the age dependence of the warp. \citet{Poggio_2026} used young giant stars to map the disk flare and spiral structure out to $\sim$8 kpc, finding an exponential flare with a scale length of $h_{fl} = 3.5 \pm 0.3$ kpc. The spatial distribution of luminous young stars has also been mapped using upper main sequence and OB stars \citep{Zari_2021}, and extended using early-type stars with kinematics from SDSS-V \citep{Zari_2025}. AGB stars have recently been used as well \citep{Kushwahaa_2026}. While powerful, many of these tracers are inherently tied to sites of recent star formation, making them not ideal for characterizing the smooth, vertical structure of the older stellar disk, which contains the bulk of mass in the disk.

Efforts to map the midplane and warp using older stellar populations have largely relied on photometric tracers, particularly red clump stars, which serve as reliable standard candles in near-infrared surveys. Early work by \citet{Lopez-Corredoira_2002} using 2MASS photometry established the large-scale warp and flare of the old disk, finding the disk begins to deviate from the plane beyond $R \sim 6$ kpc. \citet{Reyle_2009} similarly used 2MASS star counts to constrain the outer disk structure. \citet{Uppal_2024} and \citet{Khanna_2025} have extended these measurements with larger red clump samples, finding a prominent flare in the old disk population with amplitudes substantially larger than those seen in young tracers, though these estimates may be biased upward if the warp's vertical displacement is not accounted for when fitting the scale height \citep{Poggio_2026}.

More recently, \citet{Sun_2025} demonstrated that a north-south asymmetry in the \feh\ distribution of red clump stars traces the disk warp, suggesting that chemical information encodes large-scale vertical structure. However, a systematic characterization of a chemically defined midplane grounded in the vertical distribution of $\alpha$-element abundances across a wide range of Galactocentric radii has not yet been carried out. Chemical abundances offer a complementary way to map the disk. A star's surface abundances largely reflect the gas it formed from, so stars with similar abundances share a formation history and orbit the Galaxy in similar ways. The vertical structure of the abundance distribution therefore traces the plane about which the disk is symmetric, and it does so without needing a complete star-count sample or a model for the selection function, which affect direct density mapping. It also lets us trace the midplane of the older disk, for which young tracers such as classical (Type I) Cepheids are not appropriate.


In this work, we define and map the chemical midplane of the Milky Way disk using old red giant branch (RGB) stars from the SDSS-V \apogee\ survey using distances from \citet{Zhang_2023}.
We select stars in the low-$\alpha$ disk population and use the vertical distribution of \abund{Mg}{Fe} to identify the midplane as the height at which \abund{Mg}{Fe} reaches an optimum at each position in the disk.
Because surface abundances label orbits \citep[e.g.,][]{price-whelan_2021}, the location of the optimum or inflection point traces the plane of symmetry of the vertical potential and does not depend on the survey's spatial selection function (unlike most approaches that measure the location of density maximum).
We test the method on simulated Milky Way-like galaxies, fit the resulting midplane map with an $m=1$ warp model, and compare it with warp measurements from other tracers.

\section{Data}
\label{sec:data}

We use spectroscopic data from the Sloan Digital Sky Survey (SDSS) and, indirectly,
astrometric data from the \gaia\ Mission.

For spectroscopic data, we use data collected with the \apogee\ spectrographs as a part
of the \apogee\ surveys \citep[SDSS-III, SDSS-IV;][]{Majewski:2017,
Eisenstein:2011,Blanton:2017} and the \textit{Milky Way Mapper} (MWM) survey
\citep[SDSS-V;][]{Kollmeier:2026}.
The \apogee\ spectrographs are two high-resolution (resolution $R \sim 22,500$),
multi-fiber spectrographs \citep{Wilson:2019} on the 2.5 m Sloan telescope at Apache
Point Observatory \citep[APO;][]{Gunn:2006} and the 2.5 m du Pont telescope at Las
Campanas Observatory \citep[LCO;][]{Bowen:1973}.
SDSS-III and -IV \apogee\ observations used optical fibers plugged into pre-drilled aluminum plates.
SDSS-V uses a robotic fiber positioning system for observations and target acquisition \citep{Sayres:2021}.
The target selection for \apogee\ in SDSS-III and -IV is described in a series of papers
\citep{Zasowski:2013, Zasowski:2017, Beaton:2021, Santana:2021}.
For SDSS-V MWM, the target selection and targeting strategy is more complex and adaptive
\citep{DR18:2023,Blanton:2026}.

We use an internal release (0.8.0) that includes all \apogee\ spectra up to 2025-02-02 (MJD 60708), including all data from the \apogee\ surveys and the MWM \apogee\ data.
This enables precise stellar abundance measurements that we use in this work.
We use element abundances from the \textsc{ASPCAP} pipeline \citep{GarciaPerez:2016, Meszaros:2025}.

To compute stellar positions in the Galaxy, we use sky positions from the \gaia\ Mission
\citep{Gaia:2016, GaiaDR3:2023, Babusiaux:2023} and spectro-astro-photometric ``ZGR''
distance information from \citet{Zhang_2023}.
The ZGR parallaxes are more precise for giant stars at larger distances (beyond $d
\gtrsim 2~\unit{kpc}$) --- our method and results rely on modeling spatial volumes that
are beyond the horizon within which \gaia\ parallaxes alone provide precise estimates of
distance.
The ZGR parallax estimates come from a data-driven model of the \gaia\ (DR3) XP spectra
\citep{DeAngeli:2023}, \textsc{2MASS} photometry \citep{Skrutskie:2006}, and
\textsc{WISE} photometry \citep{Wright:2010,Schlafly:2019} and use the \gaia\ parallaxes
for nearby stars to effectively train their model.
For a typical RGB star at $d \approx 8~\unit{kpc}$, for example, the
median \gaia\ parallax signal-to-noise (\snr) for stars in our sample is $\varpi /
\sigma_\varpi \approx 3$ whereas the median ZGR parallax \snr\ is $\varpi / \sigma_\varpi
\approx 15$.

\subsection{Sample Selection}
\label{sec:sample}

Our parent sample contains 2,199,959 sources from Milky Way Mapper.
We select RGB stars in the low-$\alpha$ disk with high \snr\ spectra and with no
catastrophic data quality flags.
In more detail, we impose the following filters on our parent sample:

\begin{figure*}
    \centering
    \includegraphics[width=\textwidth]{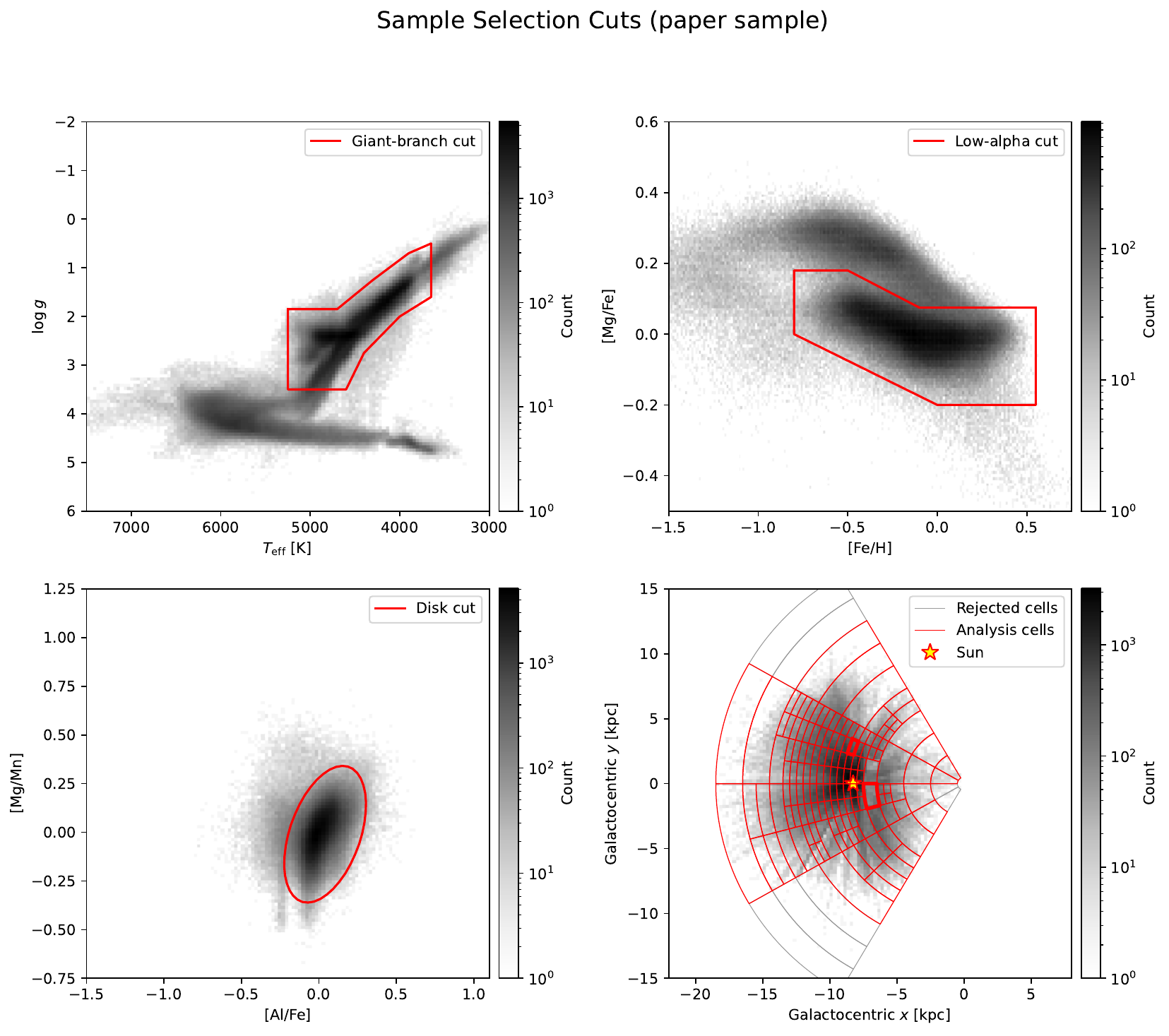}
    \caption{Sample selection cuts applied to the \apogee\ data after quality cuts. Gray shading shows star counts and red lines mark each cut. Each panel after the first includes only stars that passed the cut in the previous panel. Top left: giant-branch cut in $T_{\rm eff}$ and $\log g$, shown for stars that pass all quality cuts except this one. Top right: low-alpha cut in \mgfe\ and \feh, which selects the low-alpha sequence. Bottom left: disk cut in \mgmn\ and \alfe, which removes halo contaminants. Bottom right: Galactocentric $X$--$Y$ distribution of the sample with the analysis cells (red) and rejected cells (grey). The yellow star marks the Sun. The bolded cells are the two cells shown in Figure~\ref{fig:mgfe_vs_z_two_bins}.}
    \label{fig:sample_selection_cuts}
\end{figure*}

\begin{enumerate}
    \item To select red giant branch (RGB) stars, we use a polygon in \teff\ and \logg, as shown in the top left panel of Figure~\ref{fig:sample_selection_cuts}. The polygon spans $3650 \leq \teff \leq 5250$~K and $0.5 \leq \logg \leq 3.5$. We define this by eye using the Kiel diagram of the parent sample (after quality cuts) to include the overdensity of the red giant branch, excluding subgiants and dwarfs at higher \logg and more luminous RGB stars at lower \logg.

    \item We apply three quality requirements to the Astra ASPCAP output. (a) The spectral signal-to-noise ratio must satisfy S/N $>10$, taken from the Astra \texttt{snr} column. (b) The ASPCAP \texttt{flag\_bad} flag must not be set. This flag is built from the ASPCAP \texttt{result\_flags} and is set if the fit shows any sign of failure or unreliability, such as suspicious or unphysical parameters, a high reduced $\chi^2$, high $v\sin i$ or microturbulence, a stellar parameter at the edge of the model grid, a failed FERRE fit, low S/N, or a spectrum read error.\footnote{The full list of conditions is given in the SDSS-V Astra ASPCAP documentation (\url{https://www.sdss.org/dr20/mwm/astra/pipelines-in-astra/aspcap/}).} This is the closest analogue of the DR17 \texttt{STAR\_BAD} flag. (c) The \feh\ and \mgh\ abundances must carry no flags (\texttt{fe\_h\_flags}$=0$ and \texttt{mg\_h\_flags}$=0$), which excludes, for example, censored values, abundances at or near the edge of the model grid, and abundances flagged as unreliable at that effective temperature.

    \item We reject stars with unreliable photo-astrometric distance and extinction solutions, using the quality flags from the \citet{Zhang_2023} catalog. We remove stars with the \texttt{ln\_prior} bit set (the logarithm of the stellar-type prior at the best-fit solution is below $-7.43$) or the parallax bit set (the model parallax is more than $10\sigma$ from the \gaia\ DR3 measurement). We do not reject stars with the $\chi^2$ bit set ($\chi^2_{\rm opt}/61 > 2$). For high signal-to-noise stars, a large $\chi^2$ can come from underestimated flux uncertainties rather than a poor fit, so cutting on this goodness-of-fit could remove many usable stars (about 30\% of our sample has $\chi^2_{\rm opt}/61 > 2$).

    \item We remove stars flagged as likely members of known star clusters, using the \texttt{sdss4\_apogee\_member\_flags} bitmask from the SDSS-V Data Release 19 Astra catalog, which is inherited from the SDSS-IV \apogee\ data reduction. We keep only stars with no bit set.

    \item We select stars to be a part of the ``low-$\alpha$ disk'' using a polygonal selection in \abund{Mg}{Fe} and \abund{Fe}{H}, as visualized in the top right panel of Figure~\ref{fig:sample_selection_cuts}. The polygon spans $-0.8 < \abund{Fe}{H} < 0.55$ and $-0.2 < \abund{Mg}{Fe} < 0.18$. We set the upper edge by eye to follow the gap between the low- and high-$\alpha$ sequences. It is flat at $\abund{Mg}{Fe} = 0.18$ for $\abund{Fe}{H} < -0.5$, falls to $0.075$ at $\abund{Fe}{H} = -0.1$, and stays at $0.075$ at higher \abund{Fe}{H}.

    \item We further remove possible accreted star contaminants using an elliptical selection in \abund{Al}{Fe} and \abund{Mg}{Mn} \citep[e.g.,][]{Hawkins_2015, Horta_2021, Das_2020}, as visualized in the bottom left panel of Figure~\ref{fig:sample_selection_cuts}. We define this by eye to enclose the overdensity of in-situ disk stars in this plane.

    \item We require finite, positive ZGR parallaxes (\texttt{zgr\_plx}~$>0$) and finite Galactocentric positions.

    \item In each spatial cell (see Section~\ref{sec:spatial_binning}), we keep only stars that would still be brighter than the \gaia\ XP limit ($G = 17.5$) at the 95th percentile distance modulus and $A_G$ of that cell. This is effectively a cut on intrinsic luminosity, which makes the selection function very close to separable within in each cell (Appendix~\ref{app:separable-selfunc}).
\end{enumerate}
Our final sample size contains 438{,}499 stars.
The bottom right panel of Figure~\ref{fig:sample_selection_cuts} shows their distribution in Galactocentric $X$--$Y$ together with the final cells used in our analysis (Section~\ref{sec:spatial_binning}).

The most important quantities for our analysis below are the distance (parallax)
measurements and the \abund{Mg}{Fe} abundance measurements.

Over 99\% of our final sample has fractional parallax uncertainties below 20\%. Because 99\% of the stars lie within $d \approx 8~\unit{kpc}$, the distance uncertainties for stars below the 20\% threshold are at most $\sigma_d \approx 1.5~\unit{kpc}$. The median and 95th percentile distance uncertainties for our sample are 0.08~kpc and 0.50~kpc.
To compute distance from the ZGR parallaxes, we invert the parallaxes $d = 1/\varpi$.

We transform stellar positions to a Galactocentric frame using \texttt{astropy} with $R_\odot = 8.275~\unit{kpc}$ and place the Sun at $z=0$ (\texttt{z\_sun = 0}) in this transformation. The vertical offset of the global midplane, $z_{\rm off}$, is a parameter of our warp model, and we derive the Sun's height above the local midplane from this parameter and the fitted warp model (Section~\ref{sec:lon}).
Galactic azimuth $\phi$ is measured from the Sun--Galactic center line, with the Sun at $\phi = 0^\circ$ and $\phi$ increasing clockwise, in the direction of Galactic rotation.


\section{Methods} \label{sec:model_section}

The goal of this work is to infer the midplane position of the
low-$\alpha$ (``thin'') disk of the Milky Way as a function of Galactocentric
cylindrical radius and azimuth, $(R, \phi)$.
Commonly, the midplane is defined as the plane of symmetry of the disk's vertical
density distribution, or the location of maximum stellar density.
However, we do not measure these quantities directly from the stellar number density
because the observed density is affected by survey selection effects (especially in
spectroscopic surveys like SDSS-V) and dust extinction.
Instead, by ``midplane position,'' here we mean the height $z_0$ at which the vertical
acceleration vanishes, as we explain below (Section~\ref{sec:conceptual_framework}).
We use correlations between stellar abundances and orbital properties to define a
``chemical midplane'' that weakly depends on the spatial selection function or
extinction, under the assumptions in Section~\ref{sec:conceptual_framework}.

Our approach relies on the same principle behind Orbital Torus Imaging \citep[OTI;][]{price-whelan_2021,price-whelan_2025}. Because stellar surface abundances are effectively invariant on orbital timescales, they act as labels of orbits. In a phase-mixed disk, this means \mgfe\ can only depend on orbital action, not on orbital phase. As a result, the mean \mgfe\ as a function of vertical position $z$ depends on $z$ only through the vertical potential, and is therefore symmetric about the midplane. Here we first develop the conceptual framework for this approach and show that, under a set of assumptions, the minimum of the mean \mgfe\ profile with $z$ coincides with the midplane of the disk, and that the width of the profile relates to the vertical scale of the potential (Section~\ref{sec:conceptual_framework}). We then describe the statistical model we use to infer the midplane position and scale height from the observed \mgfe--$z$ distribution in each $(R, \phi)$ cell (Section~\ref{sec:model}).

\subsection{Conceptual Basis}
\label{sec:conceptual_framework}

Here we use a deliberately simple model to show why the minimum of the mean
\mgfe\ profile with vertical position $z$ coincides with the midplane of the disk.
We work within a single $(R, \phi)$ cell, consider only vertical kinematics, and define
$Y \equiv \mgfe$ for brevity.
We also make the following assumptions:
\begin{description}
    \item[Locally symmetric vertical potential] Vertical motion
    approximately decouples from motion in the plane, so each star conserves its
    vertical energy $E_z = \tfrac{1}{2}v_z^2 + \Phi_z(z)$ where $\Phi_z(z)$ is the local
    vertical potential. Within the cell, $\Phi_z(z)$ is mirror-symmetric about a height
    $z_0$ and increases monotonically in either direction. We take this plane of
    symmetry, where the vertical force $a_z = -\partial\Phi_z/\partial z$ vanishes, as
    the definition of the midplane.
    \item[Vertically phase mixed] The stars in our sample are
    phase-mixed in this potential, so their distribution in $(z, v_z)$ depends only on
    $E_z$.
    \item[Abundances label orbits] A star's surface \mgfe\ does not change appreciably
    over a vertical orbit. Combined with the phase-mixed assumption, the joint
    distribution of stars in $(z, v_z, Y)$ is then $f = F(E_z, Y)$ for some function
    $F$ that correlates \mgfe\ with vertical energy.
    \item[$Y$ depends monotonically on vertical energy] Stars on orbits with larger
    $E_z$ have, on average, higher \mgfe. We make no assumption about the origin of
    this correlation (whether from scattering/heating or set at formation), only that it
    exists.
    \item[Selection effects separate] The probability
    that a star enters our sample (and SDSS-V MWM) can be written as $S_z(z) \times
    S_Y(Y)$ and does not depend on $v_z$. Here $S_z$ includes the survey footprint,
    targeting, and dust extinction, and $S_Y$ includes our abundance-based sample cuts
    (Section~\ref{sec:sample}). In detail, there is a small cross-term because
    abundances do subtly affect the apparent magnitude and color of a star (the dominant
    selection criteria for \apogee\ stars), but we ignore it here (see Section~\ref{sec:assumptions}). We do not need to
    know either selection function.
\end{description}
We discuss how each of these assumptions can break down in
Section~\ref{sec:assumptions}.

Under these assumptions, the mean \mgfe, $\langle Y \rangle$, of the stars we observe
at height $z$, is
\begin{align}
    \langle Y\rangle(z)
    &= \frac{\int\! \dd v_z \int\! \dd Y\, Y\, S_z(z)\, S_Y(Y)\, F(E_z, Y)}
            {\int\! \dd v_z \int\! \dd Y\, S_z(z)\, S_Y(Y)\, F(E_z, Y)} \nonumber \\
    &= \frac{\int\! \dd v_z \int\! \dd Y\, Y\, S_Y(Y)\, F(E_z, Y)}
            {\int\! \dd v_z \int\! \dd Y\, S_Y(Y)\, F(E_z, Y)} \quad ,
    \label{eq:meanY}
\end{align}
where the spatial selection function, $S_z(z)$, cancels because it is constant at fixed
$z$.

\begin{figure}
    \centering
    \includegraphics[width=\columnwidth]{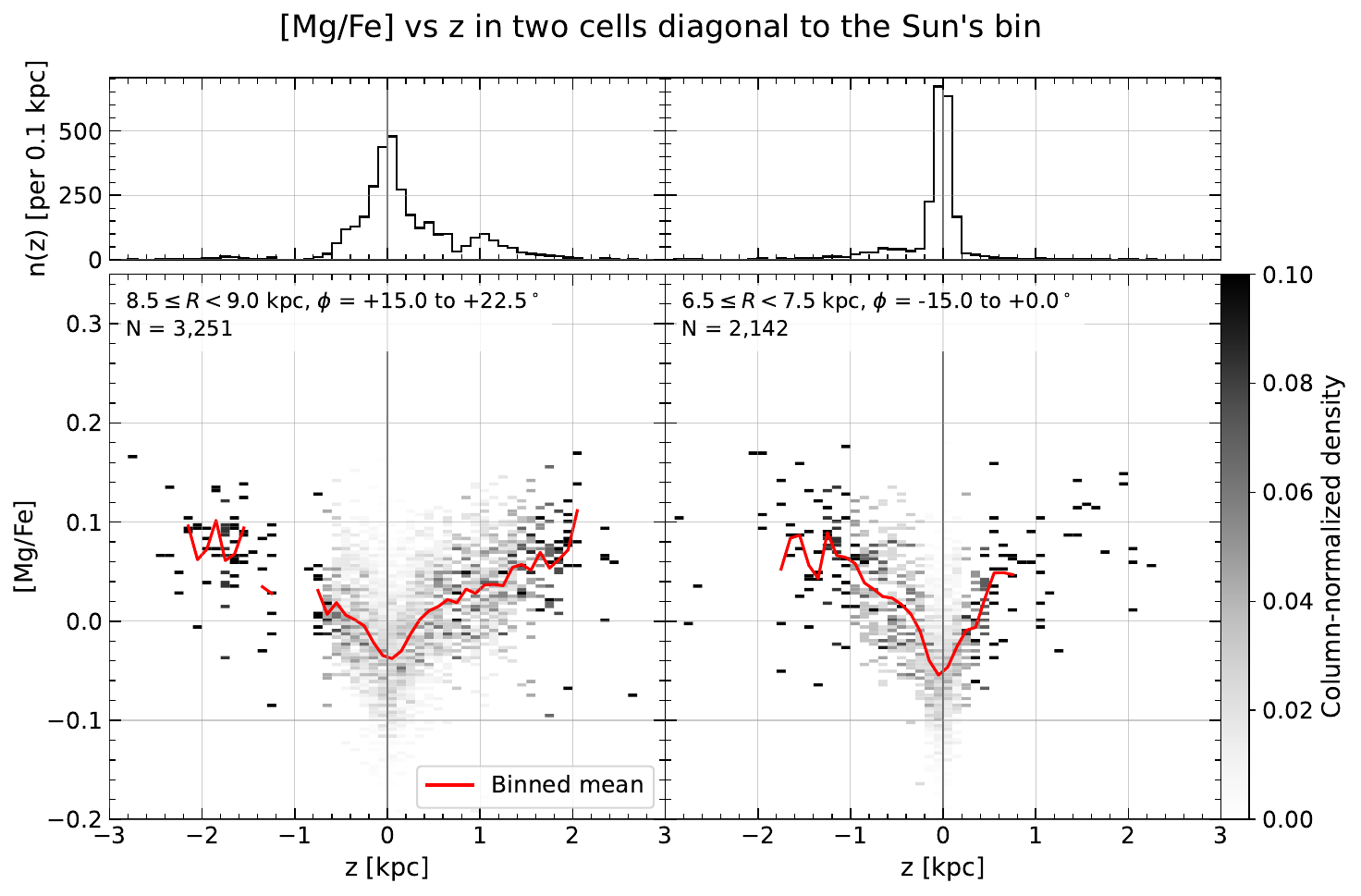}
    \caption{\mgfe\ vs $z$ in two cells diagonal to the Sun's cell. Top panels show the number of stars per 0.1~kpc bin in $z$. Bottom panels show the column-normalized stellar density in \mgfe\ for each $z$ bin (gray shading) and the binned mean \mgfe\ (red line). The gray vertical line marks $z = 0$. The $R$ and $\phi$ range and number of stars $N$ of each cell are given in the bottom panels.}
    \label{fig:mgfe_vs_z_two_bins}
\end{figure}

The right-hand side of Equation~\ref{eq:meanY} only depends on $z$ through $E_z$,
through $\Phi_z(z)$, so $\langle Y\rangle(z) = \mathcal{Y}\left(\Phi_z(z)\right)$ for
some function $\mathcal{Y}$ that is set by the chemo-dynamical history of the disk
(through $F$) and by our sample cuts (through $S_Y$).
First, because $\Phi_z$ (under our assumptions above) is symmetric about $z_0$, $\langle
Y\rangle(z)$ is also symmetric about $z_0$, so $\langle Y\rangle(z)$ has an extremum at $z_0$.
Because $Y$ increases with $E_z$, this extremum is a minimum (for \mgfe).
Second,
\begin{equation}
    \frac{\dd\langle Y\rangle}{\dd z} = \mathcal{Y}'\,\frac{\dd\Phi_z}{\dd z}
        = -\mathcal{Y}' \, a_z(z),
    \label{eq:dYdz}
\end{equation}
which vanishes at $z_0$ for any $\mathcal{Y}$.
The location of the extremum is therefore set by the potential, while the
chemo-dynamical history of the disk (through $\mathcal{Y}$) affects its depth and width.
Figure~\ref{fig:mgfe_vs_z_two_bins} shows the observed \mgfe--$z$ distribution and
binned mean in two cells diagonal to the Sun's cell. In both cells the mean \mgfe\ dips
to a minimum and rises on either side, and the minimum marks the chemical midplane.

\subsection{Spatial Binning} \label{sec:spatial_binning}

To characterize the vertical structure of the disk as a function of position, we divide our sample into spatial cells in Galactocentric radius $R$ and azimuth $\phi$.
Each cell is an annular segment ($R_{\rm lo} \le R < R_{\rm hi}$ and $\phi_{\rm lo} \le \phi < \phi_{\rm hi}$), and each star belongs to at most one cell (i.e., they are non-overlapping).

The number of stars per unit area in our sample varies by two orders of magnitude across the disk, and the distance uncertainties grow from $\sim$10~pc near the Sun to $\gtrsim$1~kpc at $d \sim 10$~kpc (Section~\ref{sec:sample}).
If we used equal bin sizes, we would either under-resolve near the Sun or over-resolve in the outer disk.
Also, in the outer disk, distance errors would scatter stars between small neighboring cells.
We instead set the cell sizes adaptively with a quadtree.
We start from root cells with size $\Delta R = 2$~kpc and $\Delta\phi = 30^\circ$ that cover $0.5 \le R < 18.5$~kpc and $|\phi| < 60^\circ$.
A cell is divided into four ``child'' cells, by halving it in both $R$ and $\phi$ if every child cell is both:
\begin{itemize}
    \item \textbf{usable:} it contains at least 32 stars (an arbitrary threshold), and its stars sample the vertical profile on both sides of the plane.
    We divide each of the ranges $-1.0 < z < -0.1$~kpc and $0.1 < z < 1.0$~kpc into four equal bins in $z$, and at least 3 of the 4 bins on each side must hold 4 or more stars.
    We only count stars with a finite, positive \mgfe\ uncertainty.
    \item \textbf{resolved:} the smaller of its radial width, $\Delta R$, and its arc length at its central radius, $\bar{R}\,\Delta\phi$, is at least the median distance uncertainty, $\sigma_d = d\,\sigma_\varpi/\varpi$, of its stars.
\end{itemize}
We allow at most two divisions, which gives a minimum cell size of 0.5~kpc by $7.5^\circ$.
The resulting cells are small near the Sun, where there are more stars observed by MWM and distances are measured precisely.
The cells are larger in the outer disk, where the \apogee\ footprint is sparse and distances are more uncertain (bottom right panel of Figure~\ref{fig:sample_selection_cuts}).
Because the criteria depend only on the positions of the stars and their distance uncertainties, the assignment of stars to cells does not depend on the quantity we measure (moments of the \mgfe\ distribution conditioned on position).
We require coverage on both sides of the plane because a cell whose stars lie mostly on one side of $z = 0$ would not enable measuring the midplane location.

We exclude cells with root cells that contain few stars in $z$; this leaves 143 cells.
Within each usable cell, the fit (Section~\ref{sec:model}) also requires at least 8 bins of 0.1~kpc in $|z| < 1.5$~kpc to have 5 or more stars (142 cells meet this).
For each cell we also record the mean position of its stars, $(\bar{R}, \bar{\phi})$, which we use in the warp fit (Section~\ref{sec:lon}).

\subsection{An effective model of the vertical abundance trend} \label{sec:model}

We trace the midplane with \mgfe\ rather than \abund{Mg}{H} or \abund{Fe}{H} because the gradient of \mgfe\ is more closely tied to the vertical orbital structure of the disk \citep{Horta_2024}, and a warp signal from the analogous [$\alpha$/M] gradient is stronger than from the metallicity gradient (Cabrera-Gadea et al., in prep.).

Within each $(R,\phi)$ cell, we bin stars in vertical height $z$ in bins of size 0.1~kpc.
We compute the per-star \mgfe\ uncertainty as $\sqrt{\sigma_{\rm [Mg/H]}^2 + \sigma_{\rm [Fe/H]}^2}$, and the \textsc{ASPCAP} uncertainties are small, with a median of 0.004~dex and a 95th percentile value of 0.010~dex.
In each $z$-bin, we compute the inverse-variance-weighted mean \mgfe.
We adopt as its uncertainty the formal error on the mean added in quadrature to the standard error from the observed scatter of the stars in the bin ($\sigma/\sqrt{N}$), so that the per-bin uncertainty reflects the intrinsic scatter of the stars contributing to it. This is likely overly conservative.
We model the resulting $[\mathrm{Mg/Fe}]$-versus-$z$ trend, $f(z)$, as the sum of two sech$^2$ components,
\begin{equation}
    f(z) = A_1\,\mathrm{sech}^2\!\left(\frac{z-z_0}{h_1}\right) + A_2\,\mathrm{sech}^2\!\left(\frac{z-z_0}{h_2}\right) + Y_0,
    \label{eq:sech2}
\end{equation}
where $A_1$ and $A_2$ are the (negative) amplitudes of the two components, $z_0$ is their shared vertical center, and $h_1$ and $h_2$ are two characteristic widths of the dip. We identify $z_0$, the height at which the model $[\mathrm{Mg/Fe}]$ reaches its minimum, as the chemical midplane height at that $(R,\phi)$ position (the density-weighted mean position of the stars within each cell).
We note that we use this model form just as an effective model. The two components are not meant to represent two disk components, as we have already selected low-$\alpha$ stars using the abundance selections described in Section~\ref{sec:sample}. Instead, including two components is a choice that allows for more flexibility in the fit.
We assume that each binned mean \mgfe\ is normally distributed about the model, conditional on height $z$ and the model parameters $\theta = (A_1, A_2, h_1, h_2, z_0, Y_0)$,
\begin{equation}
    p\left([\mathrm{Mg/Fe}] \mid z, \theta\right) = \mathcal{N}\!\left([\mathrm{Mg/Fe}] \mid f(z; \theta), \sigma_{\rm obs}^2\right),
\end{equation}
where $\sigma_{\rm obs}$ is the per-bin uncertainty described above.

\begin{figure}
    \centering
    \includegraphics[width=\columnwidth]{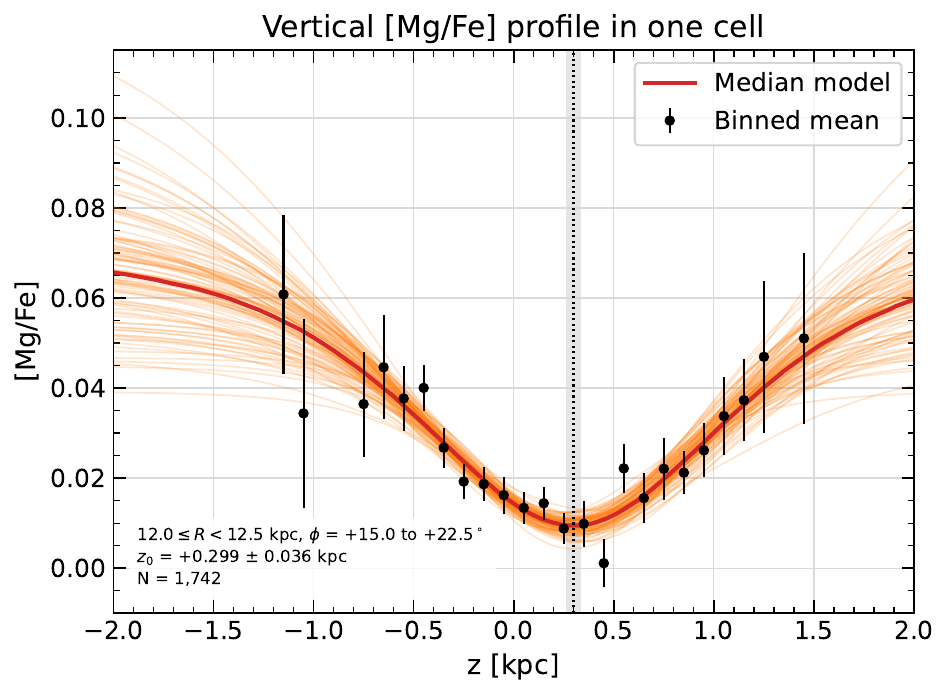}
    \caption{Vertical \mgfe\ profile in the cell at $12.0 \leq R < 12.5$~kpc and $15^\circ \leq \phi < 22.5^\circ$ ($N = 1{,}742$ stars). Black points show the inverse-variance weighted mean \mgfe\ in 0.1 kpc bins in $z$. Error bars combine the propagated measurement uncertainty with the standard error of the mean from the scatter within each bin. The red line shows the median of the two-component sech$^2$ model, and the orange lines are 100 random draws from the posterior. The dotted vertical line marks the median midplane height, $z_0 = +0.299$~kpc, and the gray band shows its $1\sigma$ uncertainty of 0.036~kpc.}
    \label{fig:sech2_fit_sun_bin}
\end{figure}

Figure~\ref{fig:sech2_fit_sun_bin} shows the fit in one cell, and Figure~\ref{fig:sech2_fits_all_bins} (Appendix~\ref{app:model-validation}) shows the fits in all cells.
In Figure~\ref{fig:sech2_fit_sun_bin} and the panels of Figure~\ref{fig:sech2_fits_all_bins}, the black markers are the data, the solid (red) line is computed from the median posterior sample values of the parameters, and the other lower-opacity (orange) lines show models computed from 100 random posterior samples; The vertical line marks the inferred (posterior) midplane location in this cell.

We do not include an outlier component in the likelihood.
In tests, a free outlier fraction tends to absorb real structure near the bottom of the mean abundance trend, especially in bins with many stars (small uncertainties).
A future extension of our model may do better by including a flexible way of modeling asymmetries, but here we only use a symmetric model as is standard in other midplane and warp-modeling work.

We adopt weakly informative priors on the model parameters, summarized in Table~\ref{tab:priors}. In particular, $h_2$ is parameterized as $h_2 = h_1 + \Delta h_2$, with $\Delta h_2 \sim \mathrm{Uniform}(0.04, 0.75)$ kpc setting the separation between the two components' characteristic widths.
We parameterize the model this way so that $h_2 > h_1$ and we avoid a degeneracy in which components 1 and 2 are swapped.
We set the upper bound of the $\Delta h_2$ prior to 0.75~kpc so that the model can capture both a sharp central dip and its broader wings. With a tighter bound, the posterior piles up against the prior boundary in some cells and the model does not reach the observed trough depth.
As we do not try to interpret the width or other properties of the mean abundance trend in this work, we feel that this is safe.

\begin{deluxetable}{lllc}
\tablecaption{Model parameters and priors \label{tab:priors}}
\tablehead{
\colhead{Parameter} & \colhead{Description} & \colhead{Prior} & \colhead{Units}
}
\startdata
$A_1$           & Amplitude of component 1   & $-\mathrm{HalfNormal}(0.2)$          & dex \\
$A_2$           & Amplitude of component 2   & $-\mathrm{HalfNormal}(0.2)$          & dex \\$h_1$         & Width of component 1       & $\mathrm{Uniform}(0.04, 1.0)$       & kpc \\
$\Delta h_2$  & $h_2 - h_1$           & $\mathrm{Uniform}(0.04, 0.75)$      & kpc \\
$z_0$         & Midplane height      & $\mathcal{N}(0.0, 0.5)$             & kpc \\
$Y_0$         & Baseline abundance      & $\mathcal{N}(0.0, 0.2)$             & dex \\
\enddata
\tablecomments{$\mathcal{N}(\mu,\sigma)$ denotes a normal distribution with mean $\mu$ and standard deviation $\sigma$. The outlier fraction $Q$ is fixed at 0 and is not treated as a free parameter (see \S\ref{sec:model}).}
\end{deluxetable}

\subsection{Parameter Inference} \label{sec:mcmc}

Because we fit each spatial cell independently, our parameter space stays small regardless of sample size.
Each fit has only six parameters (see Table~\ref{tab:priors}), which makes it practical to sample the full posterior probability distribution with Markov Chain Monte Carlo (MCMC).
We use Hamiltonian Monte Carlo, specifically the No-U-Turn Sampler \citep[NUTS;][]{hoffman_2014} as implemented in \texttt{NumPyro} \citep{Phan_2019, Bingham_2019}, which is itself built on \texttt{JAX} \citep{Bradbury_2018}, to generate posterior samples for the parameters involved in our model (Equation~\ref{eq:sech2}) for each spatial $(R, \phi)$ cell.
For each fit, we run 2 chains with 1000 warmup and 1000 post-warmup steps, using a dense mass matrix and a target acceptance probability of 0.95.
Within each spatial cell, we report the posterior median as the fiducial value of $z_0$ and the standard deviation over posterior samples as its uncertainty.
We assess convergence with the Gelman--Rubin statistic, $\hat{R}$ \citep{Gelman_1992}. For $z_0$, $\hat{R} < 1.05$ in all cells ($< 1.01$ in all but 3 of 142).

\subsection{Validation of Method with Simulations} \label{sec:sim}

To test whether the chemical midplane traces the density midplane, we apply the same analysis to two simulated Milky Way-like galaxies: a host evolved in isolation and a host with an LMC- and Sagittarius-like merger history (Appendix~\ref{app:sim-appendix}). In the isolated host, both midplanes stay close to $z = 0$. With the merger history, both are displaced coherently beyond $R \sim 10$~kpc, upward at positive azimuth and downward at negative azimuth, and the chemical midplane follows the density midplane with a median difference of $23$~pc. This supports using the chemical midplane as a proxy for the density midplane, even in a strongly perturbed disk. Fitting the twisted warp model separately to young (0--3~Gyr) and old (3+~Gyr) stars in the merger simulation gives line-of-nodes azimuths that differ by $\sim 18^\circ$, which suggests that a merger can produce an age dependence of the warp phase (Section~\ref{sec:age_dependence}).

\section{Results}
\label{sec:results}

\subsection{Chemical Midplane ($z_0$)}
\label{sec:midplane_z0}

\begin{figure*}
    \centering
    \includegraphics[width=\textwidth]{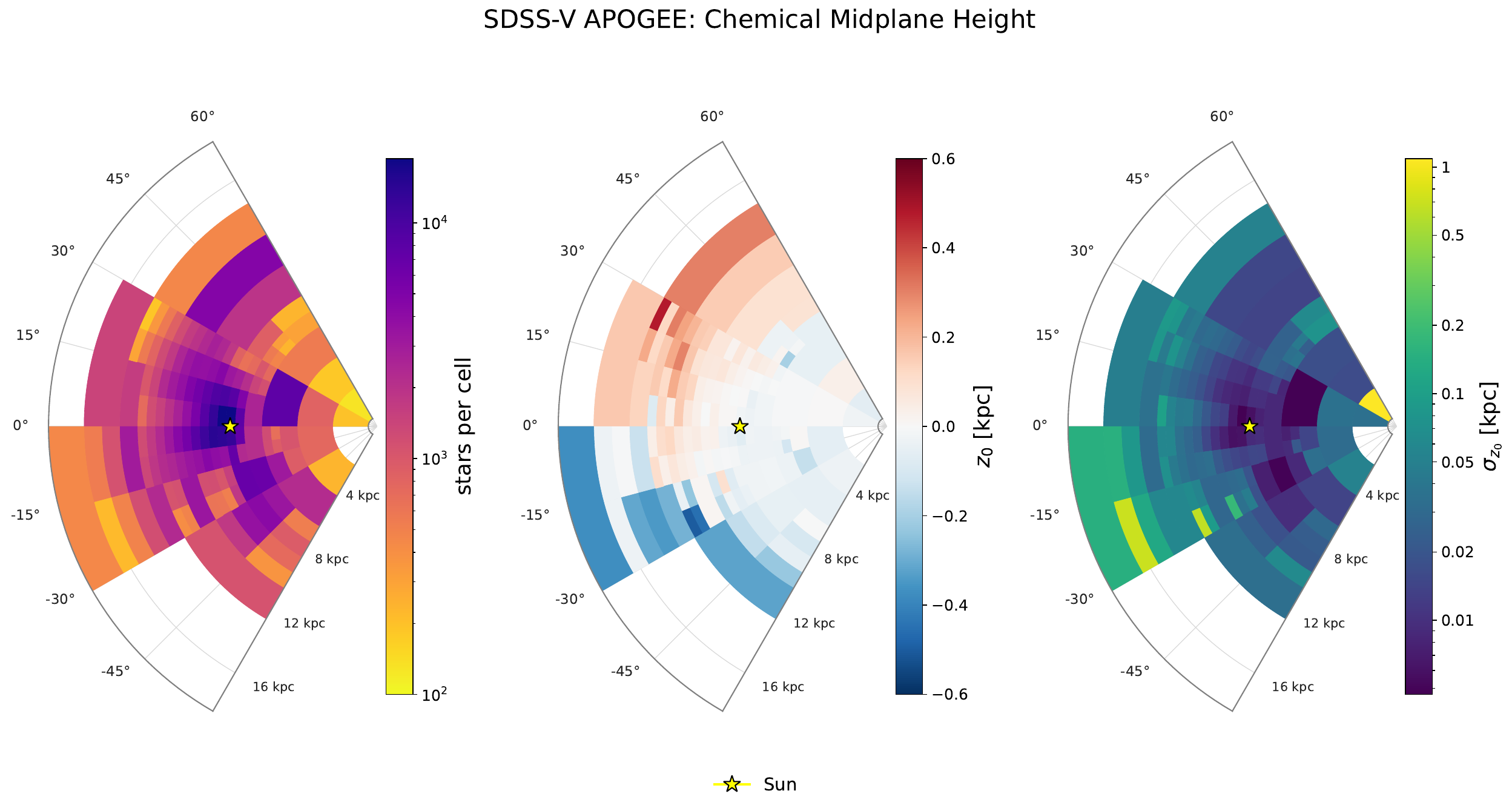}
    \caption{Number of stars in each cell (left), position of the chemical midplane, $z_0$ (middle), and its uncertainty, $\sigma_{z_0}$ (right), in adaptive cells of radius and azimuth out to $R = 18.5$~kpc and $|\phi| \leq 60^\circ$. The Sun is at $\phi = 0^\circ$, indicated by the yellow star. Positive (negative) $z_0$ indicates the midplane sits above (below) the geometric $z=0$ plane. Areas without a fitted cell are blank. The star counts and uncertainties are shown on a logarithmic scale, and the $z_0$ color scale is limited to $\pm0.6$~kpc.}

    \label{fig:z0_heatmap}
\end{figure*}

We measure the chemical midplane height $z_0$ (Section~\ref{sec:model}) in each spatial cell (Section~\ref{sec:spatial_binning}) using the 438{,}499 red giant branch stars described in Section~\ref{sec:sample}.
The left panel of Figure \ref{fig:z0_heatmap} shows the number of stars per cell.
The fitted cells contain between 128 and 18{,}728 stars (median 1{,}748, mean 2{,}976).

Figure~\ref{fig:z0_heatmap} (middle panel) shows the $z_0$ obtained in each cell.
We find a clear asymmetry in $z_0$ as a function of azimuth. All midplane heights $z_0$ are measured relative to a Galactocentric frame with the Sun at $z=0$. A nonzero $z_0$ therefore means the chemical midplane is displaced from that plane, which at large radii is the warp. At $R \gtrsim 10$ kpc, the midplane sits above $z=0$ from $\phi \sim -10^\circ$ to $\phi \sim +60^\circ$, with its highest value of $z_0 = 0.476 \pm 0.087$ kpc at $R \sim 14$ kpc and $\phi \sim 26^\circ$. It falls below $z=0$ at $\phi \lesssim -20^\circ$ out to $\phi = -60^\circ$, reaching a minimum of $z_0 = -0.501 \pm 0.648$ kpc at $R \sim 12$ kpc and $\phi \sim -26^\circ$. This pattern is broadly consistent with the expectation from the large-scale warp of the Milky Way \citep[e.g.][]{Momany:2006,Reyle_2009}. The right panel of Figure \ref{fig:z0_heatmap} shows the uncertainty $\sigma_{z_0}$ on the inferred $z_0$. For each cell, we take $z_0$ to be the median of its posterior from the MCMC fit, marginalized over the other model parameters, and we take the uncertainty, $\sigma_{z_0}$, to be the standard deviation of the posterior samples.

\subsection{Model of the Galactic Warp}
\label{sec:lon}

Our measured midplane has a clear $m=1$ asymmetry (see previous subsection), and we therefore fit a warp model to the derived midplane locations in each $(R, \phi)$ cell.
We determine the orientation of the Galactic warp's line of nodes (LoN) and vertical amplitude directly from the chemical midplane height map $z_0(R,\phi)$ measured in the adaptive cells described in Section~\ref{sec:spatial_binning} and shown in Figure \ref{fig:z0_heatmap}. 

We model $z_0$ as a single $m=1$ mode, following the classic parametric warp model \citep[e.g.][]{Poggio_2025, Skowron_2019a, Chen_2019, Dehnen_2023},
\begin{equation}
    z_0(R,\phi) = A(R)\,\sin\!\big(\phi - \phi_{\rm LoN}(R)\big) + z_{\rm off} \quad ,
\end{equation}
where $z_{\rm off}$ is a constant vertical offset that represents the vertical position of the Sun relative to the chemical midplane and the amplitude grows as a power law beyond the warp onset radius $R_w$,
\begin{equation}
    A(R) =
    \begin{cases}
        h_{w,12}\left[\dfrac{R - R_w}{12~{\rm kpc} - R_w}\right]^{\alpha_w}, & R > R_w \\[2ex]
        0, & \text{otherwise,}
    \end{cases}
\end{equation}
so that $h_{w,12}$ is the warp amplitude at $R = 12$~kpc. The choice of normalizing the warp amplitude at 12 kpc is arbitrary, and aims to facilitate the comparison with previous works in a disk region where the warp signal is strong (i.e. $ >> R_w$).
We consider two forms for the line of nodes: a \emph{straight} LoN with a constant orientation $\phi_{\rm LoN}(R) = \phi_{\rm LoN,0}$ at all radii, and a \emph{twisted} LoN that bends beyond a radius $R_t$,
\begin{equation}
    \phi_{\rm LoN}(R) =
\begin{cases}
    \phi_{\rm LoN,0} & R \le R_t \\[4pt]
    \phi_{\rm LoN,0} + \beta_w\,(R - R_t) & R > R_t \quad .
\end{cases}
\end{equation}

We add an intrinsic scatter $s$ in quadrature to the uncertainty of each cell's $z_0$, to account for scatter about the smooth model and for the uncertainties on $z_0$, which propagate from the abundance errors and may be underestimated. We fit each model to the $z_0$ values of the 142 cells evaluated at the mean position of each cell's stars, which cover $|\phi| \leq 60^\circ$ and extend to $R = 18.5$~kpc, using the No-U-Turn Sampler in \texttt{NumPyro}  \citep{Phan_2019}, using the priors listed in Table~\ref{tab:warp_priors}.

\begin{table*}
    \centering
    \caption{Priors for the $m=1$ warp model. $R_t$ and $\beta_w$ are used only in the twisted model. $\mathrm{U}(a,b)$ is a uniform distribution between $a$ and $b$.}
    \label{tab:warp_priors}
    \begin{tabular}{llcc}
        \hline\hline
                Parameter & Description & Prior & Units \\
        \hline
        $h_{w,12}$         & Warp amplitude at 12~kpc            & Uniform$(10^{-4}, 2.0)$      & kpc \\
        $R_w$              & Warp onset radius                   & Uniform$(3.0, 12.0)$         & kpc \\
        $\alpha_w$         & Amplitude growth index              & Uniform$(0.1, 5.0)$          & \nodata \\
        $\phi_{\rm LoN,0}$ & Line-of-nodes azimuth               & $\mathcal{N}(0.0, 45.0)$     & deg \\
        $R_t$              & Twist onset radius                  & Uniform$(R_w, 16.0)$         & kpc \\
        $\beta_w$          & Twist rate beyond $R_t$             & Uniform$(-30.0, 30.0)$       & deg/kpc \\
        $z_{\rm off}$ & Global vertical offset & $\mathcal{N}(0.0, 0.1)$ & kpc \\
        $s$                & Intrinsic scatter of cell $z_0$     & HalfNormal$(0.1)$            & kpc \\
        \hline
    \end{tabular}
\end{table*}

\begin{table*}
    \centering
    \caption{Median and 16th to 84th percentile range of the posterior for the $m=1$ warp model, using a straight and a twisted line of nodes (see text). $\sigma_{\varpi}$ is the standard deviation of the best-fit values over 64 resamplings of the ZGR parallaxes within their quoted errors.}
    \label{tab:lon_fit}
    \begin{tabular}{lcccc}
\hline
 & \multicolumn{2}{c}{Straight LoN} & \multicolumn{2}{c}{Twisted LoN} \\
Parameter & Value (MCMC) & $\sigma_{\varpi}$ & Value (MCMC) & $\sigma_{\varpi}$ \\
\hline
$h_{w,12}$ (kpc)         & $0.387^{+0.023}_{-0.023}$ & 0.017 & $0.376^{+0.021}_{-0.020}$ & 0.015 \\
$R_w$ (kpc)              & $6.44^{+0.82}_{-1.26}$    & 0.55  & $5.27^{+1.15}_{-1.35}$    & 0.58 \\
$\alpha_w$               & $1.57^{+0.50}_{-0.34}$    & 0.29  & $2.17^{+0.52}_{-0.46}$    & 0.28 \\
$\phi_{\rm LoN,0}$ (deg) & $-8.6^{+1.8}_{-1.9}$      & 1.5   & $-16.2^{+3.0}_{-7.3}$     & 1.8 \\
$R_t$ (kpc)              & \nodata                   & \nodata & $12.3^{+0.6}_{-2.3}$    & 0.52 \\
$\beta_w$ (deg/kpc)      & \nodata                   & \nodata & $4.15^{+1.21}_{-1.02}$  & 0.92 \\
$z_{\rm off}$ (kpc) & $-0.0242^{+0.0062}_{-0.0063}$ & 0.004 & $-0.0354^{+0.0065}_{-0.0072}$ & 0.003 \\
$s$ (kpc)                & $0.0382^{+0.0046}_{-0.0042}$ & 0.005 & $0.0327^{+0.0042}_{-0.0039}$ & 0.005 \\
\hline
\end{tabular}
\end{table*}

\begin{figure*}
    \centering
    \includegraphics[width=\textwidth]{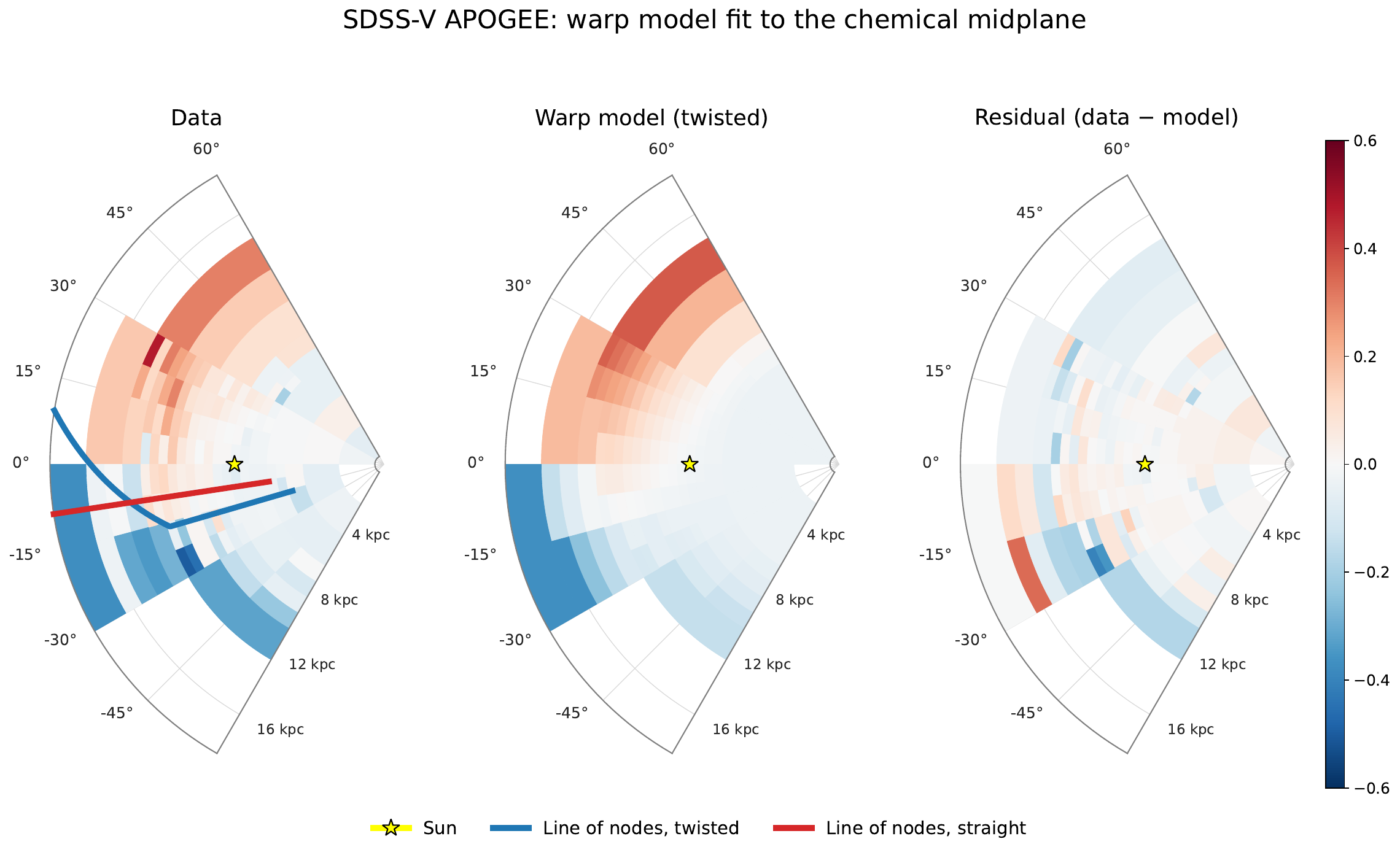}
    \caption{Chemical midplane height $z_0$ in adaptive cells of Galactocentric radius and azimuth (left), the twisted warp model evaluated at the median posterior parameters in the same cells (middle), and the residual, data minus model (right). The twisted (blue) and straight (red) lines of nodes from the model fits are drawn on the data panel. All three panels use the same color scale, limited to $\pm0.6$~kpc. The yellow star marks the Sun.}
    \label{fig:warp_fit}
\end{figure*}

Figure~\ref{fig:warp_fit} shows the observed $z_0$ midplane location map (left), the $m=1$ twisted warp model evaluated at the median posterior parameters (middle), and the residuals (right). For the twisted model we find $\phi_{\rm LoN,0} = -16.2^{\circ\,+3.0}_{\,-7.3}$ and a line of nodes that bends beyond $R_t = 12.3^{+0.6}_{-2.3}$~kpc at a rate of $\beta_w = 4.2^{+1.2}_{-1.0}\,{}^\circ\,{\rm kpc}^{-1}$, while the straight model gives $\phi_{\rm LoN,0} = -8.6^{\circ\,+1.8}_{\,-1.9}$ (Table~\ref{tab:lon_fit}). The amplitude at 12~kpc is $h_{w,12} = 0.376 \pm 0.02$~kpc for the twisted model and $0.387 \pm 0.023$~kpc for the straight model. $R_w$ and $R_t$ are only weakly constrained, with long tails toward small values. The full posteriors are shown in Appendix~\ref{app:model-validation}.

Our fitted global vertical offset is $z_{\rm off} = -24.2 \pm 6.0$~pc for the straight and $-35.4 \pm 7.0$~pc for the twisted line-of-nodes model (Table~\ref{tab:lon_fit}). Because the warp also contributes at the Sun's location, the Sun's height above the midplane is not simply $-z_{\rm off}$. Evaluating the best-fit model at the position of the Sun ($R = 8.275~\unit{kpc}$, $\phi = 0$), we find $z_\odot = 16.2^{+4.6}_{-5.2}$~pc above the local chemical midplane for the twisted model and $14.1^{+4.9}_{-4.8}$~pc for the straight model.
This is somewhat similar to other recent measurements of the solar height (e.g., $\zsun = 20.8$~pc in \citealt{Bennett_2019}).
However, we note that other measurements typically estimate the solar height using relatively local samples, whereas ours is inferred from a global model fit (covering $|\phi| \leq 60^\circ$ and $R \leq 18.5$~kpc).
Given that, it is interesting that we get a similar value.

\begin{figure*}
    \centering
    \includegraphics[width=\textwidth]{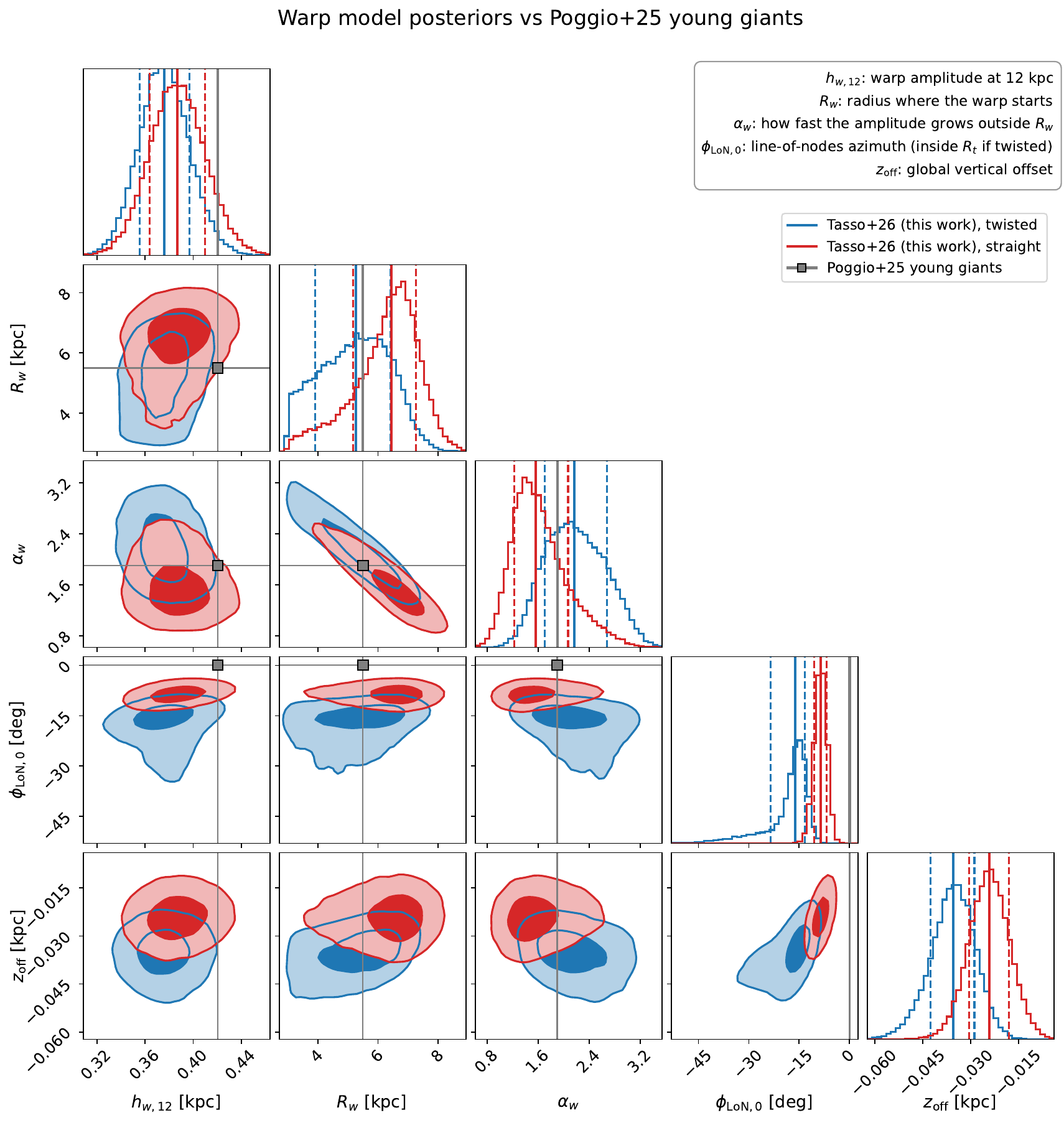}
    \caption{Posteriors of the twisted (blue) and straight (red) warp models compared with the young giant values of \citet{Poggio_2025} (gray). Panels show the warp amplitude at 12~kpc $h_w$, the warp onset radius $R_w$, the amplitude growth index $\alpha_w$, the line-of-nodes azimuth $\phi_{\rm LoN,0}$ (inside $R_t$ for the twisted model), and the global vertical position $z_{\rm off}$. Filled contours enclose 1 and 2$\sigma$. In the histograms, solid lines mark the median and dashed lines mark the 16th and 84th percentiles.}
    \label{fig:corner_poggio}
\end{figure*}

\begin{figure*}
    \centering
    \includegraphics[width=\textwidth]{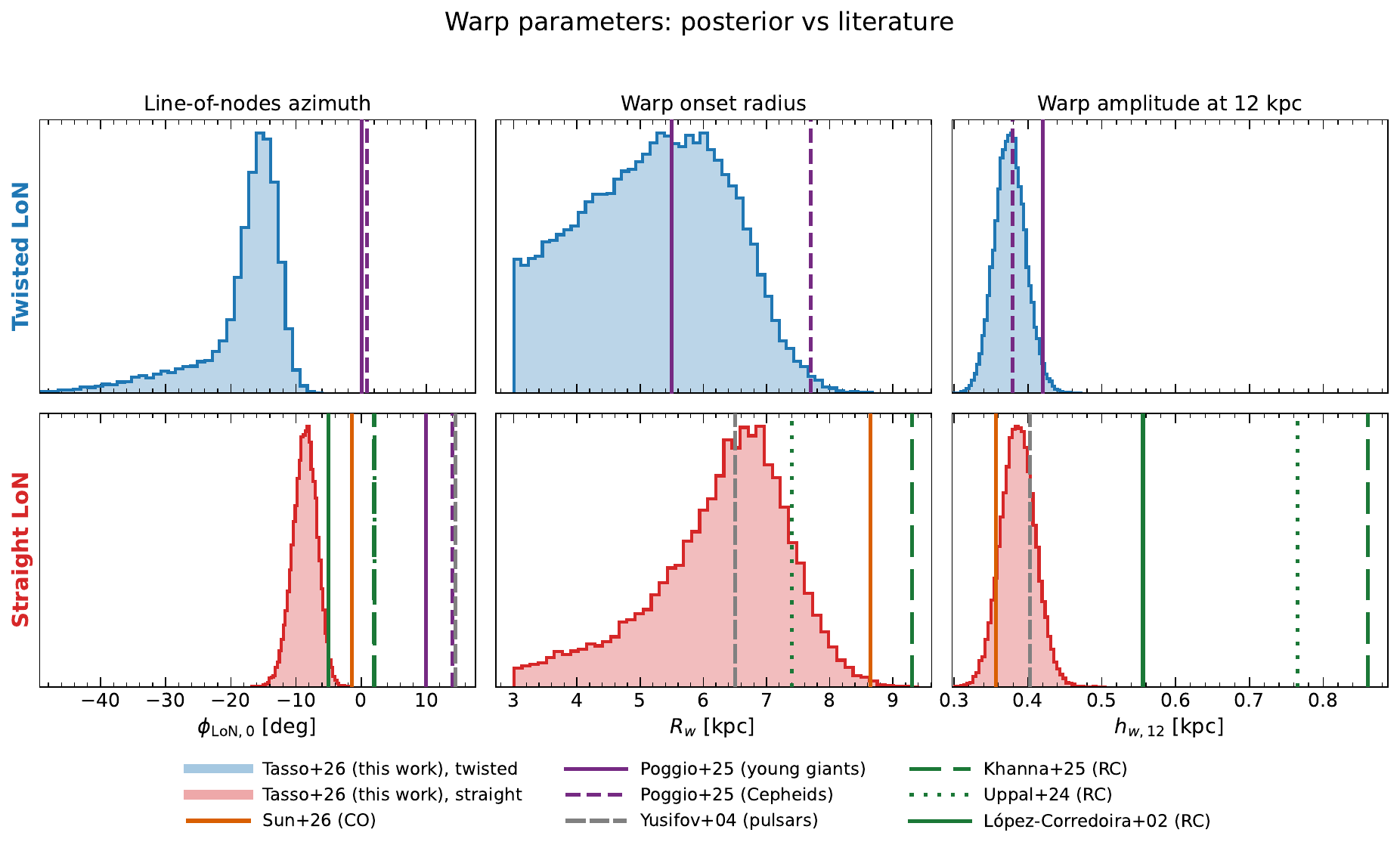}
   \caption{Posterior distributions of $\phi_{\rm LoN,0}$, $R_w$, and the warp amplitude at 12~kpc, $h_{w,12}$, from our twisted (top, blue) and straight (bottom, red) line-of-nodes fits. Vertical lines show literature values, with colors and line styles identifying each study as in the legend. Each row is compared only against studies that used the same type of model. For the literature, $h_{w,12}$ is each study's published amplitude relation evaluated at $R = 12$~kpc, using the best-fit parameters of \citet{Poggio_2025} (young giants and Cepheids), Case~I of \citet{Khanna_2025}, \citet{Uppal_2024}, \citet{Lopez-Corredoira_2002}, \citet{Yusifov_2004}, and the CO $m=1$ model of \citet{Sun_2026}.}

\label{fig:warp_hist}
\end{figure*}

To place this measurement in the context of the literature, we convert each previously reported LoN orientation \citep{Khanna_2025, Uppal_2024, Lopez-Corredoira_2002, Yusifov_2004} into this common reference frame, correcting for the differing zero points and angular-increase conventions adopted by each study.\footnote{A number of these required re-deriving the sign convention directly from each paper's stated coordinate definitions, since the direction of increasing $\phi$ is not always stated explicitly.} For the amplitude, we evaluate each study's published amplitude curve at $R = 12$~kpc, since the studies normalize $h_w$ at different radii. Figure~\ref{fig:corner_poggio} compares the posteriors of both twisted and straight LoN models with the young giant parameters of \citet{Poggio_2025}, and Figure~\ref{fig:warp_hist} compares the posteriors with the literature values, with each model compared only against studies that used the same type of line of nodes.

The right panel of Figure~\ref{fig:warp_fit} shows the residuals between the measured $z_0$ and the twisted warp model in each cell. The median residual is $+1~\unit{pc}$, so the model is unbiased overall, and the median absolute residual is $29~\unit{pc}$, comparable to the fitted intrinsic scatter $s \approx 33~\unit{pc}$. The residuals range from $-401$ to $+338~\unit{pc}$, and $13\%$ of cells deviate by more than $100~\unit{pc}$. These deviations indicate that there is additional structure in the midplane that a single $m=1$ warp does not capture.

\section{Discussion}
\label{sec:discussion}

\subsection{Returning to our assumptions and implications for our results}
\label{sec:assumptions}

Returning to the assumptions in Section~\ref{sec:conceptual_framework}, here we consider
how they may be violated in the data and in the real Milky Way, and how this could
affect our results.

First, our measurements assume that the vertical minimum of \mgfe\ traces the midplane
of the disk, which typically corresponds to the maximum of the vertical stellar density.
The chemical midplane is instead the plane where the vertical gravitational acceleration
is zero (an optimum of $\Phi_z$), rather than of the density maximum (which depends on
the tracer population and its vertical distribution).
Under our assumptions, the potential and density midplanes are equivalent.
However, the Milky Way disk is not in perfect equilibrium and the vertical kinematics of
stars are not fully phase-mixed, as evidenced by the vertical phase spiral
\citep{Antoja:2018}.
Out of equilibrium, the \mgfe\ minimum and the density maximum are both properties of
the instantaneous distribution function, so neither is guaranteed to be the same as the
potential midplane.
However, we expect the chemical midplane to be closer to the potential midplane for two
reasons.
For one, $\langle Y\rangle(z)$ is a ratio of moments (Equation~\ref{eq:meanY}), so any
factor that depends only on $z$ (like the vertical density) cancels from the ratio.
This is true if the $z$ dependence comes from survey selection or extinction (as we
discuss above), or even from vertical density variations from a perturbation (as long as
the velocity and abundance distributions are not significantly perturbed).
The density maximum will instead be biased toward locations where the tracer density is
higher.
As another consideration, our fitted $z_0$ comes from stars with a range of vertical
energies (needed to resolve the trend of $\langle Y\rangle(z)$) and therefore uses stars
that sample the potential more uniformly than the density maximum, which is dominated by
stars with low vertical energies (i.e. that are more confined to the density midplane).
So, while we expect the chemical midplane to be biased away from the potential midplane
in a disequilibrium situation, we expect it to be less biased than the density midplane,
and more representative of the midplane corresponding to the total mass distribution of
the disk.

We can see this in the simulations that we test our method on (Section~\ref{sec:sim}).
For example, in the simulation with no merger (left column of
Figure~\ref{fig:sim_chem_vs_nz_peak}), in the inner disks ($R \lesssim 16$~kpc),
both the chemical and density midplanes stay within $\approx 0.08$~kpc of $z = 0$
for 95\% of cells ($0.072$~kpc for chemistry and $0.081$~kpc for density), and
show the same weak $m=1$ pattern, which comes from small scale stochastic variation in the baryonic processes of star formation and feedback, and gas accretion which break mid-plane symmetry. Such events create short lived transient and low amplitude features compared to the influence of the satellites in the merger simulation.
Interestingly, the chemical midplane appears to be smoother than the density midplane. For $R \lesssim 16$~kpc, the scatter between neighboring cells (the standard deviation of the radial and azimuthal differences, divided by $\sqrt{2}$) is $0.013$~kpc for the chemical midplane and $0.022$~kpc for the density peak. This is what one would expect from the fact
that it is related to the potential and not the tracer density, which can have
small-scale variations that do not affect the global dynamics of stars in the disk.
In the simulation with the LMC- and Sagittarius-like satellites (right column of
Figure~\ref{fig:sim_chem_vs_nz_peak}), the chemical and density midplanes are both
strongly perturbed at $R \gtrsim 10$~kpc, and they show the same large-scale pattern of
being above $z = 0$ at positive azimuth and below $z = 0$ at negative azimuth.
However, again, the chemical midplane is smoother than the density midplane. Neighboring
cells of the density midplane differ by a median of $0.10$~kpc, compared with $0.07$~kpc
for the chemical midplane.
Still, the global warp pattern is the same from both measures, even though this
simulation has clear deviations from the assumptions in
Section~\ref{sec:conceptual_framework} (e.g., disequilibrium, local phase spirals).
This suggests that the chemical midplane is a robust measure of the global warp pattern
given the amplitude of disk perturbations we expect for the Milky Way.

Our assumption that selection effects separate into a product $S_z(z)\,S_Y(Y)$ could be violated because stars must be bright enough to have a \gaia\ XP spectrum ($G \lesssim 17.5$), and this limit interacts with extinction and with luminosity differences that correlate with \mgfe.
To remove this effect, we apply our own selection to null out any possible correlated selection effects within each spatial cell.
To do this, we include a selection in our fiducial sample that only keeps stars that would be observable with an XP spectrum anywhere in their cell (i.e., with any distance modulus in a given cell or with the 95th percentile extinction value; Section~\ref{sec:sample}).
This will produce a selection function that is separable in $z$ and $Y$ modulo uncertainties in distance and extinction, but the distance errors are small relative to the cell size, and the extinction errors are small relative to the range of extinction in each cell.
However, even if the selection function is not perfectly separable after these modifications, we expect any biases to mainly affect the warp amplitudes rather than the line of nodes (Appendix~\ref{app:separable-selfunc}).

Lastly, we have so far ignored distance errors by computing distances as $d = 1/\varpi$
from the ZGR parallaxes.
A distance error will impact both the $(R, \phi)$ cell a star is assigned to and the
uncertainty in an individual star's vertical position.
However, the ZGR parallax errors have zero mean, so for a sample of stars with a range of
vertical positions, we expect the effect of distance errors on the inferred warp
parameters to be small, as it should only increase the scatter in the observed trend of
$Y(z)$ and increase the uncertainty on $z_0$.
Most of our sample has good parallax measurements (over 99\% of of our final sample has
parallax $\snr >5$), so we expect the impact of distance errors on the inferred warp
parameters to be small.
To test this, we repeat our analysis on 64 trial resamplings of our data.
In each trial, we draw a new parallax for every star from its error distribution,
recompute positions, and assign stars to the same cells as in
Section~\ref{sec:spatial_binning}.
We fit $z_0$ in each cell by maximizing the posterior probability and we refit both warp
models with MCMC.
Because the observed parallaxes already contain one realization of the noise, each trial
has about twice the error variance of our data, so the median shift across trials is an
estimate of the bias that distance errors contribute to our measurements.
Across cells, the median shift in $z_0$ is $\approx 0~\unit{pc}$ (the median absolute shift is $7~\unit{pc}$), and the largest is $330~\unit{pc}$, in a cell with $\sigma_{z_0} = 648~\unit{pc}$, compared with a median $\sigma_{z_0}$ of $25~\unit{pc}$. The scatter of $z_0$ across trials is $22~\unit{pc}$. The warp parameters shift by at most $0.85\sigma$ (for $h_{w,12}$, which decreases by $\approx 17~\unit{pc}$), where $\sigma$ is the Table~\ref{tab:lon_fit} uncertainty. Because the scatter across trials is comparable to the MCMC uncertainties, we include it as an additional error term in Table~\ref{tab:lon_fit}.

\subsection{Comparison with other tracers of the warp}

The Galactic warp has been measured with a range of tracers. Young populations include Cepheids \citep{Skowron_2019a,Skowron_2019b, Chen_2019, Dehnen_2023, Cabrera-Gadea_2024, Poggio_2025}, young giants \citep{Poggio_2025}, OB stars \citep{Li:2019} and pulsars \citep{Yusifov_2004}. Older populations include red clump (RC) stars \citep{Lopez-Corredoira_2002, Uppal_2024, Khanna_2025}, RGB stars \citep{Momany:2006,Reyle_2009,Romero-Gomez_2019} and the general stellar populations of \citet{Cheng_2020}. These studies use different parametrizations, different fitting methods, and in some cases different reference frames, so the comparison in Figures~\ref{fig:corner_poggio} and~\ref{fig:warp_hist} is approximate. We converted each reported line of nodes to our azimuth convention, and we evaluate each amplitude relation at 12~kpc.

The red clump sample of \citet{Khanna_2025} is the closest to ours in tracer type. Both are old, giant-dominated populations observed over a broad range of radii. They obtain $R_w = 9.3$~kpc and a line of nodes within a few degrees of the Sun--Galactic center line, with the amplitude at 12~kpc around 0.9~kpc. Our warp begins at smaller radii ($R_w \sim 5$--$6$~kpc) and has a lower amplitude at 12~kpc ($\sim 0.38$~kpc). Their parameters $a_w$ (equivalent to our $\alpha_w$) and $\phi_w$ are fixed rather than fitted for, so part of the difference in amplitude and onset radius could be due to the assumed shape. Other RC studies find values that span a wide range (an amplitude at 12~kpc of $0.56$ to $0.77$~kpc) \citep{Lopez-Corredoira_2002, Uppal_2024}. The young giants and Cepheids of \citet{Poggio_2025} have amplitudes at 12~kpc of $0.42$ and $0.38$~kpc, respectively, with onset radii of $R_w = 5.5$ and $7.7$~kpc. The tracers also differ in age, which we return to in Section~\ref{sec:age_dependence}. We do not see a clear trend between amplitude and the age of the sample. Our amplitude is the lowest, followed by the young giants and Cepheids, followed by the RC samples.

Molecular gas shows a broadly similar picture. \citet{Sun_2026} fit an $m=1$ warp to more than 30{,}000 CO molecular clouds from the MWISP survey out to $R \approx 17$~kpc and find a line of nodes of $\phi_w = -1.4^{+0.7}_{-0.4}\,{}^\circ$, close to the inner line of nodes of $\approx 0^\circ$ in the twisted model of \citet{Poggio_2025}. Their line of nodes decreases with radius out to $R \approx 13$~kpc and then rises, qualitatively like our twisted model. Our line of nodes is more negative than theirs at small radii in the twisted model ($\phi_{\rm LoN,0} = -16^\circ$), but their values in individual annuli at $R \approx 11.5$--$13.5$~kpc ($\approx -6^\circ$ to $-7^\circ$) are close to our straight-model value of $-8.6^\circ$, and our twisted model reaches $\approx +1^\circ$ at $R = 16.5$~kpc, similar to the CO value at that radius. Evaluating their best-fit model at $R = 12$~kpc gives an amplitude of $\approx 0.36$~kpc, close to our $h_{w,12} = 0.38$~kpc and below the larger amplitudes inferred from red clump and Gaia/APOGEE samples. They also find vertical corrugations of $\approx 100$--$200$~pc superimposed on the warp, comparable in size to our residuals.
The warp amplitude we measure is similar to that of some other tracers, while the line of nodes differs by up to tens of degrees between populations, which may reflect differences in age, coverage or modeling. The residuals from our $m=1$ fits (Figure~\ref{fig:warp_fit}, right panel) also show clear coherent structure beyond a single warp mode, which may be related to the vertical corrugations reported in young stars and gas. A quantitative study of this structure is beyond the scope of this paper and we leave it to future work.

\subsection{Applications to future surveys}
\label{sec:future}

Our measurement covers $|\phi| \leq 60^\circ$ and $R \leq 18.5$~kpc, so it samples only part of the disk. SDSS-VI (the Hidden Galaxy Explorer) is expected to map the other side of the Milky Way. This would extend the azimuthal coverage of our method well beyond $|\phi| = 60^\circ$ and allow a direct test of whether the midplane follows a single $m=1$ warp mode around the whole disk, including whether the line of nodes twists and whether the two sides of the warp are symmetric. It would also constrain the model parameters that are currently degenerate over a limited range of azimuth, such as the line-of-nodes azimuth and its twist. In addition, more stars at large $R$ would reduce the uncertainties in the outer cells, where the warp amplitude is largest. A larger sample would also allow us to split by age and measure $z_0$ separately for young and old stars, as we did in the simulation, instead of comparing samples with different selections.

Our method may be particularly useful for SDSS-VI, DESI, WEAVE, and Gaia DR4 coverage. Because it locates the midplane from the vertical minimum of \mgfe\ and not from star counts, it is less sensitive to the survey selection function, so it can be applied where the coverage has gaps or where the stars are sparse, provided that each cell contains enough stars to define the \mgfe\ profile.

As a final point, we note that combining the chemical midplane with kinematics, such as vertical velocities \cite[e.g.,][]{Cabrera-Gadea:2026}, could constrain the dynamical evolution of the warp in a more robust and novel way than with tracer density and kinematics alone.

\section{Conclusions}
\label{sec:Conclusions}

We map the midplane of the Milky Way disk by measuring how the mean \mgfe\ of
low-$\alpha$ red giant stars from SDSS-V/\apogee\ varies with vertical position $z$ in
spatial cells across the face of the Galaxy (i.e., in cylindrical position $(R, \phi)$).
Because abundances label stellar orbits, the mean \mgfe\ in a phase-mixed disk depends
on $z$ only through the vertical potential, so its optimum (here, minimum) is the
location where the vertical acceleration is zero.
If the survey selection is separable in position and abundance, the spatial selection
function cancels from this profile and does not need to be modeled.
In our sample, we impose a luminosity cut in each spatial cell to keeps only stars
bright enough to have a \gaia\ XP spectrum anywhere in that cell to enforce this
separability.
Our data span $R \leq 18.5$~kpc and $|\phi| \leq 60^\circ$ (with the Sun at $R =
8.275$~kpc and $\phi=0$).
We find:
\begin{itemize}
    \item Beyond $R \approx 10~\unit{kpc}$, the local chemical midplane rises above $z = 0$ for $\phi \gtrsim -10^\circ$ and falls below $z=0$ for $\phi \lesssim -20^\circ$, reaching $z_0 = 476 \pm 87~\unit{pc}$ and $-501 \pm 648~\unit{pc}$ in individual spatial cells across the face of the disk (Figure~\ref{fig:z0_heatmap}, middle panel; individual fits in Figure~\ref{fig:sech2_fits_all_bins}).

    \item A single $m=1$ warp model reproduces the large-scale pattern, but there are clear high-order residuals with deviations as large as $\approx 400~\unit{pc}$ and a median of $\approx 29~\unit{pc}$ (Figure~\ref{fig:warp_fit}, right panel) that we account for here with an inferred intrinsic scatter parameter, $s \approx 33$ to $38~\unit{pc}$, in our model.
    For a straight model (with constant line of nodes, LoN), the amplitude of the warp at $R = 12~\unit{kpc}$ is $387 \pm 23~\unit{pc}$, and the line of nodes is at $\phi_{\rm LoN,0} = -8.6^{+1.8}_{-1.9}{}^\circ$. For a twisted model (radius-dependent LoN), the line of nodes is at $\phi_{\rm LoN,0} = -16.2^{+3.0}_{-7.3}$$^\circ$ with a twisting rate of $4.2^{+1.2}_{-1.0}\,{}^\circ\,\unit{kpc^{-1}}$ beyond $R_t \approx 12.3~\unit{kpc}$
    (Table~\ref{tab:lon_fit}; Figures~\ref{fig:corner_straight} and~\ref{fig:corner_twisted}).

    \item Our warp amplitude at $R=12~\unit{kpc}$, $h_{w, 12} \approx 376~\unit{pc}$, (see Figure~\ref{fig:warp_hist}, $h_{w,12}$ panels) is similar to or slightly lower than those measured previously with young giants and Cepheids ($\sim 0.4~\unit{kpc}$) and lower than those measured with red clump stars ($\sim 0.56$--$0.9~\unit{kpc}$).

    \item Our inferred line of nodes trails that of the young giants and Cepheids by $\approx 16^\circ$ in the direction of Galactic rotation (Figures~\ref{fig:corner_poggio} and~\ref{fig:warp_hist}). Since most ($\sim 75$\%) of our stars are older than 3~Gyr, this offset may reflect an age dependence of the warp phase.

    \item In a simulated galaxy with Milky Way-like satellite galaxies (LMC and Sagittarius), the chemical midplane generally agrees with the density midplane over the volume we study, $R < 18.5~\unit{kpc}$, with a median difference of $23~\unit{pc}$ (68\% of cells within $48~\unit{pc}$). In the simulated galaxy evolved in isolation, both stay close to $z = 0$, with median offsets of $11~\unit{pc}$ (chemistry) and $15~\unit{pc}$ (density) and 68\% of cells within $\approx 25~\unit{pc}$ (Figure~\ref{fig:sim_chem_vs_nz_peak}).
    For the simulation with satellites, stars younger and older than 3~Gyr develop an offset of their lines of nodes $\approx 17.6^\circ$ beyond $R \gtrsim 12~\unit{kpc}$ (Figure~\ref{fig:sim_age_bins}).
    This offset has the same sign as the offset we observe. The line of nodes of the older stars trails that of the younger stars in the direction of Galactic rotation.
\end{itemize}

Taken together, the observed offset and the merger simulation suggest that the phase of
the warp depends on stellar age, with older populations trailing younger ones in the
direction of Galactic rotation.
Because the chemical midplane needs only stellar positions and abundances, surveys that
reach the far side of the disk, such as SDSS-VI, can test whether a single $m=1$ mode
describes the warp around the full Galaxy and whether the offset from young tracers
persists at all azimuths.

\begin{acknowledgments}

JH is supported by UKRI Ernest Rutherford Fellowship ST/Z510245/1. DH is supported by the UKRI Science and Technology Facilities Council under project 101148371 as a Marie Skłodowska-Curie Research Fellowship.

This work has made use of data from the European Space Agency (ESA) mission
{\it \gaia} (\url{https://www.cosmos.esa.int/gaia}), processed by the {\it \gaia}
Data Processing and Analysis Consortium (DPAC,
\url{https://www.cosmos.esa.int/web/gaia/dpac/consortium}). Funding for the DPAC
has been provided by national institutions, in particular the institutions
participating in the {\it \gaia} Multilateral Agreement.

Funding for the Sloan Digital Sky Survey V has been provided by the Alfred P. Sloan Foundation, the Heising-Simons Foundation, the National Science Foundation, and the Participating Institutions. SDSS acknowledges support and resources from the Center for High-Performance Computing at the University of Utah. SDSS telescopes are located at Apache Point Observatory, funded by the Astrophysical Research Consortium and operated by New Mexico State University, and at Las Campanas Observatory, operated by the Carnegie Institution for Science. The SDSS web site is \url{www.sdss.org}.

SDSS is managed by the Astrophysical Research Consortium for the Participating Institutions of the SDSS Collaboration, including the Carnegie Institution for Science, Chilean National Time Allocation Committee (CNTAC) ratified researchers, Caltech, the Gotham Participation Group, Harvard University, Heidelberg University, The Flatiron Institute, The Johns Hopkins University, L'Ecole polytechnique f\'{e}d\'{e}rale de Lausanne (EPFL), Leibniz-Institut f\"{u}r Astrophysik Potsdam (AIP), Max-Planck-Institut f\"{u}r Astronomie (MPIA Heidelberg), Max-Planck-Institut f\"{u}r Extraterrestrische Physik (MPE), Nanjing University, National Astronomical Observatories of China (NAOC), New Mexico State University, The Ohio State University, Pennsylvania State University, Smithsonian Astrophysical Observatory, Space Telescope Science Institute (STScI), the Stellar Astrophysics Participation Group, Universidad Nacional Aut\'{o}noma de M\'{e}xico, University of Arizona, University of Colorado Boulder, University of Illinois at Urbana-Champaign, University of Toronto, University of Utah, University of Virginia, Yale University, and Yunnan University.

\end{acknowledgments}

\software{
    This work made use of the following software packages: \texttt{astropy} \citep{astropy:2013,astropy:2018,astropy:2022,astropy_21262391}, \texttt{Jupyter} \citep{2007CSE.....9c..21P,kluyver2016jupyter}, \texttt{matplotlib} \citep{Hunter:2007}, \texttt{numpy} \citep{numpy}, \texttt{pandas} \citep{mckinney-proc-scipy-2010,pandas_23955360}, \texttt{python} \citep{python}, \texttt{scipy} \citep{2020SciPy-NMeth,scipy_22051554}, \texttt{corner.py} \citep{corner-Foreman-Mackey-2016,corner.py_21208839}, \texttt{JAX} \citep{jax2018github}, and \texttt{numpyro} \citep{phan2019composable,bingham2019pyro}.
    Software citation information aggregated using \texttt{\href{https://flatironinstitute.github.io/software-citation-station/}{The Software Citation Station}} \citep{software-citation-station-paper,software-citation-station-zenodo}.
}

\appendix
\restartappendixnumbering

\section{Full Validation of Model Fits}
\label{app:model-validation}

\begin{figure*}
    \centering
    \includegraphics[height=0.9\textheight,keepaspectratio]{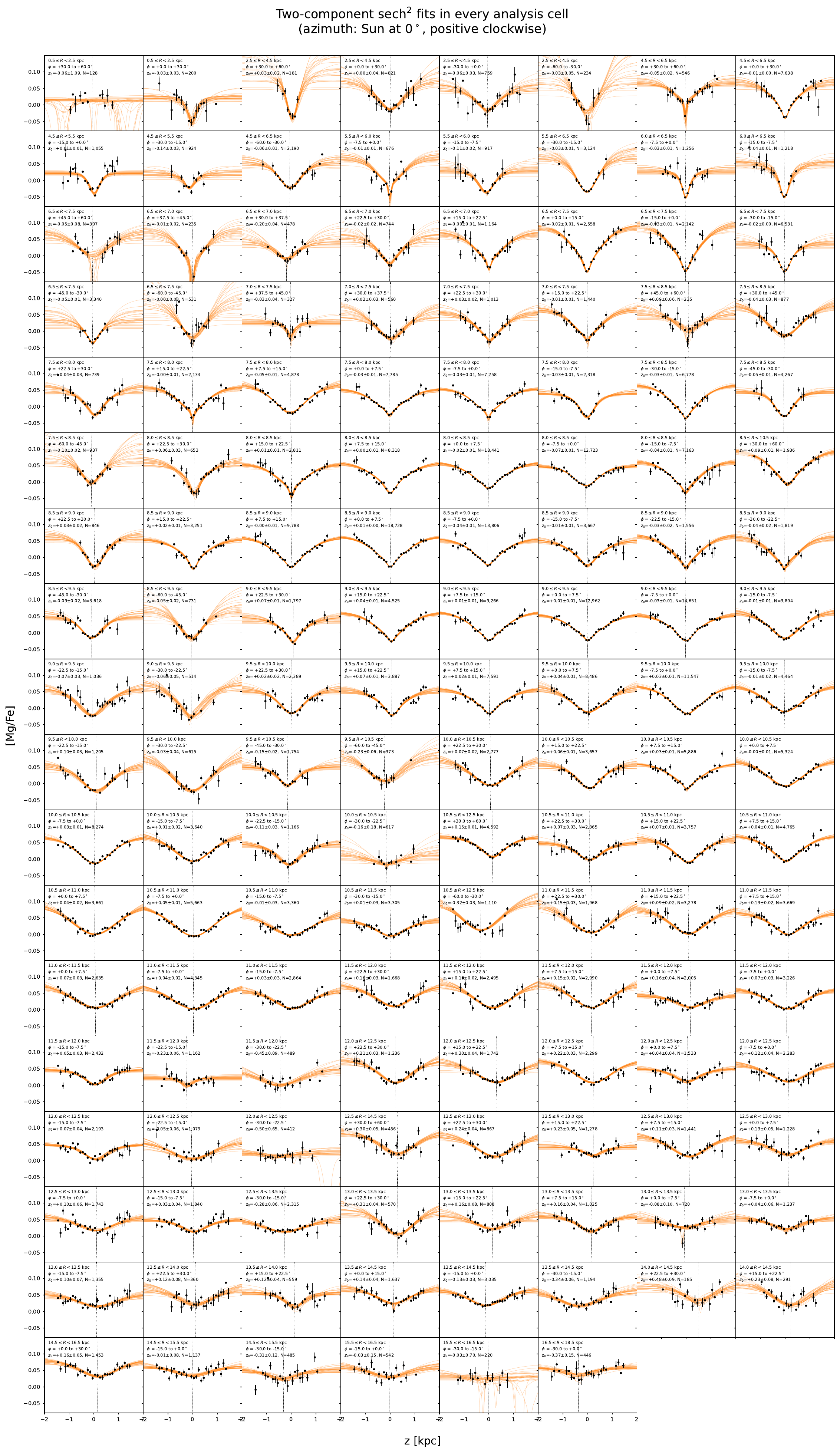}
    \caption{Two-component sech$^2$ fits in each of the 142 fitted analysis cells, ordered by radius and then azimuth. In each panel the black points are the binned mean \mgfe\ with uncertainties and the orange curves are 30 posterior draws. The vertical dotted line marks the fitted midplane height $z_0$. Each panel lists the radial and azimuthal range of the cell, $z_0$ (kpc), and the number of stars $N$. All panels span $-2 \leq z \leq 2$~kpc.}
    \label{fig:sech2_fits_all_bins}
\end{figure*}

\begin{figure*}
    \centering
    \includegraphics[width=\textwidth]{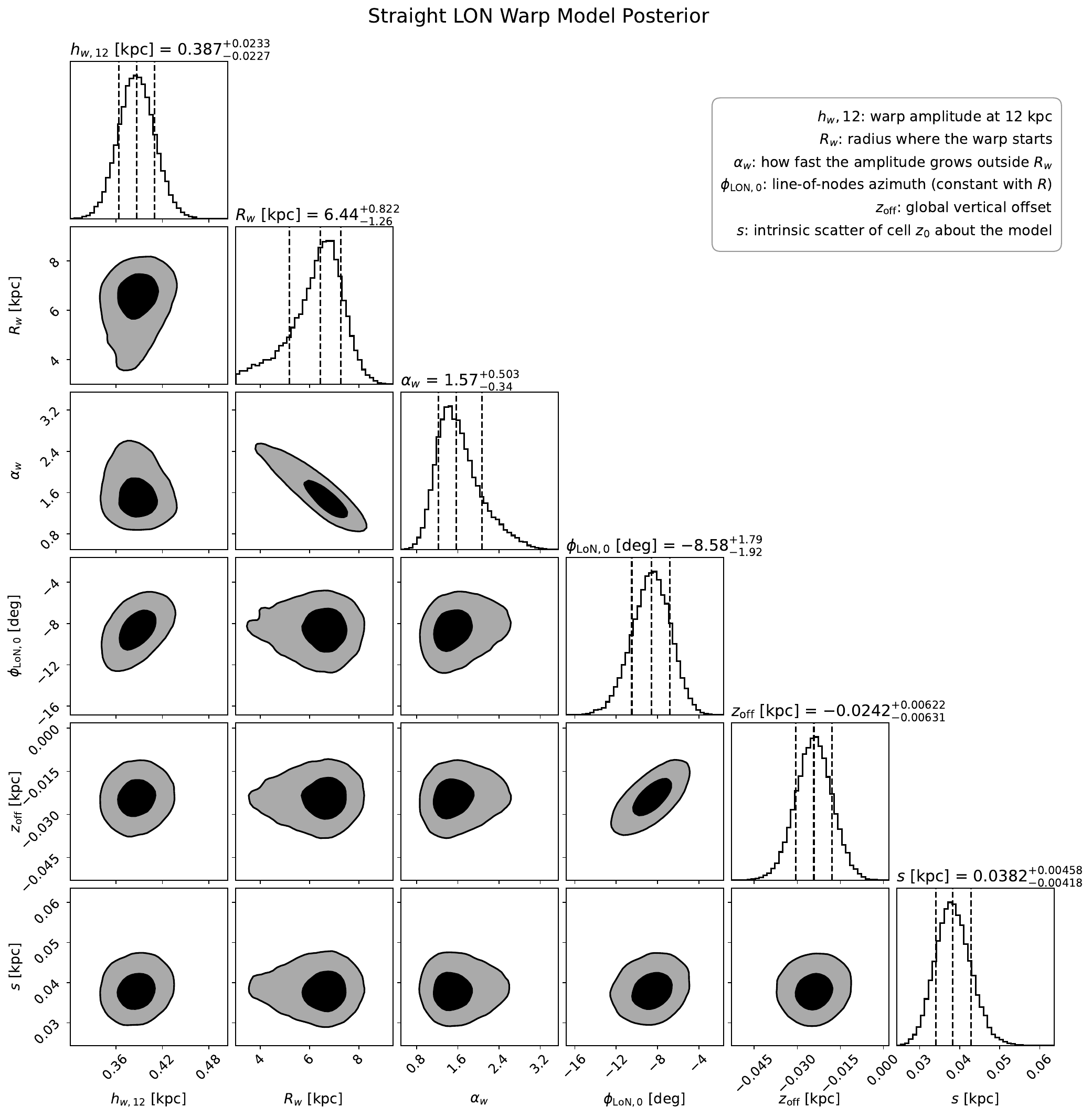}
    \caption{Posterior of the straight-LoN warp model. Filled contours enclose 1 and 2$\sigma$, and dashed lines in the histograms mark the 16th, 50th and 84th percentiles.
    }
    \label{fig:corner_straight}
\end{figure*}

\begin{figure*}
    \centering
    \includegraphics[width=\textwidth]{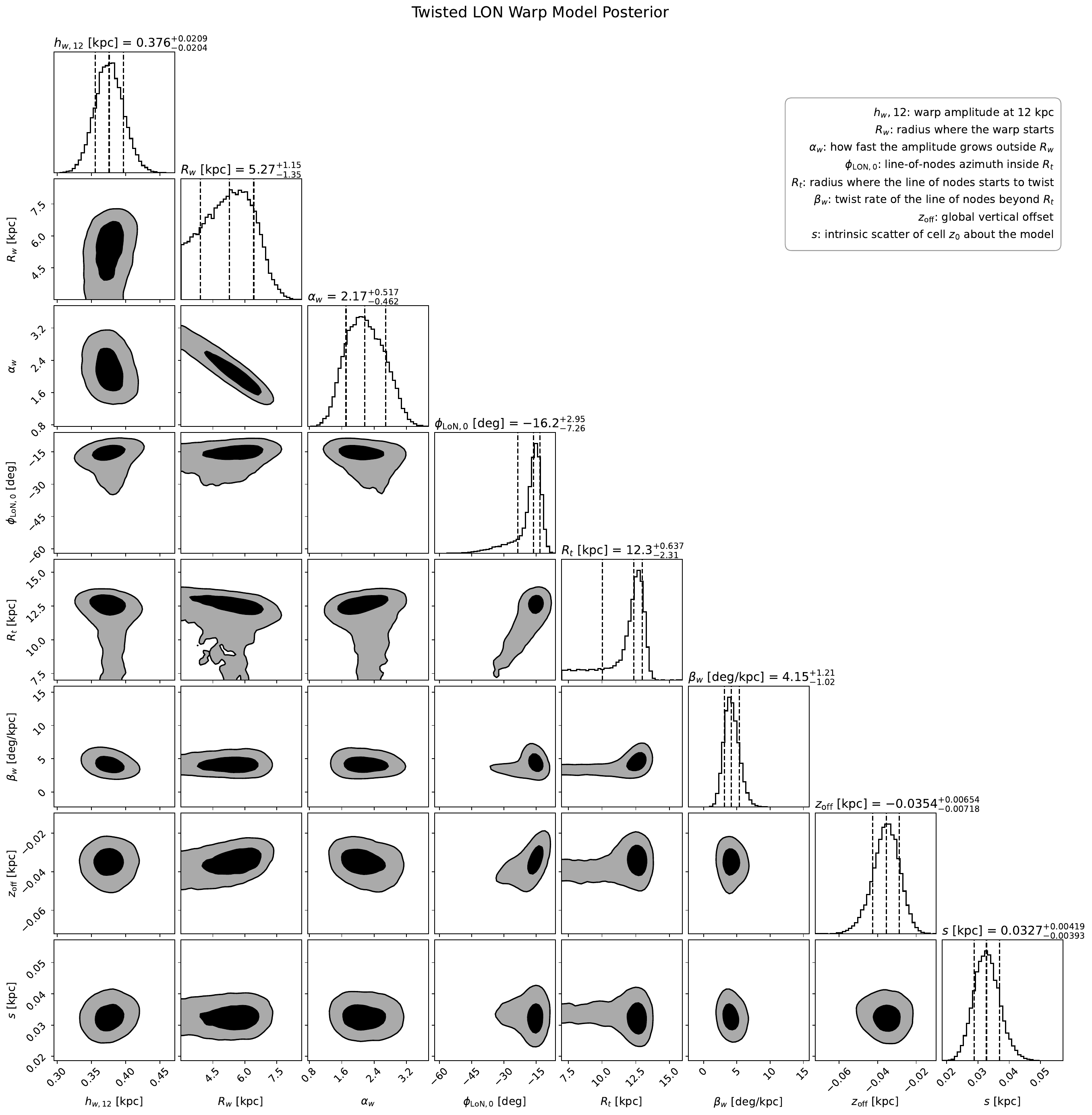}
    \caption{Same as Figure~\ref{fig:corner_straight} but for the twisted model.}
    \label{fig:corner_twisted}
\end{figure*}

Figure~\ref{fig:sech2_fits_all_bins} shows the two-component sech$^2$ fit to the vertical [Mg/Fe] profile in each of the 142 fitted analysis cells, from which we measure the midplane height $z_0$ used throughout the paper. Figures~\ref{fig:corner_straight} and~\ref{fig:corner_twisted} show the full posteriors of the warp models fit to these $z_0$ values, for the straight and twisted line-of-nodes models respectively. They include all fitted parameters, the joint distributions of each pair of parameters, and the marginal distribution of each parameter summarized in Table~\ref{tab:lon_fit}.

\section{Full Simulation Analysis}
\label{app:sim-appendix}

\subsection{Simulation Details}
\label{sim_details}
To test whether the chemical midplane traces the density midplane, we apply the same analysis to two simulated Milky Way-like galaxies: a host galaxy evolved in isolation and a host galaxy with an LMC- and Sagittarius-like merger history.
For full details on the simulation setup see Hunt et al. (in prep.), but in brief: The host galaxy is comprised of dark, stellar and gas components set up with the intitial condition generator \texttt{pgen} as described in \cite{Kawata:2014}. The dark matter halo is a $3\times10^7$ particle NFW dark matter halo with $M_{\mathrm{vir}}=0.9\times10^{12}\ M_{\odot}$ and concentration $c=9$. The stellar components are a $6\times10^6$ particle stellar bulge with $M_{\mathrm{b}}=1.2\times10^{10}\ M_{\odot}$ and scale radius 0.3 kpc, a thin stellar disk with $2.25\times10^7$ particles, $M_{\mathrm{d,thin}}=4.5\times10^{10}\ M_{\odot}$, scale radius and scale height $R_{\mathrm{d,thin}}=3.0$ and $h_{\mathrm{d,thin}}=0.3$ kpc, and a $2.5\times10^6$ particle thick disk $M_{\mathrm{d,thick}}=0.5\times10^{10}\ M_{\odot}$ with scale radius and scale height $R_{\mathrm{d,thick}}=2.0$ and $h_{\mathrm{d,thick}}=0.9$ kpc. The gas disk has $2\times10^7$ particles, with $M_{\mathrm{d,gas}}=5\times10^9\ M_{\odot}$, with a scale radius and scale height $R_{\mathrm{d,gas}}=7.0$ and $h_{\mathrm{d,gas}}=0.1$ kpc. The model also contains a $5\times10^9\ M_{\odot}$ cold gas reservoir from which it can accrete fresh gas onto the disk, in order to fuel continued star formation.

The stellar and gaseous components of the initial condition are painted with chemistry with the following relations in order to create the alpha-iron differences between the thin and thick disks. The thin (thick) disk has an \abund{Fe}{H} gradient of $-0.05$ ($-0.01$) $\unit{dex}~\unit{kpc}^{-1}$, $\abund{Fe}{H}_0=0.1$ ($-0.4$)~dex, a dispersion 0.02 (0.1)~dex, an $\abund{Fe}{H}~\unit{Gyr}^{-1}$ gradient of $-0.04$ ($0$)~$\unit{dex.Gyr^{-1}}$, \abund{\alpha}{H} gradient of $-0.16$ ($0$)~$\unit{dex}~\unit{kpc}^{-1}$, $\abund{\alpha}{H}_0 = 0.01$~(0)~dex and a dispersion 0.02 (0.1) dex, while the gas disk has \abund{Fe}{H} gradient of $-0.05~\unit{dex}~\unit{kpc}^{-1}$, $\abund{Fe}{H}_0=0~\unit{dex}$, a dispersion 0.02 dex, \abund{\alpha}{H} gradient of $0~\unit{dex}~\unit{kpc}^{-1}$, $\abund{\alpha}{H}_0=-0.1~\unit{dex}$ and a dispersion 0.02 dex. The cold gas reservoir has a uniform $\abund{Fe}{H}=-0.4$, $\abund{\alpha}{Fe}=-0.1$. The thin disk has stellar ages 0 to 2.5 Gyr, the thick disk has ages 2 to 5 Gyr and the bulge has ages 4 to 6 Gyr.

The `Host' model is evolved alone for 8.1 Gyr with the $N$-body + SPH hydrodynamics simulation code \texttt{ChaNGa} \citep{Jetley:2008,Jetley:2010,Menon:2015}, including star formation \& feedback, and chemical evolution. For the merger model, the same host is evolved again with the same parameters, also including two satellites. The first satellite galaxy is a disky dwarf again initialized with \texttt{pgen}. It is comprised of a $3.\dot3\times10^6$ particle $10^{11}\ M_{\odot}$ dark halo, a $10^6$ particle stellar disk with $M_{*}=1.41\times10^8\ M_{\odot}$, a $1.41\times10^6$ particle gas disk with $M_{\mathrm{gas}}=3.53\times10^8\ M_{\odot}$, and a $6\times10^8\ M_{\odot}$ cold gas reservoir. It is initialized at $(x,y,z,v_x,v_y,v_z)=(-244.6,0.003,-89.97,42.48,1.65,-42.0)$ kpc, km s$^{-1}$ respectively, which brings it into the disk on a polar orbit. The second satellite galaxy is a pure dark matter spheroid with $2\times10^6$ particles and $M_{\mathrm{DM}}=1.5\times10^{11}\ M_{\odot}$ generated with \texttt{agama} following $\alpha=1$, $\beta=3$, $\gamma=1$. It is initialized at $(x,y,z,v_x,v_y,v_z)=(167.0,1135.0,-164.0,-11.0,-105.0,-6.0)$ kpc, km s$^{-1}$.

We note that neither the host galaxy or the satellite merger configuration are intended to be a perfect reproduction of the Milky Way--Sagittarius--LMC system, and it should not be taken as such. While the present day is chosen to have the Sgr and LMC satellite analogs close to their real counterparts positions and orbital phase, there are discrepancies in the orbit and mass loss (see Hunt et al. in prep.) that mean we do not expect their impact on the disk to completely match the data.
Rather, it is a laboratory for exploring how such satellite interactions affect a Milky Way-like disk.

\subsection{Simulation Analysis}
\label{sim_analysis}

Rather than applying the observational sample cuts, we select all star particles tagged as thin disk from the initial condition ($2.25\times10^7$), together with all stars that form self-consistently in the simulation after that time ($\sim3.5\times10^7$), which are low-$\alpha$. This selection is analogous to the low-$\alpha$ selection of red giants in our observed sample (Section~\ref{sec:sample}). The simulations track oxygen rather than magnesium, so we use [O/Fe] as the analog of \mgfe\ for the $\alpha$-element abundance. We measure the chemical midplane height $z_0$ in polar cells of radius and azimuth using the same two-component sech$^2$ model as in Section~\ref{sec:model}, and the density midplane as the height where the stellar number density $n(z)$ peaks. In each cell, we estimate $n(z)$ over $|z| < 1.5$~kpc with a Gaussian kernel density estimate (kernel width 0.05~kpc). We locate its maximum on a fine grid with 5~pc spacing and refine the position by finding the root of the derivative of the density estimate; this height is the density midplane. These simulations are not tuned to exactly reproduce the Milky Way, so we do not expect the warp they show to match the one we measure in the data. Instead, we use them to ask whether a merger history like that of the LMC and Sagittarius can produce a coherent midplane pattern of a similar scale, and whether the chemical midplane tracks the density midplane when such a pattern is present.

For the simulations, we use fixed polar bins of radius and azimuth ($\Delta R = 0.5$~kpc and $\Delta\phi = 8^\circ$) over $R = 0.5$--$20$~kpc and $|\phi| \leq 60^\circ$. The adaptive cells used for the data are not needed here, because the simulated galaxies have no survey selection function or extinction, and every bin contains at least $1.7 \times 10^3$ star particles (median $\sim 1.8 \times 10^4$).

\begin{figure*}
    \centering
    \includegraphics[width=\textwidth]{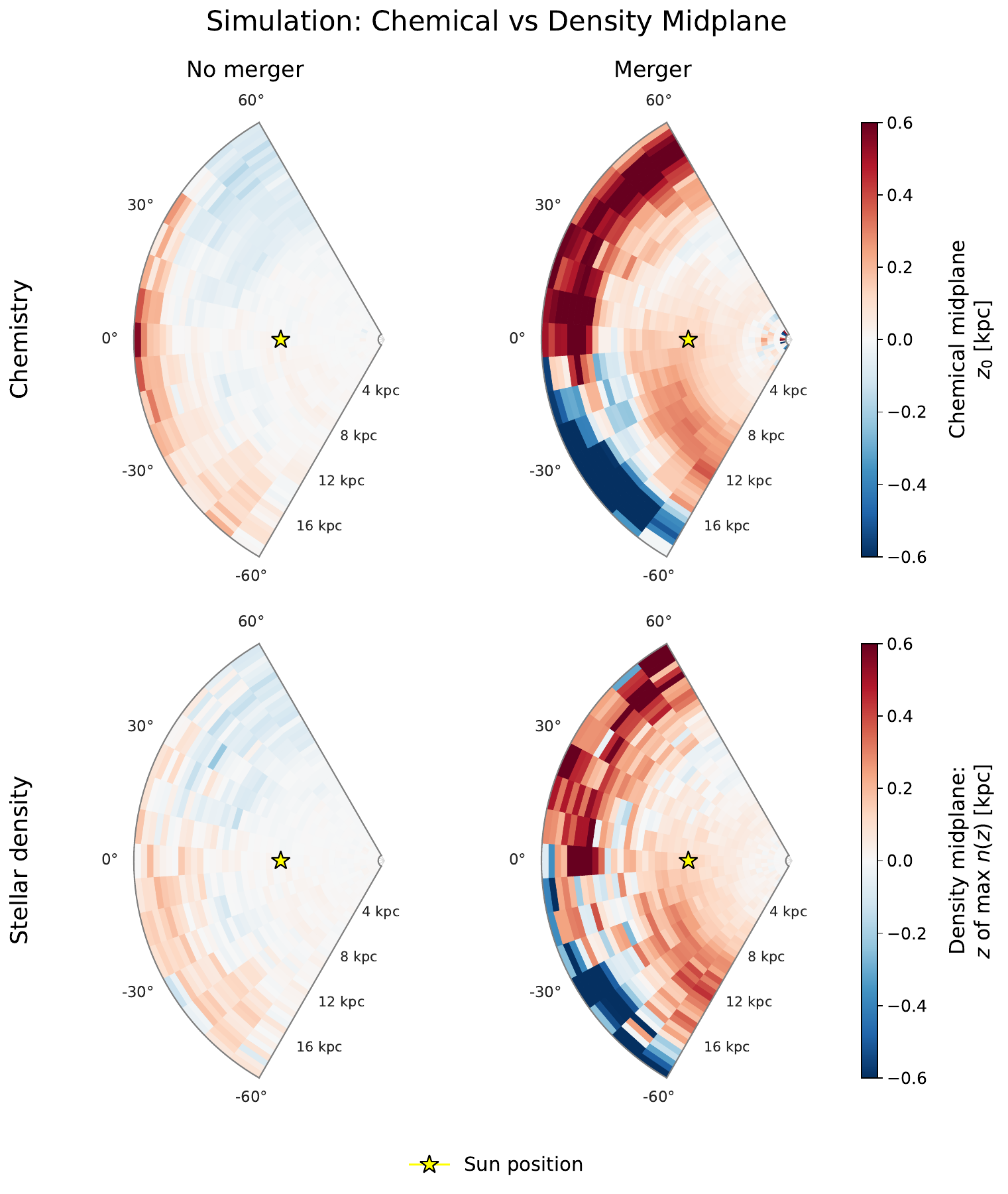}
    \caption{Simulated chemical midplane height (top row) and density midplane height (bottom row) in two simulated galaxies, shown in cells of radius and azimuth out to 20~kpc. Azimuth is measured from the Sun-equivalent position (0$^\circ$), positive in the direction of Galactic rotation, over $\pm60^\circ$. Left: the host galaxy alone, with no merger. Right: the host galaxy with an LMC- and Sagittarius-like merger history. The chemical midplane is the height $z_0$ of the minimum in the chemical tracer, and the density midplane is the height where the stellar number density $n(z)$ peaks. The color scale is limited to $\pm0.6$~kpc.}
    \label{fig:sim_chem_vs_nz_peak}
\end{figure*}

Figure~\ref{fig:sim_chem_vs_nz_peak} shows both measures in each simulation. The left column of Figure~\ref{fig:sim_chem_vs_nz_peak} shows the simulation without merger history, and the right column shows the simulation with LMC- and Sagittarius- merger history. The top row shows our chemistry method and the bottom row shows the number density method for defining the midplane location. In the host-only (no merger) simulation, both midplanes stay close to $z = 0$ across the wedge. The deviations are small, mostly below $\sim 0.2$~kpc, and they are weakly positive at negative azimuth and weakly negative at positive azimuth, mostly at large radii. With the merger history, the midplane is strongly perturbed beyond $R \sim 10$~kpc. It is displaced upward at positive azimuth and downward at negative azimuth, with a transition near $\phi \sim 0^\circ$. The displacement grows with radius and reaches the $\pm 0.6$~kpc limit of the color scale in the outer cells, while the inner disk remains close to $z = 0$. The chemical and density midplanes show the same large-scale pattern in each simulation. The density midplane is noisier from cell to cell, with a few outer cells that deviate from their neighbors.

\subsection{Dependence on stellar age}
\label{sec:age_dependence}

The literature values of the line of nodes follow a rough ordering with the age of the tracer. In the straight-line-of-nodes comparison (Figure~\ref{fig:warp_hist}), the young tracers (the young giants and Cepheids of \citet{Poggio_2025}, and the pulsars of \citet{Yusifov_2004}) have $\phi_{\rm LoN,0} \sim +10^\circ$ to $+15^\circ$, the RC studies lie lower, from $+2^\circ$ to $-5^\circ$, and our older sample gives $-8.5^\circ$.
This mixes samples with different selections, volumes, and fitting methods, so it is difficult to strongly conclude that there is a smooth trend with age.
However, the significant offset of our measurement relative to younger tracers does seem to indicate that the warp phase may depend on stellar age.

The line-of-nodes azimuth we measure differs from the value found by \citet{Poggio_2025} for young giants (Figure~\ref{fig:warp_hist}). One possibility is that the warp phase depends on stellar age.
To have approximate ages for the stars in our sample, we cross-match our sample with large catalogs of ages inferred from SDSS spectroscopy. In general, such methods train on age labels inferred from asteroseismology and then use machine learning to infer ages for red giants using spectroscopic observables including carbon, nitrogen, temperature, metallicity, surface gravity, and $\alpha$-element abundances. These ages are known to be less accurate for the oldest and youngest stars \citep{Mackereth_2019, Tayar_2026}, but in our case we are most interested in a general separation between old and young populations. For this, we used two catalogs. The first was an updated version of the \citet{Tayar_2026} neural network run on the internally available (0.8.0) release and trained using the over 30,000 stars with spectroscopy in that release that also have ages from \citet{Theodoridis_2026} derived using results from \citet{Hon_2021}. Matching to this catalog (JT26) using \texttt{sdss\_id}, we found that of
 our 438{,}499 stars, 394{,}052 (89.9\%) have an age. The median age is 4.16~Gyr, and the 16th to 84th percentile range is 2.2 to 6.7~Gyr. We repeat the match with the StarFlow catalog from \citet{Stone-Martinez_2025}, which contains age estimates that use normalizing flows trained on APOKASC-3 \citep{Pinsonneault_2025} and APO-K2 data \citep{SchonhutStasik_2024, Warfield_2024}. This catalog gives ages for 184{,}017 of our stars (42.0\%), with a median of 4.36~Gyr and a 16th to 84th percentile range of 2.3 to 6.5~Gyr. A total of 171{,}053 stars have an age in both catalogs, and 407{,}016 (92.8\%) have an age in at least one. For the 171{,}053 stars with an age in both catalogs, the StarFlow ages are younger than the JT26 ages by a median of 0.50~Gyr, with a robust scatter of 1.3~Gyr (Pearson $r = 0.79$). The ages agree to within 1~Gyr for 58\% of these stars and to within 2~Gyr for 89\%. The offset depends on age: StarFlow and JT26 agree for stars younger than 2~Gyr (median difference $+0.07$~Gyr), but StarFlow gives progressively younger ages for older stars (median difference of $-0.94$~Gyr at 6--8~Gyr and $-1.10$~Gyr at 8--14~Gyr, in bins of the JT26 age). We use these age catalogs only to place our sample on a coarse scale of ``older'' compared to the young giants and Cepheids used in other studies of the warp (e.g., \citealt{Poggio_2025, Skowron_2019a}), and we use them nowhere else in this work. None of our measurements of $z_0$ or the warp parameters depend on the ages. The two catalogs place 82.9\% of the overlapping stars on the same side of 3~Gyr. The two catalogs give similar median ages for the sample as a whole. The stars without an age are a minority in the first catalog, and we assume that the stars with ages are representative of the whole. On this scale, most of our sample is older than 3~Gyr: 72.2\% of the stars with an JT26 age are 3~Gyr or older. We therefore treat our sample as an older population than the young giants of \citet{Poggio_2025} or classical Cepheids \citep[e.g.][]{Skowron_2019a}.

As discussed above, the variation of the warp line-of-nodes based on different samples might be due (at least in part) to selection effects, extinction, distance uncertainties and different fitting techniques. However, in our Galaxy, young and old stellar populations may exhibit distinct structural properties. For instance, the geometry of young stellar populations is expected to reflect the gas clouds from which they formed. However, stars and gas may react in a different way to a given perturbation, potentially leading to a variation in the line-of-nodes across different stellar populations. Using Milky Way analogs from the TNG50 simulation, \citet{Thulasidharan_2025} propose that an age-dependent offset in the line of nodes is a dynamical signature of a past major merger. They find that galaxies with a major merger develop a persistent azimuthal misalignment between the warps of stellar populations formed before and after the merger, even after correcting for differential rotation. Galaxies without such a merger show no comparable age-dependent offset. The Milky Way has its own merger history, including the \gaia-Sausage-Enceladus event \citep{Belokurov_2018, Helmi_2018}, so the offset we find relative to the young giants of \citet{Poggio_2025} may be complicated by past merger-induced structure.

\begin{figure*}
    \centering
    \includegraphics[width=\textwidth]{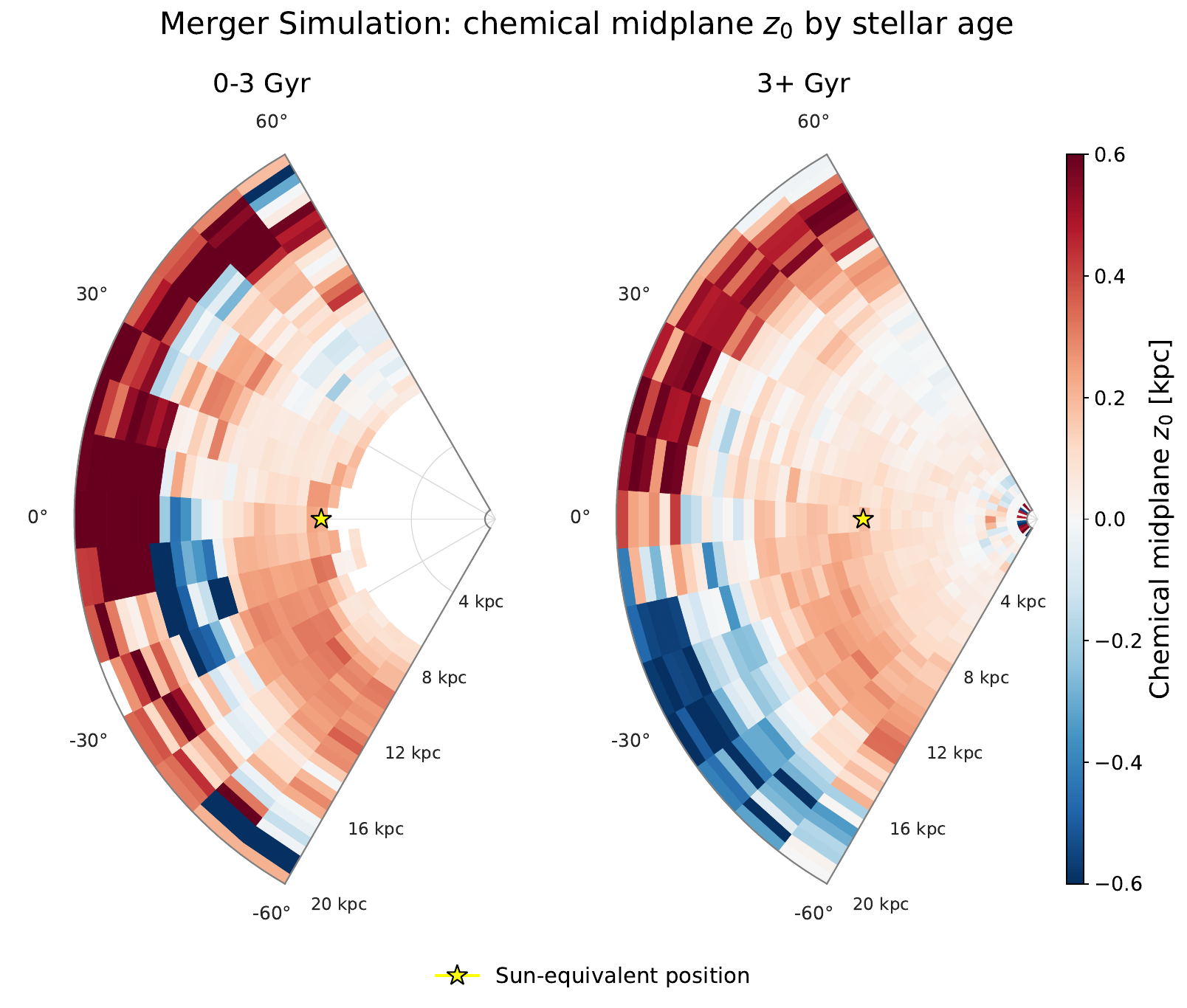}
    \caption{Chemical midplane height $z_0$ in the merger simulation for stars with ages 0--3~Gyr (left) and 3+~Gyr (right), shown in polar cells of radius and azimuth out to 20~kpc. Cells without a fit are left blank. The color scale is limited to $\pm0.6$~kpc. The yellow star marks the Sun-equivalent position.}
    \label{fig:sim_age_bins}
\end{figure*}

To see whether an age dependence of the phase could arise from a merger in our simulations, we fit the same twisted warp model to the young (0--3~Gyr) and old (3+~Gyr) stars of the merger simulation (Figure~\ref{fig:sim_age_bins}). Figure~\ref{fig:sim_age_bins} shows the chemical midplane height $z_0$ in the merger simulation comparing age bins 0-3 (left) and 3+ (right). The simulated warp is not the same as the one in our data, and we do not expect it to be. However, fitting only $R > 12$~kpc (this ignores the inner disk which looks clearly different from our own), we find $\phi_{\rm LoN,0} = 26.5^\circ$ for the young stars and $8.9^\circ$ for the old stars. The offset of $\sim 18^\circ$ has the same sign as the offset between the young giants of \citet{Poggio_2025} and our older sample, and it is of similar size to it. This suggests that a difference in line-of-nodes azimuth between young and old stars can arise in a simulated merger. However, looking at $R > 12$~kpc for this is somewhat of an arbitrary choice to try to measure the LoN in a region of the simulations that best resembles the Milky Way's observed warp, and should be treated with caution. We test only one simulation, and the young and old simulated populations are defined differently from the observed samples. We caution against over-interpreting a single comparison, and we leave a more detailed, age-resolved analysis to future work.

\section{Description of our sample with a guaranteed separable selection function}
\label{app:separable-selfunc}

Here we describe the magnitude cut we apply in Section~\ref{sec:sample}, which is designed to limit biases in our measured midplane and inferred warp model parameters from extinction combined with a magnitude-limited selection (from the requirement that stars must have a \gaia\ XP spectrum). The cut makes the selection function separable in each cell, up to uncertainties on the distances and extinctions.

Our analysis assumes that selection effects separate into a product $S_z(z)\,S_Y(Y)$, which is likely weakly violated in a magnitude-limited sample.
Stars can only appear in our sample if they are bright enough to be observed by \apogee\ and MWM and, more importantly, bright enough (in apparent magnitude) to have a \gaia\ XP spectrum.
Nearly all stars with $G \lesssim 17.5$ have XP spectra \citep{Castro-Ginard_2023, Kurbatov:2024}.
At fixed \teff\ and \logg, the luminosity of a giant depends on its metallicity and its mass (and therefore age), and both correlate with \mgfe.
Extinction varies strongly with Galactic latitude, so for a star close to the XP magnitude limit, its presence in our sample depends both on its position and on properties that correlate with \mgfe.
If we write the total selection function as $S_z(z)\,S_Y(Y)\,[1 + \epsilon(z, Y)]$ (separable up to a correction term), then to first order in $\epsilon$ the observed mean abundance is $\langle Y\rangle(z) + \delta(z)$, where $\delta(z)$ is the covariance between abundance and $\epsilon$ for stars at position $z$.
This shifts the minimum by $\Delta z_0 \approx -\delta'(z_0)/\langle Y\rangle''(z_0)$, so only the asymmetric part of $\delta$ affects our inferred midplane values (symmetric parts change the scale and width of the trend, which we do not study here).
Extinction is the most likely source of asymmetric terms in $\delta$.
Dust is concentrated toward the Galactic plane, so $\delta(z)$ is approximately symmetric about the dust layer (near $b = 0$, which is mostly nearer to the Sun), not about $z_0$ (the midplane at a given radius beyond the bulk of the dust).
In cells where the warp displaces $z_0$ from the dust layer, $\delta'(z_0) \neq 0$ and $z_0$ is pulled toward or pushed away from the dust layer.
This shift vanishes where the midplane crosses the dust layer, so it should mainly bias the warp amplitude, $h_{w,12}$, rather than the line of nodes.
To remove this effect by construction, we apply the cut described below to our fiducial sample.

The ZGR catalog reports a scalar reddening $E$ for each star, which is on the scale of
the Bayestar19 3D dust map \citep{Green_2019}.
The ZGR model then treats the observed flux of each star as $f_{\rm obs}(\lambda) =
f_0(\lambda)\,\exp[-E\,R(\lambda)]$, where $R(\lambda)$ is a universal extinction curve
that they learn from the data, so the extinction in magnitudes at wavelength $\lambda$
is $A_\lambda = (2.5/\ln 10)\,E\,R(\lambda)$.
\gaia\ $G$ is a broad band, so we compute the band-integrated extinction as
\begin{equation}
    A_G(E) = -2.5\log_{10}\left[
        \frac{\int \dd\lambda\, T_G(\lambda)\,\lambda\, f_0(\lambda)\, e^{-E\,R(\lambda)}}
             {\int \dd\lambda\, T_G(\lambda)\,\lambda\, f_0(\lambda)}
    \right] \quad ,
    \label{eq:AG}
\end{equation}
where $T_G$ is the \gaia\ DR3 $G$ passband \citep{Riello_2021} and $f_0$ is the
unreddened spectrum of the star, which we approximate as a blackbody at its \teff.
We extrapolate $R(\lambda)$ from its bluest tabulated wavelength (392~nm) to the blue
edge of the $G$ band using a quadratic fit in $1/\lambda$.
Because the effective wavelength of the band shifts to the red as extinction increases,
$A_G/E$ decreases from $\approx 2.3$ at low extinction to $\approx 1.9$ at $E = 3$
(Table~\ref{tab:AG_coeffs}).
We interpolate this table in \teff\ (from \textsc{ASPCAP}) and $E$ to compute $A_G$ for
each star.

\begin{deluxetable}{lccccc}
\tablecaption{Band-integrated \gaia\ $G$ extinction per unit ZGR reddening, $A_G/E$, as a
function of $E$ and \teff\ (Equation~\ref{eq:AG}). \label{tab:AG_coeffs}}
\tablehead{
\colhead{\teff\ (K)} & \colhead{$E=0.01$} & \colhead{$0.5$} & \colhead{$1.0$} &
\colhead{$2.0$} & \colhead{$3.0$}
}
\startdata
4000 & 2.232 & 2.134 & 2.050 & 1.921 & 1.825 \\
4500 & 2.322 & 2.211 & 2.118 & 1.974 & 1.868 \\
5000 & 2.401 & 2.279 & 2.177 & 2.020 & 1.905 \\
\enddata
\end{deluxetable}

For each star in the sample before this cut, we compute the distance modulus from the ZGR
parallax, $\mu = 5\log_{10}\left(\frac{1~\unit{mas}}{\varpi}\right) + 10$, and we
compute its $G$-band extinction $A_G$ as described above.
In each cell, we define a worst-case distance modulus, $\mu_{\max}$, as the 95th
percentile of $\mu$ over all stars in the cell, and a worst-case extinction,
$A_{G,\max}$, as the 95th percentile of $A_G$, over the stars in the cell.
We keep a star if it would remain above the XP $G$ limit at this worst-case position and
extinction,
\begin{equation}
    G + (\mu_{\max} - \mu) + (A_{G,\max} - A_G) < 17.5 \quad ,
\end{equation}
which is equivalent to requiring that its absolute, dereddened magnitude is
$M_{G,0} < 17.5 - \mu_{\max} - A_{G,\max}$ (so this is essentially a cut on the
intrinsic luminosity).

Because this is a cut on intrinsic luminosity, which is the same at all heights within a cell, the remaining sample satisfies $S = S_z(z)\,S_Y(Y)$ (up to uncertainties on the distances and extinctions).
We use this rather than a cut on extinction or reddening alone (e.g., keeping stars with low reddening) because the latter removes stars preferentially near the plane, which are the most constraining for $z_0$.
The cut retains 78.8\% of our stars, so most of the sample consists of luminous giants that would be observable anywhere in their cell given the worst-case extinction toward the outer disk, and the stars that are removed are mostly lower giant-branch stars toward the outer disk.

We compute the percentiles in cells built before the cut, and rebuild the cells afterwards, since the cut removes stars. We rebuild the cells after the cut rather than keeping the original ones, because the cut leaves some of the original cells with too few stars to fit, which would leave gaps in the map. Of the 150 cells in the final sample, 143 are identical to the cells used for the cut, and the seven that differ contain only 4\% of the stars.

\bibliography{bib}

@ARTICLE{Majewski:2004,
       author = {{Majewski}, Steven R. and {Ostheimer}, James C. and {Rocha-Pinto}, Helio J. and {Patterson}, Richard J. and {Guhathakurta}, Puragra and {Reitzel}, David},
        title = "{Detection of the Main-Sequence Turnoff of a Newly Discovered Milky Way Halo Structure in the Triangulum-Andromeda Region}",
      journal = {\apj},
         year = 2004,
        month = nov,
       volume = {615},
       number = {2},
        pages = {738-743},
          doi = {10.1086/424586},
archivePrefix = {arXiv},
       eprint = {astro-ph/0406221},
 primaryClass = {astro-ph},
       adsurl = {https://ui.adsabs.harvard.edu/abs/2004ApJ...615..738M}
}

@ARTICLE{Vislosky:2024,
       author = {{Vislosky}, E. and {Minchev}, I. and {Khoperskov}, S. and {Martig}, M. and {Buck}, T. and {Hilmi}, T. and {Ratcliffe}, B. and {Bland-Hawthorn}, J. and {Quillen}, A.~C. and {Steinmetz}, M. and {de Jong}, R.},
        title = "{Gaia DR3 data consistent with a short bar connected to a spiral arm}",
      journal = {\mnras},
         year = 2024,
        month = feb,
       volume = {528},
       number = {2},
        pages = {3576-3591},
          doi = {10.1093/mnras/stae083},
archivePrefix = {arXiv},
       eprint = {2312.03854},
 primaryClass = {astro-ph.GA},
       adsurl = {https://ui.adsabs.harvard.edu/abs/2024MNRAS.528.3576V}
}

@ARTICLE{Laporte:2020,
       author = {{Laporte}, Chervin F.~P. and {Famaey}, Benoit and {Monari}, Giacomo and {Hill}, Vanessa and {Wegg}, Christopher and {Gerhard}, Ortwin},
        title = "{Bar resonances and low angular momentum moving groups in the Galaxy revealed by their stellar ages}",
      journal = {\aap},
         year = 2020,
        month = nov,
       volume = {643},
          eid = {L3},
        pages = {L3},
          doi = {10.1051/0004-6361/202038740},
archivePrefix = {arXiv},
       eprint = {2006.13876},
 primaryClass = {astro-ph.GA},
       adsurl = {https://ui.adsabs.harvard.edu/abs/2020A&A...643L...3L}
}

@ARTICLE{Sun:2026,
       author = {{Sun}, Yan and {Zhang}, Shaobo and {Yang}, Ji and {Su}, Yang and {Yan}, Qing-Zeng and {Zhou}, Xin and {Chen}, Xuepeng and {Fang}, Min and {Du}, Fujun},
        title = "{Widespread corrugations superimposed on the Galactic warp in the outer CO disk}",
      journal = {Nature Astronomy},
         year = 2026,
        month = oct,
       volume = {10},
        pages = {1560-1569},
          doi = {10.1038/s41550-026-02938-y},
       adsurl = {https://ui.adsabs.harvard.edu/abs/2026NatAs..10.1560S}
}

@ARTICLE{Rocha-Pinto:2003,
       author = {{Rocha-Pinto}, Helio J. and {Majewski}, Steven R. and {Skrutskie}, M.~F. and {Crane}, Jeffrey D.},
        title = "{Tracing the Galactic Anticenter Stellar Stream with 2MASS M Giants}",
      journal = {\apjl},
         year = 2003,
        month = sep,
       volume = {594},
       number = {2},
        pages = {L115-L118},
          doi = {10.1086/378668},
       adsurl = {https://ui.adsabs.harvard.edu/abs/2003ApJ...594L.115R}
}

@ARTICLE{Newberg:2002,
       author = {{Newberg}, Heidi Jo and {Yanny}, Brian and {Rockosi}, Connie and {Grebel}, Eva K. and {Rix}, Hans-Walter and {Brinkmann}, Jon and {Csabai}, Istvan and {Hennessy}, Greg and {Hindsley}, Robert B. and {Ibata}, Rodrigo and {Ivezi{\'c}}, Zeljko and {Lamb}, Don and {Nash}, E. Thomas and {Odenkirchen}, Michael and {Rave}, Heather A. and {Schneider}, D.~P. and {Smith}, J. Allyn and {Stolte}, Andrea and {York}, Donald G.},
        title = "{The Ghost of Sagittarius and Lumps in the Halo of the Milky Way}",
      journal = {\apj},
         year = 2002,
        month = apr,
       volume = {569},
       number = {1},
        pages = {245-274},
          doi = {10.1086/338983},
archivePrefix = {arXiv},
       eprint = {astro-ph/0111095},
 primaryClass = {astro-ph},
       adsurl = {https://ui.adsabs.harvard.edu/abs/2002ApJ...569..245N}
}

@ARTICLE{Khanna:2019,
       author = {{Khanna}, Shourya and {Sharma}, Sanjib and {Tepper-Garcia}, Thor and {Bland-Hawthorn}, Joss and {Hayden}, Michael and {Asplund}, Martin and {Buder}, Sven and {Chen}, Boquan and {De Silva}, Gayandhi M. and {Freeman}, Ken C. and {Kos}, Janez and {Lewis}, Geraint F. and {Lin}, Jane and {Martell}, Sarah L. and {Simpson}, Jeffrey D. and {Nordlander}, Thomas and {Stello}, Dennis and {Ting}, Yuan-Sen and {Zucker}, Daniel B. and {Zwitter}, Toma{\v{z}}},
        title = "{The GALAH survey and Gaia DR2: Linking ridges, arches, and vertical waves in the kinematics of the Milky Way}",
      journal = {\mnras},
         year = 2019,
        month = nov,
       volume = {489},
       number = {4},
        pages = {4962-4979},
          doi = {10.1093/mnras/stz2462},
archivePrefix = {arXiv},
       eprint = {1902.10113},
 primaryClass = {astro-ph.GA},
       adsurl = {https://ui.adsabs.harvard.edu/abs/2019MNRAS.489.4962K}
}

@ARTICLE{Kawata:2018,
       author = {{Kawata}, Daisuke and {Baba}, Junichi and {Ciuc{\v{a}}}, Ioana and {Cropper}, Mark and {Grand}, Robert J.~J. and {Hunt}, Jason A.~S. and {Seabroke}, George},
        title = "{Radial distribution of stellar motions in Gaia DR2}",
      journal = {\mnras},
         year = 2018,
        month = sep,
       volume = {479},
       number = {1},
        pages = {L108-L112},
          doi = {10.1093/mnrasl/sly107},
archivePrefix = {arXiv},
       eprint = {1804.10175},
 primaryClass = {astro-ph.GA},
       adsurl = {https://ui.adsabs.harvard.edu/abs/2018MNRAS.479L.108K}
}

@ARTICLE{Hunt:2022,
       author = {{Hunt}, Jason A.~S. and {Price-Whelan}, Adrian M. and {Johnston}, Kathryn V. and {Darragh-Ford}, Elise},
        title = "{Multiple phase spirals suggest multiple origins in Gaia DR3}",
      journal = {\mnras},
         year = 2022,
        month = oct,
       volume = {516},
       number = {1},
        pages = {L7-L11},
          doi = {10.1093/mnrasl/slac082},
archivePrefix = {arXiv},
       eprint = {2206.06125},
 primaryClass = {astro-ph.GA},
       adsurl = {https://ui.adsabs.harvard.edu/abs/2022MNRAS.516L...7H}
}

@ARTICLE{Poggio:2020,
       author = {{Poggio}, E. and {Drimmel}, R. and {Andrae}, R. and {Bailer-Jones}, C.~A.~L. and {Fouesneau}, M. and {Lattanzi}, M.~G. and {Smart}, R.~L. and {Spagna}, A.},
        title = "{Evidence of a dynamically evolving Galactic warp}",
      journal = {Nature Astronomy},
         year = 2020,
        month = mar,
       volume = {4},
        pages = {590-596},
          doi = {10.1038/s41550-020-1017-3},
archivePrefix = {arXiv},
       eprint = {1912.10471},
 primaryClass = {astro-ph.GA},
       adsurl = {https://ui.adsabs.harvard.edu/abs/2020NatAs...4..590P}
}

@ARTICLE{JonssonMcMillan2024,
       author = {{J{\'o}nsson}, Viktor Hrannar and {McMillan}, Paul J.},
        title = "{The tangled warp of the Milky Way}",
      journal = {\aap},
         year = 2024,
        month = aug,
       volume = {688},
          eid = {A38},
        pages = {A38},
          doi = {10.1051/0004-6361/202449744},
archivePrefix = {arXiv},
       eprint = {2405.09624},
 primaryClass = {astro-ph.GA},
       adsurl = {https://ui.adsabs.harvard.edu/abs/2024A&A...688A..38J}
}

@ARTICLE{Antoja:2023,
       author = {{Antoja}, T. and {Ramos}, P. and {Garc{\'\i}a-Conde}, B. and {Bernet}, M. and {Laporte}, C.~F.~P. and {Katz}, D.},
        title = "{The phase spiral in Gaia DR3}",
      journal = {\aap},
         year = 2023,
        month = may,
       volume = {673},
          eid = {A115},
        pages = {A115},
          doi = {10.1051/0004-6361/202245518},
archivePrefix = {arXiv},
       eprint = {2212.11987},
 primaryClass = {astro-ph.GA},
       adsurl = {https://ui.adsabs.harvard.edu/abs/2023A&A...673A.115A}
}

@ARTICLE{Antoja:2018,
       author = {{Antoja}, T. and {Helmi}, A. and {Romero-G{\'o}mez}, M. and {Katz}, D. and {Babusiaux}, C. and {Drimmel}, R. and {Evans}, D.~W. and {Figueras}, F. and {Poggio}, E. and {Reyl{\'e}}, C. and {Robin}, A.~C. and {Seabroke}, G. and {Soubiran}, C.},
        title = "{A dynamically young and perturbed Milky Way disk}",
      journal = {\nat},
         year = 2018,
        month = sep,
       volume = {561},
       number = {7723},
        pages = {360-362},
          doi = {10.1038/s41586-018-0510-7},
archivePrefix = {arXiv},
       eprint = {1804.10196},
 primaryClass = {astro-ph.GA},
       adsurl = {https://ui.adsabs.harvard.edu/abs/2018Natur.561..360A}
}

@article{DR2-DPACP-36,
	Adsurl = {https://ui.adsabs.harvard.edu/abs/2018A&A...616A...1G},
	Archiveprefix = {arXiv},
	Author = {{Gaia Collaboration} and {Brown}, A.~G.~A. and {Vallenari}, A. and {Prusti}, T. and {de Bruijne}, J.~H.~J. and {Babusiaux}, C. and {Bailer-Jones}, C.~A.~L.},
	Doi = {10.1051/0004-6361/201833051},
	Eid = {A1},
	Eprint = {1804.09365},
	Journal = {\aap},
	Month = aug,
	Pages = {A1},
	Primaryclass = {astro-ph.GA},
	Title = {{Gaia Data Release 2. Summary of the contents and survey properties}},
	Volume = {616},
	Year = 2018}

@ARTICLE{Momany:2006,
       author = {{Momany}, Y. and {Zaggia}, S. and {Gilmore}, G. and {Piotto}, G. and {Carraro}, G. and {Bedin}, L.~R. and {de Angeli}, F.},
        title = "{Outer structure of the Galactic warp and flare: explaining the Canis Major over-density}",
      journal = {\aap},
         year = 2006,
        month = may,
       volume = {451},
       number = {2},
        pages = {515-538},
          doi = {10.1051/0004-6361:20054081},
archivePrefix = {arXiv},
       eprint = {astro-ph/0603385},
 primaryClass = {astro-ph},
       adsurl = {https://ui.adsabs.harvard.edu/abs/2006A&A...451..515M}
}

@ARTICLE{Li:2019,
       author = {{Li}, Chengdong and {Zhao}, Gang and {Jia}, Yunpeng and {Liao}, Shilong and {Yang}, Chengqun and {Wang}, Qixun},
        title = "{Flare and Warp of Galactic Disk with OB Stars from Gaia DR2}",
      journal = {\apj},
         year = 2019,
        month = feb,
       volume = {871},
       number = {2},
          eid = {208},
        pages = {208},
          doi = {10.3847/1538-4357/aafa17},
       adsurl = {https://ui.adsabs.harvard.edu/abs/2019ApJ...871..208L}
}

@article{Poggio_2025,
   title={The great wave: Evidence of a large-scale vertical corrugation propagating outwards in the Galactic disc},
   volume={699},
   ISSN={1432-0746},
   url={http://dx.doi.org/10.1051/0004-6361/202451668},
   DOI={10.1051/0004-6361/202451668},
   journal={\aap},
   publisher={EDP Sciences},
   author={Poggio, E. and Khanna, S. and Drimmel, R. and Zari, E. and D’Onghia, E. and Lattanzi, M. G. and Palicio, P. A. and Recio-Blanco, A. and Thulasidharan, L.},
   year={2025},
   month=July, pages={A199} }

@ARTICLE{Laporte_2022,
       author = {{Laporte}, Chervin F.~P. and {Koposov}, Sergey E. and {Belokurov}, Vasily},
        title = "{Kinematics beats dust: unveiling nested substructure in the perturbed outer disc of the Milky Way}",
      journal = {\mnras},
         year = 2022,
        month = feb,
       volume = {510},
       number = {1},
        pages = {L13-L17},
          doi = {10.1093/mnrasl/slab109},
archivePrefix = {arXiv},
       eprint = {2103.12737},
 primaryClass = {astro-ph.GA},
       adsurl = {https://ui.adsabs.harvard.edu/abs/2022MNRAS.510L..13L}
}

@ARTICLE{Horta_2024,
       author = {{Horta}, Danny and {Price-Whelan}, Adrian M. and {Hogg}, David W. and {Johnston}, Kathryn V. and {Widrow}, Lawrence and {Dalcanton}, Julianne J. and {Ness}, Melissa K. and {Hunt}, Jason A.~S.},
        title = "{Orbital Torus Imaging: Acceleration, Density, and Dark Matter in the Galactic Disk Measured with Element Abundance Gradients}",
      journal = {\apj},
         year = 2024,
        month = feb,
       volume = {962},
       number = {2},
          eid = {165},
        pages = {165},
          doi = {10.3847/1538-4357/ad16e8},
archivePrefix = {arXiv},
       eprint = {2312.07664},
 primaryClass = {astro-ph.GA},
       adsurl = {https://ui.adsabs.harvard.edu/abs/2024ApJ...962..165H}
}

@dataset{Riello_2021,
       author = {{Riello}, M. and {de Angeli}, F. and {Evans}, D.~W. and {Montegriffo}, P. and {Carrasco}, J.~M. and {Busso}, G. and {Palaversa}, L. and {Burgess}, P. and {Diener}, C. and {Davidson}, M. and {Rowell}, N. and {Fabricius}, C. and {Jordi}, C. and {Bellazzini}, M. and {Pancino}, E. and {Harrison}, D.~L. and {Cacciari}, C. and {van Leeuwen}, F. and {Hambly}, N.~C. and {Hodgkin}, S.~T. and {Osborne}, P.~J. and {Altavilla}, G. and {Barstow}, M.~A. and {Brown}, A.~G.~A. and {Castellani}, M. and {Cowell}, S. and {de Luise}, F. and {Gilmore}, G. and {Giuffrida}, G. and {Hidalgo}, S. and {Holland}, G. and {Marinoni}, S. and {Pagani}, C. and {Piersimoni}, A.~M. and {Pulone}, L. and {Ragaini}, S. and {Rainer}, M. and {Richards}, P.~J. and {Sanna}, N. and {Walton}, N.~A. and {Weiler}, M. and {Yoldas}, A.},
        title = "{VizieR Online Data Catalog: Gaia Early Data Release 3 photometric passbands (Riello+, 2021)}",
 howpublished = {VizieR On-line Data Catalog: J/A+A/649/A3. Originally published in: 2021A\&A...649A...3R},
         year = 2021,
        month = jan,
          eid = {J/A+A/649/A3},
          doi = {10.26093/cds/vizier.36490003},
    publisher = {CDS},
       adsurl = {https://ui.adsabs.harvard.edu/abs/2021yCat..36490003R}
}

@ARTICLE{Green_2019,
       author = {{Green}, Gregory M. and {Schlafly}, Edward and {Zucker}, Catherine and {Speagle}, Joshua S. and {Finkbeiner}, Douglas},
        title = "{A 3D Dust Map Based on Gaia, Pan-STARRS 1, and 2MASS}",
      journal = {\apj},
         year = 2019,
        month = dec,
       volume = {887},
       number = {1},
          eid = {93},
        pages = {93},
          doi = {10.3847/1538-4357/ab5362},
archivePrefix = {arXiv},
       eprint = {1905.02734},
 primaryClass = {astro-ph.GA},
       adsurl = {https://ui.adsabs.harvard.edu/abs/2019ApJ...887...93G}
}

@ARTICLE{Castro-Ginard_2023,
       author = {{Castro-Ginard}, Alfred and {Brown}, Anthony G.~A. and {Kostrzewa-Rutkowska}, Zuzanna and {Cantat-Gaudin}, Tristan and {Drimmel}, Ronald and {Oh}, Semyeong and {Belokurov}, Vasily and {Casey}, Andrew R. and {Fouesneau}, Morgan and {Khanna}, Shourya and {Price-Whelan}, Adrian M. and {Rix}, Hans-Walter},
        title = "{Estimating the selection function of Gaia DR3 subsamples}",
      journal = {\aap},
         year = 2023,
        month = sep,
       volume = {677},
          eid = {A37},
        pages = {A37},
          doi = {10.1051/0004-6361/202346547},
archivePrefix = {arXiv},
       eprint = {2303.17738},
 primaryClass = {astro-ph.GA},
       adsurl = {https://ui.adsabs.harvard.edu/abs/2023A&A...677A..37C}
}

@ARTICLE{Hawkins_2015,
       author = {{Hawkins}, K. and {Jofr{\'e}}, P. and {Masseron}, T. and {Gilmore}, G.},
        title = "{Using chemical tagging to redefine the interface of the Galactic disc and halo}",
      journal = {\mnras},
         year = 2015,
        month = oct,
       volume = {453},
       number = {1},
        pages = {758-774},
          doi = {10.1093/mnras/stv1586},
archivePrefix = {arXiv},
       eprint = {1507.03604},
 primaryClass = {astro-ph.GA},
       adsurl = {https://ui.adsabs.harvard.edu/abs/2015MNRAS.453..758H}
}

@ARTICLE{Cabrera-Gadea_2024,
       author = {{Cabrera-Gadea}, Mauro and {Mateu}, Cecilia and {Ramos}, Pau and {Romero-G{\'o}mez}, Merc{\'e} and {Antoja}, Teresa and {Aguilar}, Luis},
        title = "{Structure, kinematics, and time evolution of the Galactic warp from Classical Cepheids}",
      journal = {\mnras},
         year = 2024,
        month = mar,
       volume = {528},
       number = {3},
        pages = {4409-4431},
          doi = {10.1093/mnras/stae308},
archivePrefix = {arXiv},
       eprint = {2401.13736},
 primaryClass = {astro-ph.GA},
       adsurl = {https://ui.adsabs.harvard.edu/abs/2024MNRAS.528.4409C}
}

@ARTICLE{Sun_2025,
       author = {{Sun}, Weixiang and {Shen}, Han and {Jiang}, Biwei and {Liu}, Xiaowei},
        title = "{The Galactic Disk North─South Asymmetry in Metallicity May Be a New Tracer for the Disk Warp}",
      journal = {\apjl},
         year = 2025,
        month = jan,
       volume = {978},
       number = {2},
          eid = {L24},
        pages = {L24},
          doi = {10.3847/2041-8213/ada02a},
archivePrefix = {arXiv},
       eprint = {2412.12876},
 primaryClass = {astro-ph.GA},
       adsurl = {https://ui.adsabs.harvard.edu/abs/2025ApJ...978L..24S}
}

@ARTICLE{Khanna_2025,
       author = {{Khanna}, Shourya and {Yu}, Jie and {Drimmel}, Ronald and {Poggio}, Eloisa and {Cantat-Gaudin}, Tristan and {Castro-Ginard}, Alfred and {Kurbatov}, Evgeny and {Belokurov}, Vasily and {Brown}, Anthony and {Fouesneau}, Morgan and {Casey}, Andrew and {Rix}, Hans-Walter},
        title = "{GaiaUnlimited: The old stellar disc of the Milky Way as traced by the red clump}",
      journal = {\aap},
         year = 2025,
        month = sep,
       volume = {701},
          eid = {A270},
        pages = {A270},
          doi = {10.1051/0004-6361/202452798},
archivePrefix = {arXiv},
       eprint = {2410.22036},
 primaryClass = {astro-ph.GA},
       adsurl = {https://ui.adsabs.harvard.edu/abs/2025A&A...701A.270K}
}

@ARTICLE{Cabrera-Gadea_2025,
       author = {{Cabrera-Gadea}, Mauro and {Mateu}, Cecilia and {Ramos}, Pau},
        title = "{RR Lyrae stars trace the Milky Way warp}",
      journal = {\aap},
         year = 2025,
        month = sep,
       volume = {701},
          eid = {A136},
        pages = {A136},
          doi = {10.1051/0004-6361/202452736},
archivePrefix = {arXiv},
       eprint = {2410.22427},
 primaryClass = {astro-ph.GA},
       adsurl = {https://ui.adsabs.harvard.edu/abs/2025A&A...701A.136C}
}

@ARTICLE{Uppal_2024,
       author = {{Uppal}, Namita and {Ganesh}, Shashikiran and {Schultheis}, Mathias},
        title = "{Warp and flare of the old Galactic disc as traced by the red clump stars}",
      journal = {\mnras},
         year = 2024,
        month = jan,
       volume = {527},
       number = {3},
        pages = {4863-4873},
          doi = {10.1093/mnras/stad3525},
archivePrefix = {arXiv},
       eprint = {2311.09616},
 primaryClass = {astro-ph.GA},
       adsurl = {https://ui.adsabs.harvard.edu/abs/2024MNRAS.527.4863U}
}

@misc{Kushwahaa_2026,
      title={Milky Way's warped disc traced by AGB stars},
      author={Tanya Kushwahaa and Mikako Matsuura and Jason A. S. Hunt and Daisuke Kawata and Roger Wesson and Timothy A. Davis},
      year={2026},
      eprint={2606.10235},
      archivePrefix={arXiv},
      primaryClass={astro-ph.GA},
      url={https://arxiv.org/abs/2606.10235},
}

@ARTICLE{Romero-Gomez_2019,
       author = {{Romero-G{\'o}mez}, M. and {Mateu}, C. and {Aguilar}, L. and {Figueras}, F. and {Castro-Ginard}, A.},
        title = "{Gaia kinematics reveal a complex lopsided and twisted Galactic disc warp}",
      journal = {\aap},
         year = 2019,
        month = jul,
       volume = {627},
          eid = {A150},
        pages = {A150},
          doi = {10.1051/0004-6361/201834908},
archivePrefix = {arXiv},
       eprint = {1812.07576},
 primaryClass = {astro-ph.GA},
       adsurl = {https://ui.adsabs.harvard.edu/abs/2019A&A...627A.150R}
}

@ARTICLE{Reyle_2009,
       author = {{Reyl{\'e}}, C. and {Marshall}, D.~J. and {Robin}, A.~C. and {Schultheis}, M.},
        title = "{The Milky Way's external disc constrained by 2MASS star counts}",
      journal = {\aap},
         year = 2009,
        month = mar,
       volume = {495},
       number = {3},
        pages = {819-826},
          doi = {10.1051/0004-6361/200811341},
archivePrefix = {arXiv},
       eprint = {0812.3739},
 primaryClass = {astro-ph},
       adsurl = {https://ui.adsabs.harvard.edu/abs/2009A&A...495..819R}
}

@ARTICLE{Lopez-Corredoira_2002,
       author = {{L{\'o}pez-Corredoira}, M. and {Cabrera-Lavers}, A. and {Garz{\'o}n}, F. and {Hammersley}, P.~L.},
        title = "{Old stellar Galactic disc in near-plane regions according to 2MASS: Scales, cut-off, flare and warp}",
      journal = {\aap},
         year = 2002,
        month = nov,
       volume = {394},
        pages = {883-899},
          doi = {10.1051/0004-6361:20021175},
archivePrefix = {arXiv},
       eprint = {astro-ph/0208236},
 primaryClass = {astro-ph},
       adsurl = {https://ui.adsabs.harvard.edu/abs/2002A&A...394..883L}
}

@ARTICLE{Zari_2025,
       author = {{Zari}, Eleonora and {Villase{\~n}or}, Jaime and {Kounkel}, Marina and {Rix}, Hans-Walter and {Frankel}, Neige and {Tkachenko}, Andrew and {Khoperskov}, Sergey and {D'Onghia}, Elena and {Roman-Lopes}, Alexandre and {Rom{\'a}n-Z{\'u}{\~n}iga}, Carlos and {Stringfellow}, Guy S. and {Tan}, Jonathan C. and {Wofford}, Aida and {Bizyaev}, Dmitry and {Donor}, John and {Fern{\'a}ndez-Trincado}, Jos{\'e} G. and {Morrison}, Sean and {Pan}, Kaike and {Sanchez}, Sebastian F. and {Saydjari}, Andrew},
        title = "{The large-scale kinematics of young stars in the Milky Way disc: First results from SDSS-V}",
      journal = {\aap},
         year = 2025,
        month = nov,
       volume = {703},
          eid = {A303},
        pages = {A303},
          doi = {10.1051/0004-6361/202556444},
archivePrefix = {arXiv},
       eprint = {2509.10387},
 primaryClass = {astro-ph.GA},
       adsurl = {https://ui.adsabs.harvard.edu/abs/2025A&A...703A.303Z}
}

@ARTICLE{Das_2020,
       author = {{Das}, Payel and {Hawkins}, Keith and {Jofr{\'e}}, Paula},
        title = "{Ages and kinematics of chemically selected, accreted Milky Way halo stars}",
      journal = {\mnras},
         year = 2020,
        month = apr,
       volume = {493},
       number = {4},
        pages = {5195-5207},
          doi = {10.1093/mnras/stz3537},
archivePrefix = {arXiv},
       eprint = {1903.09320},
 primaryClass = {astro-ph.GA},
       adsurl = {https://ui.adsabs.harvard.edu/abs/2020MNRAS.493.5195D}
}

@ARTICLE{Horta_2021,
       author = {{Horta}, Danny and {Schiavon}, Ricardo P. and {Mackereth}, J. Ted and {Pfeffer}, Joel and {Mason}, Andrew C. and {Kisku}, Shobhit and {Fragkoudi}, Francesca and {Allende Prieto}, Carlos and {Cunha}, Katia and {Hasselquist}, Sten and {Holtzman}, Jon and {Majewski}, Steven R. and {Nataf}, David and {O'Connell}, Robert W. and {Schultheis}, Mathias and {Smith}, Verne V.},
        title = "{Evidence from APOGEE for the presence of a major building block of the halo buried in the inner Galaxy}",
      journal = {\mnras},
         year = 2021,
        month = jan,
       volume = {500},
       number = {1},
        pages = {1385-1403},
          doi = {10.1093/mnras/staa2987},
archivePrefix = {arXiv},
       eprint = {2007.10374},
 primaryClass = {astro-ph.GA},
       adsurl = {https://ui.adsabs.harvard.edu/abs/2021MNRAS.500.1385H}
}

@ARTICLE{Zari_2021,
       author = {{Zari}, E. and {Rix}, H.-W. and {Frankel}, N. and {Xiang}, M. and {Poggio}, E. and {Drimmel}, R. and {Tkachenko}, A.},
        title = "{Mapping luminous hot stars in the Galaxy}",
      journal = {\aap},
         year = 2021,
        month = jun,
       volume = {650},
          eid = {A112},
        pages = {A112},
          doi = {10.1051/0004-6361/202039726},
archivePrefix = {arXiv},
       eprint = {2102.08684},
 primaryClass = {astro-ph.GA},
       adsurl = {https://ui.adsabs.harvard.edu/abs/2021A&A...650A.112Z}
}

@misc{Poggio_2026,
  title = {The flare and spiral structure of the Milky Way's disc as traced by young giant stars},
  author = {Poggio, E. and Drimmel, R. and Khanna, S. and Andrae, R. and Lattanzi, M. G.},
  year = {2026},
  eprint = {2605.12603},
  archivePrefix = {arXiv},
  primaryClass = {astro-ph.GA},
  url = {https://arxiv.org/abs/2605.12603}
}

@ARTICLE{Zhang_2023,
       author = {{Zhang}, Xiangyu and {Green}, Gregory M. and {Rix}, Hans-Walter},
        title = "{Parameters of 220 million stars from Gaia BP/RP spectra}",
      journal = {\mnras},
         year = 2023,
        month = sep,
       volume = {524},
       number = {2},
        pages = {1855-1884},
          doi = {10.1093/mnras/stad1941},
archivePrefix = {arXiv},
       eprint = {2303.03420},
 primaryClass = {astro-ph.SR},
       adsurl = {https://ui.adsabs.harvard.edu/abs/2023MNRAS.524.1855Z}
}

@ARTICLE{Tayar_2026,
       author = {{Tayar}, Jamie and {Mankowski}, Carli and {Tunca}, Lara and {Jordan}, Dante and {Severino}, Mia and {McArthur}, Sydney and {Benton}, Zeina and {Armstrong}, Sophia and {Leddy}, Alexa and {Bower}, Emily and {Sanchez}, Zabdiel and {Avery}, Colin and {Cummings}, Emily and {Donley}, Joshua and {Freeman}, Rachel and {Fulcher}, David R. and {Hervie}, Vanessa and {Ivey}, James and {Kenney}, Hyde and {MacMillan}, William and {Mahoney}, Jake and {Maramba}, Eve and {Philip}, Erin and {Simpson}, Yazmeen and {Strojie}, Ethan},
        title = "{Evaluating the Sensitivity of the Age Inferences of Red Giant Stars to Machine Learning Methodology}",
      journal = {arXiv e-prints},
         year = 2026,
        month = jun,
          eid = {arXiv:2606.24383},
        pages = {arXiv:2606.24383},
          doi = {10.48550/arXiv.2606.24383},
archivePrefix = {arXiv},
       eprint = {2606.24383},
 primaryClass = {astro-ph.SR},
       adsurl = {https://ui.adsabs.harvard.edu/abs/2026arXiv260624383T}
}

@ARTICLE{Stone-Martinez_2025,
       author = {{Stone-Martinez}, Alexander and {Holtzman}, Jon A. and {Lu}, Yuxi(Lucy) and {Hasselquist}, Sten and {Imig}, Julie and {Griffith}, Emily J. and {Bellinger}, Earl P. and {Saydjari}, Andrew K.},
        title = "{StarFlow: Leveraging Normalizing Flows for Stellar Age Estimation in SDSS-V DR19}",
      journal = {\aj},
         year = 2025,
        month = aug,
       volume = {170},
       number = {2},
          eid = {66},
        pages = {66},
          doi = {10.3847/1538-3881/addd18},
archivePrefix = {arXiv},
       eprint = {2503.03138},
 primaryClass = {astro-ph.SR},
       adsurl = {https://ui.adsabs.harvard.edu/abs/2025AJ....170...66S}
}

@misc{Phan_2019,
      title={Composable Effects for Flexible and Accelerated Probabilistic Programming in NumPyro},
      author={Du Phan and Neeraj Pradhan and Martin Jankowiak},
      year={2019},
      eprint={1912.11554},
      archivePrefix={arXiv},
      primaryClass={stat.ML},
      url={https://arxiv.org/abs/1912.11554},
}

@article{Bingham_2019,
  author  = {Eli Bingham and Jonathan P. Chen and Martin Jankowiak and Fritz Obermeyer and Neeraj Pradhan and Theofanis Karaletsos and Rohit Singh and Paul Szerlip and Paul Horsfall and Noah D. Goodman},
  title   = {Pyro: Deep Universal Probabilistic Programming},
  journal = {Journal of Machine Learning Research},
  year    = {2019},
  volume  = {20},
  number  = {28},
  pages   = {1--6},
  url     = {http://jmlr.org/papers/v20/18-403.html}
}

@software{Bradbury_2018,
    author      = {James Bradbury and Roy Frostig and Peter Hawkins and Matthew James Johnson and Chris Leary and Dougal Maclaurin and George Necula and Adam Paszke and Jake Vander{P}las and Skye Wanderman-{M}ilne and Qiao Zhang},
    title       = {{JAX}: composable transformations of {P}ython+{N}um{P}y programs},
    url         = {http://github.com/google/jax},
    version     = {0.2.5},
    year        = {2018}
}

@article{hoffman_2014,
  author  = {Matthew D. Hoffman and Andrew Gelman},
  title   = {The No-U-Turn Sampler: Adaptively Setting Path Lengths in Hamiltonian Monte Carlo},
  journal = {Journal of Machine Learning Research},
  year    = {2014},
  volume  = {15},
  number  = {47},
  pages   = {1593--1623},
  url     = {http://jmlr.org/papers/v15/hoffman14a.html}
}

@ARTICLE{price-whelan_2021,
       author = {{Price-Whelan}, Adrian M. and {Hogg}, David W. and {Johnston}, Kathryn V. and {Ness}, Melissa K. and {Rix}, Hans-Walter and {Beaton}, Rachael L. and {Brownstein}, Joel R. and {Garc{\'\i}a-Hern{\'a}ndez}, D.~A. and {Hasselquist}, Sten and {Hayes}, Christian R. and {Lane}, Richard R. and {Shetrone}, Matthew and {Sobeck}, Jennifer and {Zasowski}, Gail},
        title = "{Orbital Torus Imaging: Using Element Abundances to Map Orbits and Mass in the Milky Way}",
      journal = {\apj},
         year = 2021,
        month = mar,
       volume = {910},
       number = {1},
          eid = {17},
        pages = {17},
          doi = {10.3847/1538-4357/abe1b7},
archivePrefix = {arXiv},
       eprint = {2012.00015},
 primaryClass = {astro-ph.GA},
       adsurl = {https://ui.adsabs.harvard.edu/abs/2021ApJ...910...17P}
}

@ARTICLE{price-whelan_2025,
       author = {{Price-Whelan}, Adrian M. and {Hunt}, Jason A.~S. and {Horta}, Danny and {Oeur}, Micah and {Hogg}, David W. and {Johnston}, Kathryn and {Widrow}, Lawrence},
        title = "{Data-driven Dynamics with Orbital Torus Imaging: A Flexible Model of the Vertical Phase Space of the Galaxy}",
      journal = {\apj},
         year = 2025,
        month = feb,
       volume = {979},
       number = {2},
          eid = {115},
        pages = {115},
          doi = {10.3847/1538-4357/ad969a},
archivePrefix = {arXiv},
       eprint = {2401.07903},
 primaryClass = {astro-ph.GA},
       adsurl = {https://ui.adsabs.harvard.edu/abs/2025ApJ...979..115P}
}

@ARTICLE{Skowron_2019a,
       author = {{Skowron}, Dorota M. and {Skowron}, Jan and {Mr{\'o}z}, Przemek and {Udalski}, Andrzej and {Pietrukowicz}, Pawe{\l} and {Soszy{\'n}ski}, Igor and {Szyma{\'n}ski}, Micha{\l} K. and {Poleski}, Rados{\l}aw and {Koz{\l}owski}, Szymon and {Ulaczyk}, Krzysztof and {Rybicki}, Krzysztof and {Iwanek}, Patryk},
        title = "{A three-dimensional map of the Milky Way using classical Cepheid variable stars}",
      journal = {Science},
         year = 2019,
        month = aug,
       volume = {365},
       number = {6452},
        pages = {478-482},
          doi = {10.1126/science.aau3181},
archivePrefix = {arXiv},
       eprint = {1806.10653},
 primaryClass = {astro-ph.GA},
       adsurl = {https://ui.adsabs.harvard.edu/abs/2019Sci...365..478S}
}

@ARTICLE{Gelman_1992,
       author = {{Gelman}, Andrew and {Rubin}, Donald B.},
        title = "{Inference from Iterative Simulation Using Multiple Sequences}",
      journal = {Statistical Science},
         year = 1992,
        month = jan,
       volume = {7},
        pages = {457-472},
          doi = {10.1214/ss/1177011136},
       adsurl = {https://ui.adsabs.harvard.edu/abs/1992StaSc...7..457G}
}

@ARTICLE{Bennett_2019,
       author = {{Bennett}, Morgan and {Bovy}, Jo},
        title = "{Vertical waves in the solar neighbourhood in Gaia DR2}",
      journal = {\mnras},
         year = 2019,
        month = jan,
       volume = {482},
       number = {1},
        pages = {1417-1425},
          doi = {10.1093/mnras/sty2813},
archivePrefix = {arXiv},
       eprint = {1809.03507},
 primaryClass = {astro-ph.GA},
       adsurl = {https://ui.adsabs.harvard.edu/abs/2019MNRAS.482.1417B}
}

@ARTICLE{Skowron_2019b,
       author = {{Skowron}, D.~M. and {Skowron}, J. and {Mr{\'o}z}, P. and {Udalski}, A. and {Pietrukowicz}, P. and {Soszy{\'n}ski}, I. and {Szyma{\'n}ski}, M.~K. and {Poleski}, R. and {Koz{\l}owski}, S. and {Ulaczyk}, K. and {Rybicki}, K. and {Iwanek}, P. and {. Wrona}, M. and {Gromadzki}, M.},
        title = "{Mapping the Northern Galactic Disk Warp with Classical Cepheids}",
      journal = {\actaa},
         year = 2019,
        month = dec,
       volume = {69},
       number = {4},
        pages = {305-320},
          doi = {10.32023/0001-5237/69.4.1},
archivePrefix = {arXiv},
       eprint = {1912.11142},
 primaryClass = {astro-ph.GA},
       adsurl = {https://ui.adsabs.harvard.edu/abs/2019AcA....69..305S}
}

@ARTICLE{Chen_2019,
       author = {{Chen}, Xiaodian and {Wang}, Shu and {Deng}, Licai and {de Grijs}, Richard and {Liu}, Chao and {Tian}, Hao},
        title = "{An intuitive 3D map of the Galactic warp's precession traced by classical Cepheids}",
      journal = {Nature Astronomy},
         year = 2019,
        month = feb,
       volume = {3},
        pages = {320-325},
          doi = {10.1038/s41550-018-0686-7},
archivePrefix = {arXiv},
       eprint = {1902.00998},
 primaryClass = {astro-ph.GA},
       adsurl = {https://ui.adsabs.harvard.edu/abs/2019NatAs...3..320C}
}

@ARTICLE{Dehnen_2023,
       author = {{Dehnen}, Walter and {Semczuk}, Marcin and {Sch{\"o}nrich}, Ralph},
        title = "{A twisted and precessing Cepheid warp in the outer Milky Way disc}",
      journal = {\mnras},
         year = 2023,
        month = jul,
       volume = {523},
       number = {1},
        pages = {1556-1564},
          doi = {10.1093/mnras/stad1502},
archivePrefix = {arXiv},
       eprint = {2305.09343},
 primaryClass = {astro-ph.GA},
       adsurl = {https://ui.adsabs.harvard.edu/abs/2023MNRAS.523.1556D}
}

@ARTICLE{Cheng_2020,
       author = {{Cheng}, Xinlun and {Anguiano}, Borja and {Majewski}, Steven R. and {Hayes}, Christian and {Arras}, Phil and {Chiappini}, Cristina and {Hasselquist}, Sten and {de Andrade Queiroz}, Anna B{\'a}rbara and {Nitschelm}, Christian and {Garc{\'\i}a-Hern{\'a}ndez}, Domingo An{\'\i}bal and {Lane}, Richard R. and {Roman-Lopes}, Alexandre and {Frinchaboy}, Peter M.},
        title = "{Exploring the Galactic Warp through Asymmetries in the Kinematics of the Galactic Disk}",
      journal = {\apj},
         year = 2020,
        month = dec,
       volume = {905},
       number = {1},
          eid = {49},
        pages = {49},
          doi = {10.3847/1538-4357/abc3c2},
archivePrefix = {arXiv},
       eprint = {2010.10398},
 primaryClass = {astro-ph.GA},
       adsurl = {https://ui.adsabs.harvard.edu/abs/2020ApJ...905...49C}
}

@INPROCEEDINGS{Yusifov_2004,
       author = {{Yusifov}, I.},
        title = "{Pulsars and the Warp of the Galaxy}",
    booktitle = {The Magnetized Interstellar Medium},
         year = 2004,
       editor = {{Uyaniker}, B. and {Reich}, W. and {Wielebinski}, R.},
        month = feb,
        pages = {165-169},
          doi = {10.48550/arXiv.astro-ph/0405517},
archivePrefix = {arXiv},
       eprint = {astro-ph/0405517},
 primaryClass = {astro-ph},
       adsurl = {https://ui.adsabs.harvard.edu/abs/2004mim..proc..165Y}
}

@ARTICLE{Thulasidharan_2025,
       author = {{Thulasidharan}, Lekshmi and {D'Onghia}, Elena and {Benjamin}, Robert},
        title = "{Azimuthal Misalignments in Stellar Warp Structure as Dynamical Tracers of Mergers in Milky Way─like Galaxies}",
      journal = {\apjl},
         year = 2025,
        month = nov,
       volume = {993},
       number = {1},
          eid = {L28},
        pages = {L28},
          doi = {10.3847/2041-8213/ae11b4},
archivePrefix = {arXiv},
       eprint = {2510.04194},
 primaryClass = {astro-ph.GA},
       adsurl = {https://ui.adsabs.harvard.edu/abs/2025ApJ...993L..28T}
}

@article{Belokurov_2018,
   title={Co-formation of the disc and the stellar halo★},
   volume={478},
   ISSN={1365-2966},
   url={http://dx.doi.org/10.1093/mnras/sty982},
   DOI={10.1093/mnras/sty982},
   number={1},
   journal={Monthly Notices of the Royal Astronomical Society},
   publisher={Oxford University Press (OUP)},
   author={Belokurov, V and Erkal, D and Evans, N W and Koposov, S E and Deason, A J},
   year={2018},
   month=June, pages={611–619} }

@article{Helmi_2018,
   title={The merger that led to the formation of the Milky Way’s inner stellar halo and thick disk},
   volume={563},
   ISSN={1476-4687},
   url={http://dx.doi.org/10.1038/s41586-018-0625-x},
   DOI={10.1038/s41586-018-0625-x},
   number={7729},
   journal={Nature},
   publisher={Springer Science and Business Media LLC},
   author={Helmi, Amina and Babusiaux, Carine and Koppelman, Helmer H. and Massari, Davide and Veljanoski, Jovan and Brown, Anthony G. A.},
   year={2018},
   month=Oct, pages={85–88} }

@ARTICLE{Eisenstein:2011,
       author = {{Eisenstein}, Daniel J. and {Weinberg}, David H. and {Agol}, Eric and {Aihara}, Hiroaki and {Allende Prieto}, Carlos and {Anderson}, Scott F. and {Arns}, James A. and {Aubourg}, {\'E}ric and {Bailey}, Stephen and {Balbinot}, Eduardo and et al.},
        title = "{SDSS-III: Massive Spectroscopic Surveys of the Distant Universe, the Milky Way, and Extra-Solar Planetary Systems}",
      journal = {\aj},
         year = 2011,
        month = sep,
       volume = {142},
       number = {3},
          eid = {72},
        pages = {72},
          doi = {10.1088/0004-6256/142/3/72},
archivePrefix = {arXiv},
       eprint = {1101.1529},
 primaryClass = {astro-ph.IM},
       adsurl = {https://ui.adsabs.harvard.edu/abs/2011AJ....142...72E}
}

@ARTICLE{Blanton:2017,
       author = {{Blanton}, Michael R. and {Bershady}, Matthew A. and {Abolfathi}, Bela and {Albareti}, Franco D. and {Allende Prieto}, Carlos and {Almeida}, Andres and {Alonso-Garc{\'\i}a}, Javier and {Anders}, Friedrich and {Anderson}, Scott F. and {Andrews}, Brett and et al.},
        title = "{Sloan Digital Sky Survey IV: Mapping the Milky Way, Nearby Galaxies, and the Distant Universe}",
      journal = {\aj},
         year = 2017,
        month = jul,
       volume = {154},
       number = {1},
          eid = {28},
        pages = {28},
          doi = {10.3847/1538-3881/aa7567},
archivePrefix = {arXiv},
       eprint = {1703.00052},
 primaryClass = {astro-ph.GA},
       adsurl = {https://ui.adsabs.harvard.edu/abs/2017AJ....154...28B}
}

@ARTICLE{Kollmeier:2026,
       author = {{Kollmeier}, Juna A. and {Rix}, Hans-Walter and {Aerts}, Conny and {Aird}, James and {Vera Alfaro}, Pablo and {Almeida}, Andr{\'e}s and {Anderson}, Scott F. and {Arseneau}, Stefan M. and {Assef}, Roberto J. and {Aviram}, Shir and et al.},
        title = "{Sloan Digital Sky Survey. V. Pioneering Panoptic Spectroscopy}",
      journal = {\aj},
         year = 2026,
        month = jan,
       volume = {171},
       number = {1},
          eid = {52},
        pages = {52},
          doi = {10.3847/1538-3881/ae0576},
archivePrefix = {arXiv},
       eprint = {2507.06989},
 primaryClass = {astro-ph.IM},
       adsurl = {https://ui.adsabs.harvard.edu/abs/2026AJ....171...52K}
}

@ARTICLE{Wilson:2019,
       author = {{Wilson}, J.~C. and {Hearty}, F.~R. and {Skrutskie}, M.~F. and {Majewski}, S.~R. and {Holtzman}, J.~A. and {Eisenstein}, D. and {Gunn}, J. and {Blank}, B. and {Henderson}, C. and {Smee}, S. and et al.},
        title = "{The Apache Point Observatory Galactic Evolution Experiment (APOGEE) Spectrographs}",
      journal = {\pasp},
         year = 2019,
        month = may,
       volume = {131},
       number = {999},
        pages = {055001},
          doi = {10.1088/1538-3873/ab0075},
archivePrefix = {arXiv},
       eprint = {1902.00928},
 primaryClass = {astro-ph.IM},
       adsurl = {https://ui.adsabs.harvard.edu/abs/2019PASP..131e5001W}
}

@ARTICLE{Gunn:2006,
       author = {{Gunn}, James E. and {Siegmund}, Walter A. and {Mannery}, Edward J. and {Owen}, Russell E. and {Hull}, Charles L. and {Leger}, R. French and {Carey}, Larry N. and {Knapp}, Gillian R. and {York}, Donald G. and {Boroski}, William N. and et al.},
        title = "{The 2.5 m Telescope of the Sloan Digital Sky Survey}",
      journal = {\aj},
         year = 2006,
        month = apr,
       volume = {131},
       number = {4},
        pages = {2332-2359},
          doi = {10.1086/500975},
archivePrefix = {arXiv},
       eprint = {astro-ph/0602326},
 primaryClass = {astro-ph},
       adsurl = {https://ui.adsabs.harvard.edu/abs/2006AJ....131.2332G}
}

@ARTICLE{Bowen:1973,
       author = {{Bowen}, I.~S. and {Vaughan}, Jr., A.~H.},
        title = "{The optical design of the 40-in. telescope and of the Ir{\'e}n{\'e}e DuPont telescope at Las Campanas Observatory, Chile.}",
      journal = {\ao},
         year = 1973,
        month = jan,
       volume = {12},
        pages = {1430-1434},
          doi = {10.1364/AO.12.001430},
       adsurl = {https://ui.adsabs.harvard.edu/abs/1973ApOpt..12.1430B}
}

@ARTICLE{Majewski:2017,
       author = {{Majewski}, Steven R. and {Schiavon}, Ricardo P. and {Frinchaboy}, Peter M. and {Allende Prieto}, Carlos and {Barkhouser}, Robert and {Bizyaev}, Dmitry and {Blank}, Basil and {Brunner}, Sophia and {Burton}, Adam and {Carrera}, Ricardo and et al.},
        title = "{The Apache Point Observatory Galactic Evolution Experiment (APOGEE)}",
      journal = {\aj},
         year = 2017,
        month = sep,
       volume = {154},
       number = {3},
          eid = {94},
        pages = {94},
          doi = {10.3847/1538-3881/aa784d},
archivePrefix = {arXiv},
       eprint = {1509.05420},
 primaryClass = {astro-ph.IM},
       adsurl = {https://ui.adsabs.harvard.edu/abs/2017AJ....154...94M}
}

@ARTICLE{Zasowski:2017,
       author = {{Zasowski}, G. and {Cohen}, R.~E. and {Chojnowski}, S.~D. and {Santana}, F. and {Oelkers}, R.~J. and {Andrews}, B. and {Beaton}, R.~L. and {Bender}, C. and {Bird}, J.~C. and {Bovy}, J. and et al.},
        title = "{Target Selection for the SDSS-IV APOGEE-2 Survey}",
      journal = {\aj},
         year = 2017,
        month = nov,
       volume = {154},
       number = {5},
          eid = {198},
        pages = {198},
          doi = {10.3847/1538-3881/aa8df9},
archivePrefix = {arXiv},
       eprint = {1708.00155},
 primaryClass = {astro-ph.GA},
       adsurl = {https://ui.adsabs.harvard.edu/abs/2017AJ....154..198Z}
}

@ARTICLE{Zasowski:2013,
       author = {{Zasowski}, G. and {Johnson}, Jennifer A. and {Frinchaboy}, P.~M. and {Majewski}, S.~R. and {Nidever}, D.~L. and {Rocha Pinto}, H.~J. and {Girardi}, L. and {Andrews}, B. and {Chojnowski}, S.~D. and {Cudworth}, K.~M. and et al.},
        title = "{Target Selection for the Apache Point Observatory Galactic Evolution Experiment (APOGEE)}",
      journal = {\aj},
         year = 2013,
        month = oct,
       volume = {146},
       number = {4},
          eid = {81},
        pages = {81},
          doi = {10.1088/0004-6256/146/4/81},
archivePrefix = {arXiv},
       eprint = {1308.0351},
 primaryClass = {astro-ph.GA},
       adsurl = {https://ui.adsabs.harvard.edu/abs/2013AJ....146...81Z}
}

@ARTICLE{Beaton:2021,
       author = {{Beaton}, Rachael L. and {Oelkers}, Ryan J. and {Hayes}, Christian R. and {Covey}, Kevin R. and {Chojnowski}, S.~D. and {De Lee}, Nathan and {Sobeck}, Jennifer S. and {Majewski}, Steven R. and {Cohen}, Roger E. and {Fern{\'a}ndez-Trincado}, Jos{\'e} and et al.},
        title = "{Final Targeting Strategy for the Sloan Digital Sky Survey IV Apache Point Observatory Galactic Evolution Experiment 2 North Survey}",
      journal = {\aj},
         year = 2021,
        month = dec,
       volume = {162},
       number = {6},
          eid = {302},
        pages = {302},
          doi = {10.3847/1538-3881/ac260c},
archivePrefix = {arXiv},
       eprint = {2108.11907},
 primaryClass = {astro-ph.GA},
       adsurl = {https://ui.adsabs.harvard.edu/abs/2021AJ....162..302B}
}

@ARTICLE{Santana:2021,
       author = {{Santana}, Felipe A. and {Beaton}, Rachael L. and {Covey}, Kevin R. and {O'Connell}, Julia E. and {Longa-Pe{\~n}a}, Pen{\'e}lope and {Cohen}, Roger and {Fern{\'a}ndez-Trincado}, Jos{\'e} G. and {Hayes}, Christian R. and {Zasowski}, Gail and {Sobeck}, Jennifer S. and et al.},
        title = "{Final Targeting Strategy for the SDSS-IV APOGEE-2S Survey}",
      journal = {\aj},
         year = 2021,
        month = dec,
       volume = {162},
       number = {6},
          eid = {303},
        pages = {303},
          doi = {10.3847/1538-3881/ac2cbc},
archivePrefix = {arXiv},
       eprint = {2108.11908},
 primaryClass = {astro-ph.GA},
       adsurl = {https://ui.adsabs.harvard.edu/abs/2021AJ....162..303S}
}

@ARTICLE{DR18:2023,
       author = {{Almeida}, Andr{\'e}s and {Anderson}, Scott F. and {Argudo-Fern{\'a}ndez}, Maria and {Badenes}, Carles and {Barger}, Kat and {Barrera-Ballesteros}, Jorge K. and {Bender}, Chad F. and {Benitez}, Erika and {Besser}, Felipe and {Bird}, Jonathan C. and {Bizyaev}, Dmitry and {Blanton}, Michael R. and {Bochanski}, John and {Bovy}, Jo and {Brandt}, William Nielsen and {Brownstein}, Joel R. and {Buchner}, Johannes and {Bulbul}, Esra and {Burchett}, Joseph N. and {Cano D{\'\i}az}, Mariana and {Carlberg}, Joleen K. and {Casey}, Andrew R. and {Chandra}, Vedant and {Cherinka}, Brian and {Chiappini}, Cristina and {Coker}, Abigail A. and {Comparat}, Johan and {Conroy}, Charlie and {Contardo}, Gabriella and {Cortes}, Arlin and {Covey}, Kevin and {Crane}, Jeffrey D. and {Cunha}, Katia and {Dabbieri}, Collin and {Davidson}, James W. and {Davis}, Megan C. and {de Andrade Queiroz}, Anna Barbara and {De Lee}, Nathan and {M{\'e}ndez Delgado}, Jos{\'e} Eduardo and {Demasi}, Sebastian and {Di Mille}, Francesco and {Donor}, John and {Dow}, Peter and {Dwelly}, Tom and {Eracleous}, Mike and {Eriksen}, Jamey and {Fan}, Xiaohui and {Farr}, Emily and {Frederick}, Sara and {Fries}, Logan and {Frinchaboy}, Peter and {G{\"a}nsicke}, Boris T. and {Ge}, Junqiang and {Gonz{\'a}lez {\'A}vila}, Consuelo and {Grabowski}, Katie and {Grier}, Catherine and {Guiglion}, Guillaume and {Gupta}, Pramod and {Hall}, Patrick and {Hawkins}, Keith and {Hayes}, Christian R. and {Hermes}, J.~J. and {Hern{\'a}ndez-Garc{\'\i}a}, Lorena and {Hogg}, David W. and {Holtzman}, Jon A. and {Ibarra-Medel}, Hector Javier and {Ji}, Alexander and {Jofre}, Paula and {Johnson}, Jennifer A. and {Jones}, Amy M. and {Kinemuchi}, Karen and {Kluge}, Matthias and {Koekemoer}, Anton and {Kollmeier}, Juna A. and {Kounkel}, Marina and {Krishnarao}, Dhanesh and {Krumpe}, Mirko and {Lacerna}, Ivan and {Lago}, Paulo Jakson Assuncao and {Laporte}, Chervin and {Liu}, Chao and {Liu}, Ang and {Liu}, Xin and {Lopes}, Alexandre Roman and {Macktoobian}, Matin and {Majewski}, Steven R. and {Malanushenko}, Viktor and {Maoz}, Dan and {Masseron}, Thomas and {Masters}, Karen L. and {Matijevic}, Gal and {McBride}, Aidan and {Medan}, Ilija and {Merloni}, Andrea and {Morrison}, Sean and {Myers}, Natalie and {M{\'e}sz{\'a}ros}, Szabolcs and {Negrete}, C. Alenka and {Nidever}, David L. and {Nitschelm}, Christian and {Oravetz}, Daniel and {Oravetz}, Audrey and {Pan}, Kaike and {Peng}, Yingjie and {Pinsonneault}, Marc H. and {Pogge}, Rick and {Qiu}, Dan and {Ramirez}, Solange V. and {Rix}, Hans-Walter and {Fern{\'a}ndez Rosso}, Daniela and {Runnoe}, Jessie and {Salvato}, Mara and {Sanchez}, Sebastian F. and {Santana}, Felipe A. and {Saydjari}, Andrew and {Sayres}, Conor and {Schlaufman}, Kevin C. and {Schneider}, Donald P. and {Schwope}, Axel and {Serna}, Javier and {Shen}, Yue and {Sobeck}, Jennifer and {Song}, Ying-Yi and {Souto}, Diogo and {Spoo}, Taylor and {Stassun}, Keivan G. and {Steinmetz}, Matthias and {Straumit}, Ilya and {Stringfellow}, Guy and {S{\'a}nchez-Gallego}, Jos{\'e} and {Taghizadeh-Popp}, Manuchehr and {Tayar}, Jamie and {Thakar}, Ani and {Tissera}, Patricia B. and {Tkachenko}, Andrew and {Hernandez Toledo}, Hector and {Trakhtenbrot}, Benny and {Fern{\'a}ndez-Trincado}, Jos{\'e} G. and {Troup}, Nicholas and {Trump}, Jonathan R. and {Tuttle}, Sarah and {Ulloa}, Natalie and {Vazquez-Mata}, Jose Antonio and {Vera Alfaro}, Pablo and {Villanova}, Sandro and {Wachter}, Stefanie and {Weijmans}, Anne-Marie and {Wheeler}, Adam and {Wilson}, John and {Wojno}, Leigh and {Wolf}, Julien and {Xue}, Xiang-Xiang and {Ybarra}, Jason E. and {Zari}, Eleonora and {Zasowski}, Gail},
        title = "{The Eighteenth Data Release of the Sloan Digital Sky Surveys: Targeting and First Spectra from SDSS-V}",
      journal = {\apjs},
         year = 2023,
        month = aug,
       volume = {267},
       number = {2},
          eid = {44},
        pages = {44},
          doi = {10.3847/1538-4365/acda98},
archivePrefix = {arXiv},
       eprint = {2301.07688},
 primaryClass = {astro-ph.GA},
       adsurl = {https://ui.adsabs.harvard.edu/abs/2023ApJS..267...44A}
}

@ARTICLE{Blanton:2026,
       author = {{Blanton}, Michael R. and {Carlberg}, Joleen K. and {Dwelly}, Tom and {Medan}, Ilija and {Chojnowski}, S. Drew and {Covey}, Kevin and {Davis}, Megan C. and {Donor}, John and {Gupta}, Pramod and {Ji}, Alexander P. and {Johnson}, Jennifer A. and {Kollmeier}, Juna A. and {Sanchez-Gallego}, Jos{\'e} and {Sayres}, Conor and {Zari}, Eleonora},
        title = "{robostrategy: Field and Target Assignment Optimization in the Sloan Digital Sky Survey V}",
      journal = {\aj},
         year = 2026,
        month = jan,
       volume = {171},
       number = {1},
          eid = {37},
        pages = {37},
          doi = {10.3847/1538-3881/ae196d},
archivePrefix = {arXiv},
       eprint = {2505.21328},
 primaryClass = {astro-ph.IM},
       adsurl = {https://ui.adsabs.harvard.edu/abs/2026AJ....171...37B}
}

@ARTICLE{Sayres:2021,
       author = {{Sayres}, Conor and {S{\'a}nchez-Gallego}, Jos{\'e} R. and {Blanton}, Michael R. and {Araujo}, Ricardo and {Bouri}, Mohamed and {Grossen}, Lo{\"\i}c and {Kneib}, Jean-Paul and {Kollmeier}, Juna A. and {Kronig}, Luzius and {Pogge}, Richard W. and {Tuttle}, Sarah},
        title = "{SDSS-V Algorithms: Fast, Collision-free Trajectory Planning for Heavily Overlapping Robotic Fiber Positioners}",
      journal = {\aj},
         year = 2021,
        month = feb,
       volume = {161},
       number = {2},
          eid = {92},
        pages = {92},
          doi = {10.3847/1538-3881/abd0f2},
archivePrefix = {arXiv},
       eprint = {2012.04721},
 primaryClass = {cs.RO},
       adsurl = {https://ui.adsabs.harvard.edu/abs/2021AJ....161...92S}
}

@ARTICLE{GarciaPerez:2016,
       author = {{Garc{\'\i}a P{\'e}rez}, Ana E. and {Allende Prieto}, Carlos and {Holtzman}, Jon A. and {Shetrone}, Matthew and {M{\'e}sz{\'a}ros}, Szabolcs and {Bizyaev}, Dmitry and {Carrera}, Ricardo and {Cunha}, Katia and {Garc{\'\i}a-Hern{\'a}ndez}, D.~A. and {Johnson}, Jennifer A. and et al.},
        title = "{ASPCAP: The APOGEE Stellar Parameter and Chemical Abundances Pipeline}",
      journal = {\aj},
         year = 2016,
        month = jun,
       volume = {151},
       number = {6},
          eid = {144},
        pages = {144},
          doi = {10.3847/0004-6256/151/6/144},
archivePrefix = {arXiv},
       eprint = {1510.07635},
 primaryClass = {astro-ph.SR},
       adsurl = {https://ui.adsabs.harvard.edu/abs/2016AJ....151..144G}
}

@ARTICLE{Meszaros:2025,
       author = {{M{\'e}sz{\'a}ros}, Szabolcs and {Jofr{\'e}}, Paula and {Johnson}, Jennifer A. and {Bird}, Jonathan C. and {Bovy}, Jo and {Casey}, Andrew R. and {Chanam{\'e}}, Julio and {Cunha}, Katia and {De Lee}, Nathan and {Frinchaboy}, Peter and et al.},
        title = "{SDSS-V Milky Way Mapper (MWM): ASPCAP Stellar Parameters and Abundances in SDSS-V Data Release 19}",
      journal = {\aj},
         year = 2025,
        month = aug,
       volume = {170},
       number = {2},
          eid = {96},
        pages = {96},
          doi = {10.3847/1538-3881/ade4b9},
archivePrefix = {arXiv},
       eprint = {2506.07845},
 primaryClass = {astro-ph.SR},
       adsurl = {https://ui.adsabs.harvard.edu/abs/2025AJ....170...96M}
}

@ARTICLE{Gaia:2016,
       author = {{Gaia Collaboration} and {Prusti}, T. and {de Bruijne}, J.~H.~J. and {Brown}, A.~G.~A. and {Vallenari}, A. and {Babusiaux}, C. and {Bailer-Jones}, C.~A.~L. and {Bastian}, U. and {Biermann}, M. and {Evans}, D.~W. and et al.},
        title = "{The Gaia mission}",
      journal = {\aap},
         year = 2016,
        month = nov,
       volume = {595},
          eid = {A1},
        pages = {A1},
          doi = {10.1051/0004-6361/201629272},
archivePrefix = {arXiv},
       eprint = {1609.04153},
 primaryClass = {astro-ph.IM},
       adsurl = {https://ui.adsabs.harvard.edu/abs/2016A&A...595A...1G}
}

@ARTICLE{GaiaDR3:2023,
       author = {{Gaia Collaboration} and {Vallenari}, A. and {Brown}, A.~G.~A. and {Prusti}, T. and {de Bruijne}, J.~H.~J. and {Arenou}, F. and {Babusiaux}, C. and {Biermann}, M. and {Creevey}, O.~L. and {Ducourant}, C. and et al.},
        title = "{Gaia Data Release 3. Summary of the content and survey properties}",
      journal = {\aap},
         year = 2023,
        month = jun,
       volume = {674},
          eid = {A1},
        pages = {A1},
          doi = {10.1051/0004-6361/202243940},
archivePrefix = {arXiv},
       eprint = {2208.00211},
 primaryClass = {astro-ph.GA},
       adsurl = {https://ui.adsabs.harvard.edu/abs/2023A&A...674A...1G}
}

@ARTICLE{DeAngeli:2023,
       author = {{De Angeli}, F. and {Weiler}, M. and {Montegriffo}, P. and {Evans}, D.~W. and {Riello}, M. and {Andrae}, R. and {Carrasco}, J.~M. and {Busso}, G. and {Burgess}, P.~W. and {Cacciari}, C. and {Davidson}, M. and {Harrison}, D.~L. and {Hodgkin}, S.~T. and {Jordi}, C. and {Osborne}, P.~J. and {Pancino}, E. and {Altavilla}, G. and {Barstow}, M.~A. and {Bailer-Jones}, C.~A.~L. and {Bellazzini}, M. and {Brown}, A.~G.~A. and {Castellani}, M. and {Cowell}, S. and {Delchambre}, L. and {De Luise}, F. and {Diener}, C. and {Fabricius}, C. and {Fouesneau}, M. and {Fr{\'e}mat}, Y. and {Gilmore}, G. and {Giuffrida}, G. and {Hambly}, N.~C. and {Hidalgo}, S. and {Holland}, G. and {Kostrzewa-Rutkowska}, Z. and {van Leeuwen}, F. and {Lobel}, A. and {Marinoni}, S. and {Miller}, N. and {Pagani}, C. and {Palaversa}, L. and {Piersimoni}, A.~M. and {Pulone}, L. and {Ragaini}, S. and {Rainer}, M. and {Richards}, P.~J. and {Rixon}, G.~T. and {Ruz-Mieres}, D. and {Sanna}, N. and {Sarro}, L.~M. and {Rowell}, N. and {Sordo}, R. and {Walton}, N.~A. and {Yoldas}, A.},
        title = "{Gaia Data Release 3. Processing and validation of BP/RP low-resolution spectral data}",
      journal = {\aap},
         year = 2023,
        month = jun,
       volume = {674},
          eid = {A2},
        pages = {A2},
          doi = {10.1051/0004-6361/202243680},
archivePrefix = {arXiv},
       eprint = {2206.06143},
 primaryClass = {astro-ph.IM},
       adsurl = {https://ui.adsabs.harvard.edu/abs/2023A&A...674A...2D}
}

@ARTICLE{Skrutskie:2006,
       author = {{Skrutskie}, M.~F. and {Cutri}, R.~M. and {Stiening}, R. and {Weinberg}, M.~D. and {Schneider}, S. and {Carpenter}, J.~M. and {Beichman}, C. and {Capps}, R. and {Chester}, T. and {Elias}, J. and et al.},
        title = "{The Two Micron All Sky Survey (2MASS)}",
      journal = {\aj},
         year = 2006,
        month = feb,
       volume = {131},
       number = {2},
        pages = {1163-1183},
          doi = {10.1086/498708},
       adsurl = {https://ui.adsabs.harvard.edu/abs/2006AJ....131.1163S}
}

@ARTICLE{Wright:2010,
       author = {{Wright}, Edward L. and {Eisenhardt}, Peter R.~M. and {Mainzer}, Amy K. and {Ressler}, Michael E. and {Cutri}, Roc M. and {Jarrett}, Thomas and {Kirkpatrick}, J. Davy and {Padgett}, Deborah and {McMillan}, Robert S. and {Skrutskie}, Michael and et al.},
        title = "{The Wide-field Infrared Survey Explorer (WISE): Mission Description and Initial On-orbit Performance}",
      journal = {\aj},
         year = 2010,
        month = dec,
       volume = {140},
       number = {6},
        pages = {1868-1881},
          doi = {10.1088/0004-6256/140/6/1868},
archivePrefix = {arXiv},
       eprint = {1008.0031},
 primaryClass = {astro-ph.IM},
       adsurl = {https://ui.adsabs.harvard.edu/abs/2010AJ....140.1868W}
}

@ARTICLE{Schlafly:2019,
       author = {{Schlafly}, Edward F. and {Meisner}, Aaron M. and {Green}, Gregory M.},
        title = "{The unWISE Catalog: Two Billion Infrared Sources from Five Years of WISE Imaging}",
      journal = {\apjs},
         year = 2019,
        month = feb,
       volume = {240},
       number = {2},
          eid = {30},
        pages = {30},
          doi = {10.3847/1538-4365/aafbea},
archivePrefix = {arXiv},
       eprint = {1901.03337},
 primaryClass = {astro-ph.IM},
       adsurl = {https://ui.adsabs.harvard.edu/abs/2019ApJS..240...30S}
}

@ARTICLE{Kawata:2014,
       author = {{Kawata}, Daisuke and {Hunt}, Jason A.~S. and {Grand}, Robert J.~J. and {Pasetto}, Stefano and {Cropper}, Mark},
        title = "{Gas and stellar motions and observational signatures of corotating spiral arms}",
      journal = {\mnras},
         year = 2014,
        month = sep,
       volume = {443},
       number = {3},
        pages = {2757-2765},
          doi = {10.1093/mnras/stu1292},
archivePrefix = {arXiv},
       eprint = {1404.2620},
 primaryClass = {astro-ph.GA},
       adsurl = {https://ui.adsabs.harvard.edu/abs/2014MNRAS.443.2757K}
}

@ARTICLE{Menon:2015,
       author = {{Menon}, Harshitha and {Wesolowski}, Lukasz and {Zheng}, Gengbin and {Jetley}, Pritish and {Kale}, Laxmikant and {Quinn}, Thomas and {Governato}, Fabio},
        title = "{Adaptive techniques for clustered N-body cosmological simulations}",
      journal = {Computational Astrophysics and Cosmology},
         year = 2015,
        month = mar,
       volume = {2},
          eid = {1},
        pages = {1},
          doi = {10.1186/s40668-015-0007-9},
archivePrefix = {arXiv},
       eprint = {1409.1929},
 primaryClass = {astro-ph.IM},
       adsurl = {https://ui.adsabs.harvard.edu/abs/2015ComAC...2....1M}
}

@ARTICLE{Sun_2026,
       author = {{Sun}, Yan and {Zhang}, Shaobo and {Yang}, Ji and {Su}, Yang and {Yan}, Qing-Zeng and {Zhou}, Xin and {Chen}, Xuepeng and {Fang}, Min and {Du}, Fujun},
        title = "{Widespread corrugations superimposed on the Galactic warp in the outer CO disk}",
      journal = {Nature Astronomy},
         year = 2026,
        month = oct,
       volume = {10},
        pages = {1560-1569},
          doi = {10.1038/s41550-026-02938-y},
       adsurl = {https://scixplorer.org/abs/2026NatAs..10.1560S}
}

@INPROCEEDINGS{Jetley:2008,
  author={Jetley, Pritish and Gioachin, Filippo and Mendes, Celso and Kale, Laxmikant V. and Quinn, Thomas},
  booktitle={2008 IEEE International Symposium on Parallel and Distributed Processing},
  title={Massively parallel cosmological simulations with ChaNGa},
  year={2008},
  volume={},
  number={},
  pages={1-12},
  doi={10.1109/IPDPS.2008.4536319}}

@INPROCEEDINGS{Jetley:2010,
       author = {{Jetley}, Pritish and {Wesolowski}, Lukasz and {Gioachin}, Filippo and {Kal{\'e}}, Laxmikant V. and {Quinn}, Thomas R.},
        title = "{Scaling Hierarchical N-body Simulations on GPU Clusters}",
    booktitle = {2010 ACM/IEEE International Conference for High Performance Computing, Networking, Storage and Analysis},
         year = 2010,
        month = nov,
    publisher = {IEEE},
          eid = {46},
        pages = {46},
          doi = {10.1109/SC.2010.49},
       adsurl = {https://ui.adsabs.harvard.edu/abs/2010sc...conf...46J}
}

@ARTICLE{Drimmel:2023,
       author = {{Gaia Collaboration} and {Drimmel}, R. and {Romero-G{\'o}mez}, M. and {Chemin}, L. and {Ramos}, P. and {Poggio}, E. and {Ripepi}, V. and {Andrae}, R. and {Blomme}, R. and {Cantat-Gaudin}, T. and et al.},
        title = "{Gaia Data Release 3. Mapping the asymmetric disc of the Milky Way}",
      journal = {\aap},
         year = 2023,
        month = jun,
       volume = {674},
          eid = {A37},
        pages = {A37},
          doi = {10.1051/0004-6361/202243797},
archivePrefix = {arXiv},
       eprint = {2206.06207},
 primaryClass = {astro-ph.GA},
       adsurl = {https://ui.adsabs.harvard.edu/abs/2023A&A...674A..37G}
}

@ARTICLE{Morganson:2016,
       author = {{Morganson}, Eric and {Conn}, Blair and {Rix}, Hans-Walter and {Bell}, Eric F. and {Burgett}, William S. and {Chambers}, Kenneth and {Dolphin}, Andrew and {Draper}, Peter W. and {Flewelling}, Heather and {Hodapp}, Klaus and et al.},
        title = "{Mapping the Monoceros Ring in 3D with Pan-STARRS1}",
      journal = {\apj},
         year = 2016,
        month = jul,
       volume = {825},
       number = {2},
          eid = {140},
        pages = {140},
          doi = {10.3847/0004-637X/825/2/140},
archivePrefix = {arXiv},
       eprint = {1604.07501},
 primaryClass = {astro-ph.GA},
       adsurl = {https://ui.adsabs.harvard.edu/abs/2016ApJ...825..140M}
}

@ARTICLE{Price-Whelan:2015,
       author = {{Price-Whelan}, Adrian M. and {Johnston}, Kathryn V. and {Sheffield}, Allyson A. and {Laporte}, Chervin F.~P. and {Sesar}, Branimir},
        title = "{A reinterpretation of the Triangulum-Andromeda stellar clouds: a population of halo stars kicked out of the Galactic disc}",
      journal = {\mnras},
         year = 2015,
        month = sep,
       volume = {452},
       number = {1},
        pages = {676-685},
          doi = {10.1093/mnras/stv1324},
archivePrefix = {arXiv},
       eprint = {1503.08780},
 primaryClass = {astro-ph.GA},
       adsurl = {https://ui.adsabs.harvard.edu/abs/2015MNRAS.452..676P}
}

@ARTICLE{Xu:2015,
       author = {{Xu}, Yan and {Newberg}, Heidi Jo and {Carlin}, Jeffrey L. and {Liu}, Chao and {Deng}, Licai and {Li}, Jing and {Sch{\"o}nrich}, Ralph and {Yanny}, Brian},
        title = "{Rings and Radial Waves in the Disk of the Milky Way}",
      journal = {\apj},
         year = 2015,
        month = mar,
       volume = {801},
       number = {2},
          eid = {105},
        pages = {105},
          doi = {10.1088/0004-637X/801/2/105},
archivePrefix = {arXiv},
       eprint = {1503.00257},
 primaryClass = {astro-ph.GA},
       adsurl = {https://ui.adsabs.harvard.edu/abs/2015ApJ...801..105X}
}

@ARTICLE{Widrow:2012,
       author = {{Widrow}, Lawrence M. and {Gardner}, Susan and {Yanny}, Brian and {Dodelson}, Scott and {Chen}, Hsin-Yu},
        title = "{Galactoseismology: Discovery of Vertical Waves in the Galactic Disk}",
      journal = {\apjl},
         year = 2012,
        month = may,
       volume = {750},
       number = {2},
          eid = {L41},
        pages = {L41},
          doi = {10.1088/2041-8205/750/2/L41},
archivePrefix = {arXiv},
       eprint = {1203.6861},
 primaryClass = {astro-ph.GA},
       adsurl = {https://ui.adsabs.harvard.edu/abs/2012ApJ...750L..41W}
}

@ARTICLE{Sheffield:2018,
       author = {{Sheffield}, Allyson A. and {Price-Whelan}, Adrian M. and {Tzanidakis}, Anastasios and {Johnston}, Kathryn V. and {Laporte}, Chervin F.~P. and {Sesar}, Branimir},
        title = "{A Disk Origin for the Monoceros Ring and A13 Stellar Overdensities}",
      journal = {\apj},
         year = 2018,
        month = feb,
       volume = {854},
       number = {1},
          eid = {47},
        pages = {47},
          doi = {10.3847/1538-4357/aaa4b6},
archivePrefix = {arXiv},
       eprint = {1801.01171},
 primaryClass = {astro-ph.GA},
       adsurl = {https://ui.adsabs.harvard.edu/abs/2018ApJ...854...47S}
}

@ARTICLE{Li:2017,
       author = {{Li}, Ting S. and {Sheffield}, Allyson A. and {Johnston}, Kathryn V. and {Marshall}, Jennifer L. and {Majewski}, Steven R. and {Price-Whelan}, Adrian M. and {Damke}, Guillermo J. and {Beaton}, Rachael L. and {Bernard}, Edouard J. and {Richardson}, Whitney and et al.},
        title = "{Exploring Halo Substructure with Giant Stars. XV. Discovery of a Connection between the Monoceros Ring and the Triangulum-Andromeda Overdensity?}",
      journal = {\apj},
         year = 2017,
        month = jul,
       volume = {844},
       number = {1},
          eid = {74},
        pages = {74},
          doi = {10.3847/1538-4357/aa7a0d},
archivePrefix = {arXiv},
       eprint = {1703.05384},
 primaryClass = {astro-ph.GA},
       adsurl = {https://ui.adsabs.harvard.edu/abs/2017ApJ...844...74L}
}

@ARTICLE{Babusiaux:2023,
       author = {{Babusiaux}, C. and {Fabricius}, C. and {Khanna}, S. and {Muraveva}, T. and {Reyl{\'e}}, C. and {Spoto}, F. and {Vallenari}, A. and {Luri}, X. and {Arenou}, F. and {{\'A}lvarez}, M.~A. and {Anders}, F. and {Antoja}, T. and {Balbinot}, E. and {Barache}, C. and {Bauchet}, N. and {Bossini}, D. and {Busonero}, D. and {Cantat-Gaudin}, T. and {Carrasco}, J.~M. and {Dafonte}, C. and {Diakit{\'e}}, S. and {Figueras}, F. and {Garcia-Gutierrez}, A. and {Garofalo}, A. and {Helmi}, A. and {Jim{\'e}nez-Arranz}, {\'O}. and {Jordi}, C. and {Kervella}, P. and {Kostrzewa-Rutkowska}, Z. and {Leclerc}, N. and {Licata}, E. and {Manteiga}, M. and {Masip}, A. and {Mongui{\'o}}, M. and {Ramos}, P. and {Robichon}, N. and {Robin}, A.~C. and {Romero-G{\'o}mez}, M. and {S{\'a}ez}, A. and {Santove{\~n}a}, R. and {Spina}, L. and {Torralba Elipe}, G. and {Weiler}, M.},
        title = "{Gaia Data Release 3. Catalogue validation}",
      journal = {\aap},
         year = 2023,
        month = jun,
       volume = {674},
          eid = {A32},
        pages = {A32},
          doi = {10.1051/0004-6361/202243790},
archivePrefix = {arXiv},
       eprint = {2206.05989},
 primaryClass = {astro-ph.SR},
       adsurl = {https://ui.adsabs.harvard.edu/abs/2023A&A...674A..32B}
}

@ARTICLE{Mackereth_2019,
       author = {{Mackereth}, J. Ted and {Bovy}, Jo and {Leung}, Henry W. and {Schiavon}, Ricardo P. and {Trick}, Wilma H. and {Chaplin}, William J. and {Cunha}, Katia and {Feuillet}, Diane K. and {Majewski}, Steven R. and {Martig}, Marie and {Miglio}, Andrea and {Nidever}, David and {Pinsonneault}, Marc H. and {Aguirre}, Victor Silva and {Sobeck}, Jennifer and {Tayar}, Jamie and {Zasowski}, Gail},
        title = "{Dynamical heating across the Milky Way disc using APOGEE and Gaia}",
      journal = {\mnras},
         year = 2019,
        month = oct,
       volume = {489},
       number = {1},
        pages = {176-195},
          doi = {10.1093/mnras/stz1521},
archivePrefix = {arXiv},
       eprint = {1901.04502},
 primaryClass = {astro-ph.GA},
       adsurl = {https://ui.adsabs.harvard.edu/abs/2019MNRAS.489..176M}
}

@ARTICLE{Warfield_2024,
       author = {{Warfield}, Jack T. and {Zinn}, Joel C. and {Schonhut-Stasik}, Jessica and {Johnson}, James W. and {Pinsonneault}, Marc H. and {Johnson}, Jennifer A. and {Stello}, Dennis and {Beaton}, Rachael L. and {Elsworth}, Yvonne and {Garc{\'\i}a}, Rafael A. and {Mathur}, Savita and {Mosser}, Beno{\^\i}t and {Serenelli}, Aldo and {Tayar}, Jamie},
        title = "{The APO-K2 Catalog. II. Accurate Stellar Ages for Red Giant Branch Stars across the Milky Way}",
      journal = {\aj},
         year = 2024,
        month = may,
       volume = {167},
       number = {5},
          eid = {208},
        pages = {208},
          doi = {10.3847/1538-3881/ad33bb},
archivePrefix = {arXiv},
       eprint = {2403.16250},
 primaryClass = {astro-ph.GA},
       adsurl = {https://ui.adsabs.harvard.edu/abs/2024AJ....167..208W}
}

@ARTICLE{SchonhutStasik_2024,
       author = {{Schonhut-Stasik}, Jessica and {Zinn}, Joel C. and {Stassun}, Keivan G. and {Pinsonneault}, Marc and {Johnson}, Jennifer A. and {Warfield}, Jack T. and {Stello}, Dennis and {Elsworth}, Yvonne and {Garc{\'\i}a}, Rafael A. and {Mathur}, Savita and {Mosser}, Benoit and {Hon}, Marc and {Tayar}, Jamie and {Stringfellow}, Guy S. and {Beaton}, Rachael L. and {J{\"o}nsson}, Henrik and {Minniti}, Dante},
        title = "{The APO-K2 Catalog. I. 7500 Red Giants with Fundamental Stellar Parameters from APOGEE DR17 Spectroscopy and K2-GAP Asteroseismology}",
      journal = {\aj},
         year = 2024,
        month = feb,
       volume = {167},
       number = {2},
          eid = {50},
        pages = {50},
          doi = {10.3847/1538-3881/ad0b13},
archivePrefix = {arXiv},
       eprint = {2304.10654},
 primaryClass = {astro-ph.SR},
       adsurl = {https://ui.adsabs.harvard.edu/abs/2024AJ....167...50S}
}

@ARTICLE{Pinsonneault_2025,
       author = {{Pinsonneault}, Marc H. and {Zinn}, Joel C. and {Tayar}, Jamie and {Serenelli}, Aldo and {Garc{\'\i}a}, Rafael A. and {Mathur}, Savita and {Vrard}, Mathieu and {Elsworth}, Yvonne P. and {Mosser}, Benoit and {Stello}, Dennis and {Bell}, Keaton J. and {Bugnet}, Lisa and {Corsaro}, Enrico and {Gaulme}, Patrick and {Hekker}, Saskia and {Hon}, Marc and {Huber}, Daniel and {Kallinger}, Thomas and {Cao}, Kaili and {Johnson}, Jennifer A. and {Liagre}, Bastien and {Patton}, Rachel A. and {Santos}, {\^A}ngela R.~G. and {Basu}, Sarbani and {Beck}, Paul G. and {Beers}, Timothy C. and {Chaplin}, William J. and {Cunha}, Katia and {Frinchaboy}, Peter M. and {Girardi}, L{\'e}o and {Godoy-Rivera}, Diego and {Holtzman}, Jon A. and {J{\"o}nsson}, Henrik and {M{\'e}sz{\'a}ros}, Szabolcs and {Reyes}, Claudia and {Rix}, Hans-Walter and {Shetrone}, Matthew and {Smith}, Verne V. and {Spoo}, Taylor and {Stassun}, Keivan G. and {Wang}, Ji},
        title = "{APOKASC-3: The Third Joint Spectroscopic and Asteroseismic Catalog for Evolved Stars in the Kepler Fields}",
      journal = {\apjs},
         year = 2025,
        month = feb,
       volume = {276},
       number = {2},
          eid = {69},
        pages = {69},
          doi = {10.3847/1538-4365/ad9fef},
archivePrefix = {arXiv},
       eprint = {2410.00102},
 primaryClass = {astro-ph.SR},
       adsurl = {https://ui.adsabs.harvard.edu/abs/2025ApJS..276...69P}
}

@ARTICLE{Theodoridis_2026,
       author = {{Theodoridis}, Artemis Theano and {Morales}, Leslie and {Tayar}, Jamie},
        title = "{Asteroseismically Inferred Ages of 132,000 Red Giants with TESS}",
      journal = {\apjs},
         year = 2026,
        month = mar,
       volume = {283},
       number = {1},
          eid = {4},
        pages = {4},
          doi = {10.3847/1538-4365/ae315a},
archivePrefix = {arXiv},
       eprint = {2505.08109},
 primaryClass = {astro-ph.SR},
       adsurl = {https://ui.adsabs.harvard.edu/abs/2026ApJS..283....4T}
}

@ARTICLE{Hon_2021,
       author = {{Hon}, Marc and {Huber}, Daniel and {Kuszlewicz}, James S. and {Stello}, Dennis and {Sharma}, Sanjib and {Tayar}, Jamie and {Zinn}, Joel C. and {Vrard}, Mathieu and {Pinsonneault}, Marc H.},
        title = "{A ``Quick Look'' at All-sky Galactic Archeology with TESS: 158,000 Oscillating Red Giants from the MIT Quick-look Pipeline}",
      journal = {\apj},
         year = 2021,
        month = oct,
       volume = {919},
       number = {2},
          eid = {131},
        pages = {131},
          doi = {10.3847/1538-4357/ac14b1},
archivePrefix = {arXiv},
       eprint = {2108.01241},
 primaryClass = {astro-ph.SR},
       adsurl = {https://ui.adsabs.harvard.edu/abs/2021ApJ...919..131H}
}

@ARTICLE{Kurbatov:2024,
       author = {{Kurbatov}, Evgeny P. and {Belokurov}, Vasily and {Koposov}, Sergey and {Kravtsov}, Andrey and {Davies}, Elliot Y. and {Brown}, Anthony G.~A. and {Cantat-Gaudin}, Tristan and {Castro-Ginard}, Alfred and {Casey}, Andrew R. and {Drimmel}, Ronald and et al.},
        title = "{The realm of Aurora. Density distribution of metal-poor giants in the heart of the Galaxy}",
      journal = {arXiv e-prints},
         year = 2024,
        month = oct,
          eid = {arXiv:2410.22250},
        pages = {arXiv:2410.22250},
          doi = {10.48550/arXiv.2410.22250},
archivePrefix = {arXiv},
       eprint = {2410.22250},
 primaryClass = {astro-ph.GA},
       adsurl = {https://scixplorer.org/abs/2024arXiv241022250K}
}

@software{astropy_21262391,
  author       = {Astropy Collaboration},
  title        = {Astropy},
  month        = jul,
  year         = 2026,
  publisher    = {Zenodo},
  version      = {v8.0.1},
  doi          = {10.5281/zenodo.21262391},
  url          = {https://doi.org/10.5281/zenodo.21262391},
}

@article{astropy:2013,
  adsurl        = {http://adsabs.harvard.edu/abs/2013A\%26A...558A..33A},
  archiveprefix = {arXiv},
  author        = {{Astropy Collaboration} and {Robitaille}, T.~P. and {Tollerud}, E.~J. and {Greenfield}, P. and {Droettboom}, M. and {Bray}, E. and {Aldcroft}, T. and {Davis}, M. and {Ginsburg}, A. and {Price-Whelan}, A.~M. and {Kerzendorf}, W.~E. and {Conley}, A. and {Crighton}, N. and {Barbary}, K. and {Muna}, D. and {Ferguson}, H. and {Grollier}, F. and {Parikh}, M.~M. and {Nair}, P.~H. and {Unther}, H.~M. and {Deil}, C. and {Woillez}, J. and {Conseil}, S. and {Kramer}, R. and {Turner}, J.~E.~H. and {Singer}, L. and {Fox}, R. and {Weaver}, B.~A. and {Zabalza}, V. and {Edwards}, Z.~I. and {Azalee Bostroem}, K. and {Burke}, D.~J. and {Casey}, A.~R. and {Crawford}, S.~M. and {Dencheva}, N. and {Ely}, J. and {Jenness}, T. and {Labrie}, K. and {Lim}, P.~L. and {Pierfederici}, F. and {Pontzen}, A. and {Ptak}, A. and {Refsdal}, B. and {Servillat}, M. and {Streicher}, O.},
  doi           = {10.1051/0004-6361/201322068},
  eid           = {A33},
  eprint        = {1307.6212},
  journal       = {\aap},
  month         = oct,
  pages         = {A33},
  primaryclass  = {astro-ph.IM},
  title         = {{Astropy: A community Python package for astronomy}},
  volume        = 558,
  year          = 2013
}

@article{astropy:2018,
  author        = {{Astropy Collaboration} and {Price-Whelan}, A.~M. and {Sip{\H{o}}cz}, B.~M. and {G{\"u}nther}, H.~M. and {Lim}, P.~L. and {Crawford}, S.~M. and {Conseil}, S. and {Shupe}, D.~L. and {Craig}, M.~W. and {Dencheva}, N. and {Ginsburg}, A. and {Vand erPlas}, J.~T. and {Bradley}, L.~D. and {P{\'e}rez-Su{\'a}rez}, D. and {de Val-Borro}, M. and {Aldcroft}, T.~L. and {Cruz}, K.~L. and {Robitaille}, T.~P. and {Tollerud}, E.~J. and {Ardelean}, C. and {Babej}, T. and {Bach}, Y.~P. and {Bachetti}, M. and {Bakanov}, A.~V. and {Bamford}, S.~P. and {Barentsen}, G. and {Barmby}, P. and {Baumbach}, A. and {Berry}, K.~L. and {Biscani}, F. and {Boquien}, M. and {Bostroem}, K.~A. and {Bouma}, L.~G. and {Brammer}, G.~B. and {Bray}, E.~M. and {Breytenbach}, H. and {Buddelmeijer}, H. and {Burke}, D.~J. and {Calderone}, G. and {Cano Rodr{\'\i}guez}, J.~L. and {Cara}, M. and {Cardoso}, J.~V.~M. and {Cheedella}, S. and {Copin}, Y. and {Corrales}, L. and {Crichton}, D. and {D'Avella}, D. and {Deil}, C. and {Depagne}, {\'E}. and {Dietrich}, J.~P. and {Donath}, A. and {Droettboom}, M. and {Earl}, N. and {Erben}, T. and {Fabbro}, S. and {Ferreira}, L.~A. and {Finethy}, T. and {Fox}, R.~T. and {Garrison}, L.~H. and {Gibbons}, S.~L.~J. and {Goldstein}, D.~A. and {Gommers}, R. and {Greco}, J.~P. and {Greenfield}, P. and {Groener}, A.~M. and {Grollier}, F. and {Hagen}, A. and {Hirst}, P. and {Homeier}, D. and {Horton}, A.~J. and {Hosseinzadeh}, G. and {Hu}, L. and {Hunkeler}, J.~S. and {Ivezi{\'c}}, {\v{Z}}. and {Jain}, A. and {Jenness}, T. and {Kanarek}, G. and {Kendrew}, S. and {Kern}, N.~S. and {Kerzendorf}, W.~E. and {Khvalko}, A. and {King}, J. and {Kirkby}, D. and {Kulkarni}, A.~M. and {Kumar}, A. and {Lee}, A. and {Lenz}, D. and {Littlefair}, S.~P. and {Ma}, Z. and {Macleod}, D.~M. and {Mastropietro}, M. and {McCully}, C. and {Montagnac}, S. and {Morris}, B.~M. and {Mueller}, M. and {Mumford}, S.~J. and {Muna}, D. and {Murphy}, N.~A. and {Nelson}, S. and {Nguyen}, G.~H. and {Ninan}, J.~P. and {N{\"o}the}, M. and {Ogaz}, S. and {Oh}, S. and {Parejko}, J.~K. and {Parley}, N. and {Pascual}, S. and {Patil}, R. and {Patil}, A.~A. and {Plunkett}, A.~L. and {Prochaska}, J.~X. and {Rastogi}, T. and {Reddy Janga}, V. and {Sabater}, J. and {Sakurikar}, P. and {Seifert}, M. and {Sherbert}, L.~E. and {Sherwood-Taylor}, H. and {Shih}, A.~Y. and {Sick}, J. and {Silbiger}, M.~T. and {Singanamalla}, S. and {Singer}, L.~P. and {Sladen}, P.~H. and {Sooley}, K.~A. and {Sornarajah}, S. and {Streicher}, O. and {Teuben}, P. and {Thomas}, S.~W. and {Tremblay}, G.~R. and {Turner}, J.~E.~H. and {Terr{\'o}n}, V. and {van Kerkwijk}, M.~H. and {de la Vega}, A. and {Watkins}, L.~L. and {Weaver}, B.~A. and {Whitmore}, J.~B. and {Woillez}, J. and {Zabalza}, V. and {Astropy Contributors}},
  title         = "{The Astropy Project: Building an Open-science Project and Status of the v2.0 Core Package}",
  journal       = {\aj},
  year          = 2018,
  month         = sep,
  volume        = {156},
  number        = {3},
  eid           = {123},
  pages         = {123},
  doi           = {10.3847/1538-3881/aabc4f},
  archiveprefix = {arXiv},
  eprint        = {1801.02634},
  primaryclass  = {astro-ph.IM},
  adsurl        = {https://ui.adsabs.harvard.edu/abs/2018AJ....156..123A}
}

@article{astropy:2022,
  author        = {{Astropy Collaboration} and {Price-Whelan}, Adrian M. and {Lim}, Pey Lian and {Earl}, Nicholas and {Starkman}, Nathaniel and {Bradley}, Larry and {Shupe}, David L. and {Patil}, Aarya A. and {Corrales}, Lia and {Brasseur}, C.~E. and {N{"o}the}, Maximilian and {Donath}, Axel and {Tollerud}, Erik and {Morris}, Brett M. and {Ginsburg}, Adam and {Vaher}, Eero and {Weaver}, Benjamin A. and {Tocknell}, James and {Jamieson}, William and {van Kerkwijk}, Marten H. and {Robitaille}, Thomas P. and {Merry}, Bruce and {Bachetti}, Matteo and {G{"u}nther}, H. Moritz and {Aldcroft}, Thomas L. and {Alvarado-Montes}, Jaime A. and {Archibald}, Anne M. and {B{'o}di}, Attila and {Bapat}, Shreyas and {Barentsen}, Geert and {Baz{'a}n}, Juanjo and {Biswas}, Manish and {Boquien}, M{'e}d{'e}ric and {Burke}, D.~J. and {Cara}, Daria and {Cara}, Mihai and {Conroy}, Kyle E. and {Conseil}, Simon and {Craig}, Matthew W. and {Cross}, Robert M. and {Cruz}, Kelle L. and {D'Eugenio}, Francesco and {Dencheva}, Nadia and {Devillepoix}, Hadrien A.~R. and {Dietrich}, J{"o}rg P. and {Eigenbrot}, Arthur Davis and {Erben}, Thomas and {Ferreira}, Leonardo and {Foreman-Mackey}, Daniel and {Fox}, Ryan and {Freij}, Nabil and {Garg}, Suyog and {Geda}, Robel and {Glattly}, Lauren and {Gondhalekar}, Yash and {Gordon}, Karl D. and {Grant}, David and {Greenfield}, Perry and {Groener}, Austen M. and {Guest}, Steve and {Gurovich}, Sebastian and {Handberg}, Rasmus and {Hart}, Akeem and {Hatfield-Dodds}, Zac and {Homeier}, Derek and {Hosseinzadeh}, Griffin and {Jenness}, Tim and {Jones}, Craig K. and {Joseph}, Prajwel and {Kalmbach}, J. Bryce and {Karamehmetoglu}, Emir and {Ka{l}uszy{'n}ski}, Miko{l}aj and {Kelley}, Michael S.~P. and {Kern}, Nicholas and {Kerzendorf}, Wolfgang E. and {Koch}, Eric W. and {Kulumani}, Shankar and {Lee}, Antony and {Ly}, Chun and {Ma}, Zhiyuan and {MacBride}, Conor and {Maljaars}, Jakob M. and {Muna}, Demitri and {Murphy}, N.~A. and {Norman}, Henrik and {O'Steen}, Richard and {Oman}, Kyle A. and {Pacifici}, Camilla and {Pascual}, Sergio and {Pascual-Granado}, J. and {Patil}, Rohit R. and {Perren}, Gabriel I. and {Pickering}, Timothy E. and {Rastogi}, Tanuj and {Roulston}, Benjamin R. and {Ryan}, Daniel F. and {Rykoff}, Eli S. and {Sabater}, Jose and {Sakurikar}, Parikshit and {Salgado}, Jes{'u}s and {Sanghi}, Aniket and {Saunders}, Nicholas and {Savchenko}, Volodymyr and {Schwardt}, Ludwig and {Seifert-Eckert}, Michael and {Shih}, Albert Y. and {Jain}, Anany Shrey and {Shukla}, Gyanendra and {Sick}, Jonathan and {Simpson}, Chris and {Singanamalla}, Sudheesh and {Singer}, Leo P. and {Singhal}, Jaladh and {Sinha}, Manodeep and {Sip{H{o}}cz}, Brigitta M. and {Spitler}, Lee R. and {Stansby}, David and {Streicher}, Ole and {{{S}}umak}, Jani and {Swinbank}, John D. and {Taranu}, Dan S. and {Tewary}, Nikita and {Tremblay}, Grant R. and {Val-Borro}, Miguel de and {Van Kooten}, Samuel J. and {Vasovi{'c}}, Zlatan and {Verma}, Shresth and {de Miranda Cardoso}, Jos{'e} Vin{'i}cius and {Williams}, Peter K.~G. and {Wilson}, Tom J. and {Winkel}, Benjamin and {Wood-Vasey}, W.~M. and {Xue}, Rui and {Yoachim}, Peter and {Zhang}, Chen and {Zonca}, Andrea and {Astropy Project Contributors}},
  title         = "{The Astropy Project: Sustaining and Growing a Community-oriented Open-source Project and the Latest Major Release (v5.0) of the Core Package}",
  journal       = {\apj},
  year          = 2022,
  month         = aug,
  volume        = {935},
  number        = {2},
  eid           = {167},
  pages         = {167},
  doi           = {10.3847/1538-4357/ac7c74},
  archiveprefix = {arXiv},
  eprint        = {2206.14220},
  primaryclass  = {astro-ph.IM},
  adsurl        = {https://ui.adsabs.harvard.edu/abs/2022ApJ...935..167A}
}

@article{2007CSE.....9c..21P,
  author        = {{Perez}, Fernando and {Granger}, Brian E.},
  title         = "{IPython: A System for Interactive Scientific Computing}",
  journal       = {Computing in Science and Engineering},
  year          = "2007",
  month         = "Jan",
  volume        = {9},
  number        = {3},
  pages         = {21--29},
  doi           = {10.1109/MCSE.2007.53},
  adsurl        = {https://ui.adsabs.harvard.edu/abs/2007CSE.....9c..21P}
}

@conference{kluyver2016jupyter,
  title         = {Jupyter Notebooks -- a publishing format for reproducible computational workflows},
  author        = {Thomas Kluyver and Benjamin Ragan-Kelley and Fernando P{\'e}rez and Brian Granger and Matthias Bussonnier and Jonathan Frederic and Kyle Kelley and Jessica Hamrick and Jason Grout and Sylvain Corlay and Paul Ivanov and Dami{\'a}n Avila and Safia Abdalla and Carol Willing},
  booktitle     = {Positioning and Power in Academic Publishing: Players, Agents and Agendas},
  editor        = {F. Loizides and B. Schmidt},
  organization  = {IOS Press},
  pages         = {87--90},
  year          = {2016}
}

@article{Hunter:2007,
  author        = {Hunter, J. D.},
  title         = {Matplotlib: A 2D graphics environment},
  journal       = {Computing in Science \& Engineering},
  volume        = {9},
  number        = {3},
  pages         = {90--95},
  publisher     = {IEEE COMPUTER SOC},
  doi           = {10.1109/MCSE.2007.55},
  year          = 2007
}

@article{numpy,
  title         = {Array programming with {NumPy}},
  author        = {Charles R. Harris and K. Jarrod Millman and St{\'{e}}fan J. van der Walt and Ralf Gommers and Pauli Virtanen and David Cournapeau and Eric Wieser and Julian Taylor and Sebastian Berg and Nathaniel J. Smith and Robert Kern and Matti Picus and Stephan Hoyer and Marten H. van Kerkwijk and Matthew Brett and Allan Haldane and Jaime Fern{\'{a}}ndez del R{\'{i}}o and Mark Wiebe and Pearu Peterson and Pierre G{\'{e}}rard-Marchant and Kevin Sheppard and Tyler Reddy and Warren Weckesser and Hameer Abbasi and Christoph Gohlke and Travis E. Oliphant},
  year          = {2020},
  month         = sep,
  journal       = {Nature},
  volume        = {585},
  number        = {7825},
  pages         = {357--362},
  doi           = {10.1038/s41586-020-2649-2},
  publisher     = {Springer Science and Business Media {LLC}},
  url           = {https://doi.org/10.1038/s41586-020-2649-2}
}

@software{pandas_23955360,
  author       = {The pandas development team},
  title        = {pandas-dev/pandas: Pandas},
  month        = nov,
  year         = 2026,
  publisher    = {Zenodo},
  version      = {v3.1.1},
  doi          = {10.5281/zenodo.23955360},
  url          = {https://doi.org/10.5281/zenodo.23955360},
}

@inproceedings{mckinney-proc-scipy-2010,
  author        = {{W}es {M}c{K}inney},
  title         = {{D}ata {S}tructures for {S}tatistical {C}omputing in {P}ython},
  booktitle     = {{P}roceedings of the 9th {P}ython in {S}cience {C}onference},
  pages         = {56--61},
  year          = {2010},
  editor        = {{S}t\'efan van der {W}alt and {J}arrod {M}illman},
  doi           = {10.25080/Majora-92bf1922-00a}
}

@book{python,
  author        = {Van Rossum, Guido and Drake, Fred L.},
  title         = {Python 3 Reference Manual},
  year          = {2009},
  isbn          = {1441412697},
  publisher     = {CreateSpace},
  address       = {Scotts Valley, CA}
}

@software{scipy_22051554,
  author       = {Ralf Gommers and
                  Pauli Virtanen and
                  Matt Haberland and
                  Evgeni Burovski and
                  Tyler Reddy and
                  Warren Weckesser and
                  Andrew Nelson and
                  Travis E. Oliphant and
                  David Cournapeau and
                  Ilhan Polat and
                  alexbrc and
                  Lucas Colley and
                  Pamphile Roy and
                  Pearu Peterson and
                  Jake Bowhay and
                  Josh Wilson and
                  endolith and
                  Nikolay Mayorov and
                  Albert Steppi and
                  Stefan van der Walt and
                  Matthew Brett and
                  Denis Laxalde and
                  Eric Larson and
                  Atsushi Sakai and
                  Jarrod Millman and
                  Lars and
                  peterbell10 and
                  CJ Carey and
                  Paul van Mulbregt and
                  Daniel Schmitz},
  title        = {scipy/scipy: SciPy 1.18.1},
  month        = aug,
  year         = 2026,
  publisher    = {Zenodo},
  version      = {v1.18.1},
  doi          = {10.5281/zenodo.22051554},
  url          = {https://doi.org/10.5281/zenodo.22051554},
  swhid        = {swh:1:dir:8fdbb1dcd4b987826bb26d75683f82301049990f
                   ;origin=https://doi.org/10.5281/zenodo.595738;visi
                   t=swh:1:snp:db9c289760efc4c9759248bd3e143d9d750177
                   52;anchor=swh:1:rel:109b0ee932559daf612f204df3e2af
                   422b399f11;path=scipy-scipy-c2df8ac
                  },
}

@article{2020SciPy-NMeth,
  author        = {Virtanen, Pauli and Gommers, Ralf and Oliphant, Travis E. and Haberland, Matt and Reddy, Tyler and Cournapeau, David and Burovski, Evgeni and Peterson, Pearu and Weckesser, Warren and Bright, Jonathan and {van der Walt}, St{\'e}fan J. and Brett, Matthew and Wilson, Joshua and Millman, K. Jarrod and Mayorov, Nikolay and Nelson, Andrew R. J. and Jones, Eric and Kern, Robert and Larson, Eric and Carey, C J and Polat, {\.I}lhan and Feng, Yu and Moore, Eric W. and {VanderPlas}, Jake and Laxalde, Denis and Perktold, Josef and Cimrman, Robert and Henriksen, Ian and Quintero, E. A. and Harris, Charles R. and Archibald, Anne M. and Ribeiro, Ant{\^o}nio H. and Pedregosa, Fabian and {van Mulbregt}, Paul and {SciPy 1.0 Contributors}},
  title         = {{{SciPy} 1.0: Fundamental Algorithms for Scientific Computing in Python}},
  journal       = {Nature Methods},
  year          = {2020},
  volume        = {17},
  pages         = {261--272},
  adsurl        = {https://rdcu.be/b08Wh},
  doi           = {10.1038/s41592-019-0686-2}
}

@software{corner.py_21208839,
  author       = {Dan Foreman-Mackey and
                  Adrian Price-Whelan and
                  Will Vousden and
                  Geoffrey Ryan and
                  Matt Pitkin and
                  Thomas Vandal and
                  Víctor Zabalza and
                  jsheyl and
                  Emily Rice and
                  Michael Smith and
                  Leo Singer and
                  Gregory Ashton and
                  Arfon Smith and
                  James Tocknell and
                  Johannes Buchner and
                  Justin Gerber and
                  Kyle Barbary and
                  Marlon and
                  Remy Prechelt and
                  Sebastian M. Castillo Hair and
                  Stephan Hoyer and
                  Zhaozhou Li and
                  Thomas A Caswell and
                  Wolfgang Kerzendorf and
                  Yodo9000 and
                  Zach Weiner and
                  Kelle Cruz and
                  James Matthews},
  title        = {dfm/corner.py: v2.3.0rc3},
  month        = jul,
  year         = 2026,
  publisher    = {Zenodo},
  version      = {v2.3.0rc3},
  doi          = {10.5281/zenodo.21208839},
  url          = {https://doi.org/10.5281/zenodo.21208839},
}

@article{corner-Foreman-Mackey-2016,
  author        = {{Foreman-Mackey}, Daniel},
  title         = "{corner.py: Scatterplot matrices in Python}",
  journal       = {The Journal of Open Source Software},
  year          = 2016,
  month         = jun,
  volume        = {1},
  pages         = {24},
  doi           = {10.21105/joss.00024},
  adsurl        = {https://ui.adsabs.harvard.edu/abs/2016JOSS....1...24F}
}

@software{jax2018github,
  author        = {James Bradbury and Roy Frostig and Peter Hawkins and Matthew James Johnson and Chris Leary and Dougal Maclaurin and George Necula and Adam Paszke and Jake Vander{P}las and Skye Wanderman-{M}ilne and Qiao Zhang},
  title         = {{JAX}: composable transformations of {P}ython+{N}um{P}y programs},
  url           = {http://github.com/google/jax},
  version       = {0.3.13},
  year          = {2018}
}

@article{phan2019composable,
  author        = {{Phan}, Du and {Pradhan}, Neeraj and {Jankowiak}, Martin},
  title         = "{Composable Effects for Flexible and Accelerated Probabilistic Programming in NumPyro}",
  journal       = {arXiv e-prints},
  year          = 2019,
  month         = dec,
  eid           = {arXiv:1912.11554},
  pages         = {arXiv:1912.11554},
  doi           = {10.48550/arXiv.1912.11554},
  archiveprefix = {arXiv},
  eprint        = {1912.11554},
  primaryclass  = {stat.ML},
  adsurl        = {https://ui.adsabs.harvard.edu/abs/2019arXiv191211554P}
}

@article{bingham2019pyro,
  author        = {Eli Bingham and Jonathan P. Chen and Martin Jankowiak and Fritz Obermeyer and Neeraj Pradhan and Theofanis Karaletsos and Rohit Singh and Paul A. Szerlip and Paul Horsfall and Noah D. Goodman},
  title         = {Pyro: Deep Universal Probabilistic Programming},
  journal       = {J. Mach. Learn. Res.},
  volume        = {20},
  pages         = {28:1--28:6},
  year          = {2019},
  url           = {http://jmlr.org/papers/v20/18-403.html}
}

@article{software-citation-station-paper,
  author        = {{Wagg}, Tom and {Broekgaarden}, Floor S.},
  title         = "{Streamlining and standardizing software citations with The Software Citation Station}",
  journal       = {arXiv e-prints},
  year          = 2024,
  month         = jun,
  eid           = {arXiv:2406.04405},
  pages         = {arXiv:2406.04405},
  archiveprefix = {arXiv},
  eprint        = {2406.04405},
  primaryclass  = {astro-ph.IM},
  adsurl        = {https://ui.adsabs.harvard.edu/abs/2024arXiv240604405W}
}

@software{software-citation-station-zenodo,
  author       = {Tom Wagg and
                  Floor Broekgaarden and
                  Phil Van-Lane and
                  Kai Wu and
                  Kayhan Gültekin},
  title        = {TomWagg/software-citation-station: v1.4},
  month        = nov,
  year         = 2025,
  publisher    = {Zenodo},
  version      = {v1.4},
  doi          = {10.5281/zenodo.17654855},
  url          = {https://doi.org/10.5281/zenodo.17654855},
  swhid        = {swh:1:dir:9a009430037c791424a572f542e9a5d5c1fb44ff
                   ;origin=https://doi.org/10.5281/zenodo.13225526;vi
                   sit=swh:1:snp:ef11f058d718d691f0661c9445c2251328cb
                   ac95;anchor=swh:1:rel:84cde4e532032537b7f014c4467c
                   102430a53fa2;path=TomWagg-software-citation-
                   station-61a588a
                  },
}

@ARTICLE{Cabrera-Gadea:2026,
       author = {{Cabrera-Gadea}, Mauro and {Horta}, Danny and {Mateu}, Cecilia and {Tasso}, Emma and {Price-Whelan}, Adrian M.},
        title = "{The chemical side of the Milky Way warp. The [α/M] abundance from Red Giant Branch stars in Gaia DR3 reveal the warped distribution of the disc.}",
         year = {in prep.},
        month = dec,
archivePrefix = {arXiv},
       eprint = {xxxx.xxxxx},
 primaryClass = {astro-ph.GA},
}
\bibliographystyle{aasjournal}

\end{document}